\documentclass[10pt,
aps, 
reprint, 
onecolumn,
superscriptaddress, 
notitlepage,
amsmath,amssymb
]{revtex4-2}
\usepackage{siunitx} 
\usepackage[paperwidth=210mm,paperheight=297mm, hmargin=2.85cm,vmargin=2cm]{geometry}

\usepackage{caption}
\usepackage{hyperref}
\usepackage{rotating}
\usepackage{scrlayer-scrpage}
\usepackage{braket}
\usepackage{amsmath, mathtools}
\usepackage{amssymb}
\usepackage{tabularx}
\usepackage{bbm}
\usepackage[ngerman,english]{babel}
\usepackage{float}
\usepackage[table]{xcolor}
\usepackage{csquotes}
\usepackage[T1]{fontenc}
\usepackage{lmodern}
\usepackage{grffile} 
\usepackage{wrapfig}
\usepackage{graphicx}
\usepackage{subcaption}                  

\usepackage{mdframed}
\usepackage[printonlyused, withpage]{acronym}
\usepackage[sort&compress]{natbib}
\usepackage{overpic}
\usepackage{multirow}

\newcommand{\ts}{t_\mathrm{s}}

\newcommand{\rmax}{{r}_\mathrm{max}} 
\newcommand{\rmaxt}{\tilde{r}_\mathrm{max}} 
\newcommand{\dtm}{\Delta t_\mathrm{max}} 
\newcommand{\oav}{\langle o \rangle}

\newcommand{\rb}{{r}_\mathrm{best}}
\newcommand{\rbt}{\tilde{r}_\mathrm{best}}
\newcommand{\rglob}{\tilde{r}_\mathrm{glob}}
\newcommand{\rloc}{\tilde{r}_\mathrm{loc}}
\newcommand{\lav}{\langle\ell\rangle}

\newcommand{\tdriv}{t_\mathrm{driv}}

\newcommand{\twait}{t_\mathrm{wait}}

\newcommand{\dwalk}{d_\mathrm{walk}}
\newcommand{\twalk}{t_\mathrm{walk}}

\newcommand{\lde}{\Delta L_\mathrm{d}}
\newcommand{\lo}{\Delta L_\mathrm{o}}

\newcommand{\dmax}{d_\mathrm{max}}

\newcommand{\vp}{v_\mathrm{u}}
\newcommand{\vv}{v_\mathrm{b}}

\newcommand{\rt}{\tilde{r}}

\newcommand{\tind}{t_\mathrm{car}}

\usepackage{tikz}
\newcommand{\sfl}[2]{\tikz\node[inner sep=0pt,label={[anchor=north west]north west:\textbf{#1}}] {#2};}

\usepackage{float}
\floatstyle{boxed} 
\restylefloat*{figure}
\restylefloat*{table}

\begin{document}

\title{Taming Travel Time Fluctuations through Adaptive Stop Pooling}

\author{Charlotte Lotze}
\affiliation{Chair of Network Dynamics, Institute of Theoretical Physics and Center for Advancing Electronics Dresden (cfaed), TUD Dresden University of Technology, 01062 Dresden, Germany}
\email{charlotte.lotze@tu-dresden.de}

\author{Philip Marszal}
\affiliation{Chair of Network Dynamics, Institute of Theoretical Physics and Center for Advancing Electronics Dresden (cfaed), TUD Dresden University of Technology, 01062 Dresden, Germany} 

\author{Malte Schröder} 
\affiliation{Chair of Network Dynamics, Institute of Theoretical Physics and Center for Advancing Electronics Dresden (cfaed), TUD Dresden University of Technology, 01062 Dresden, Germany}

\author{Marc Timme}
\affiliation{Chair of Network Dynamics, Institute of Theoretical Physics and Center for Advancing Electronics Dresden (cfaed), TUD Dresden University of Technology, 01062 Dresden, Germany}
\affiliation{Lakeside Labs, 9020 Klagenfurt am W{\"o}rthersee, Austria}
\email{marc.timme@tu-dresden.de}

\begin{abstract}
Ride sharing services combine trips of multiple users in the same vehicle and may provide more sustainable transport than private cars. As mobility demand varies during the day, the travel times experienced by passengers may substantially vary as well, making the service quality unreliable.
We show through model simulations that such travel time fluctuations may be drastically reduced by stop pooling. Having users walk to meet at joint locations for pick-up or drop-off allows buses to travel more direct routes by avoiding frequent door-to-door detours, especially during high demand.
We in particular propose \textit{adaptive} stop pooling by adjusting the maximum walking distance to the temporally and spatially varying demand.
The results highlight that adaptive stop pooling may substantially reduce travel time fluctuations while even improving the average travel time of ride sharing services, especially for high demand. Such quality improvements may in turn increase the acceptance and adoption of ride sharing services.
\end{abstract}
\maketitle

\section{Introduction}

\begin{figure}[b]
   			\begin{subfigure}[t]{7.5 cm}
   				\begin{center}\sfl{a}{
   						\includegraphics[width=7cm]{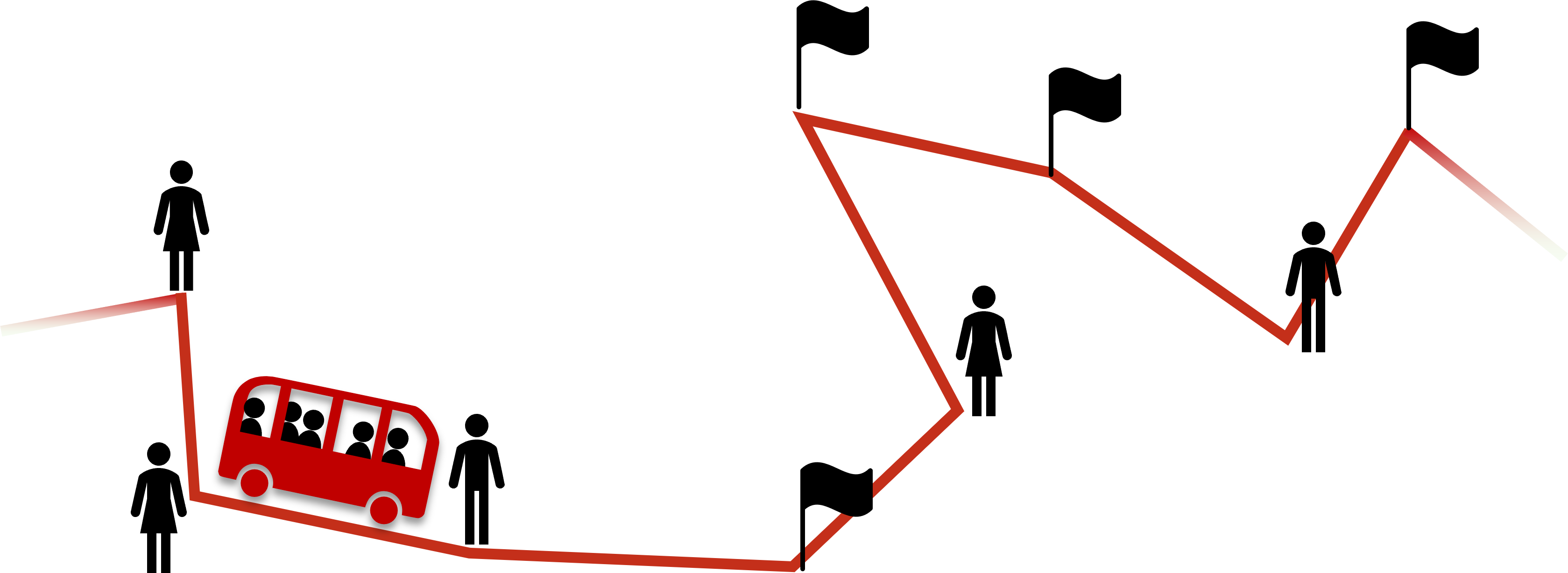}}
   				\end{center}
   			\end{subfigure}   
   			\hfill
   			\begin{subfigure}[t]{7.5 cm}
   				\begin{center}
   					\sfl{b}{
   						\includegraphics[width=7cm]{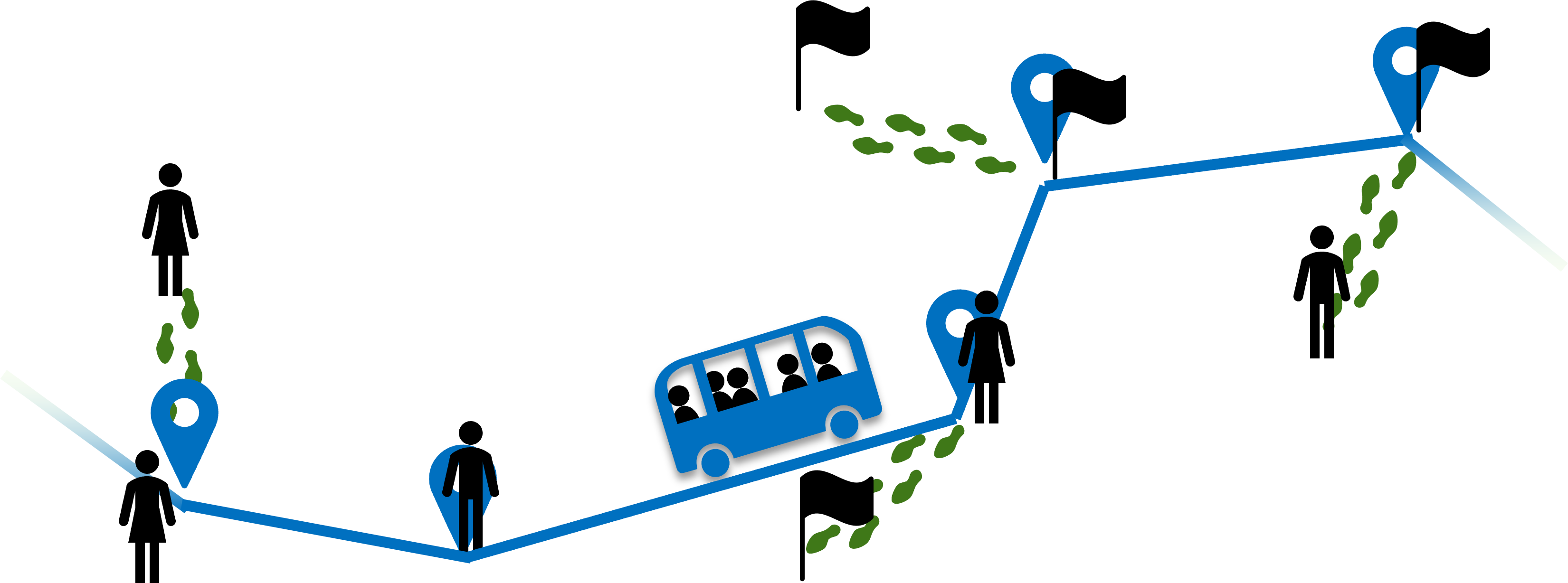}}
   				\end{center}
   			\end{subfigure} 
   			\caption[Ride sharing buses might save detours when users accept short walks.]{\small  
                \textbf{Ride sharing buses might reduce detours when users accept short walks.}
   				(a) Door-to-door ride sharing routes directly visiting every stop contain large detours when many users share one bus. 
                (b) With stop pooling, buses might serve multiple users at shared stops. Some users walk a short distance to a nearby stop such that the bus route is more direct.
            }
   			\label{fig:transport_modes}    
   		\end{figure}

In ride sharing systems, on-demand shuttles simultaneously transport multiple users in the same vehicle. Ride sharing services thus require fewer vehicles and may be ecologically and economically more sustainable than transport by private cars  \cite{Beckmann.2013,Lotze.2022,Santi.2014,Tachet.2017, Muhle.2023, Lotze.2023b}. 
Yet users incur detours and travel longer than in private cars, especially if many users share the same vehicle, see Fig.~\ref{fig:transport_modes}a for an illustration. These detours may be reduced by \textit{stop pooling} \cite{Mounesan.2021, Fielbaum.2021b, Lotze.2022}, where some users walk a short distance to a neighboring stop such that two or more users are served together at one pooled stop (Fig.\ref{fig:transport_modes}b). With stop pooling the ride sharing vehicles, often (mini)buses, take routes that are more direct, avoiding detours and thereby improving both user experience and service efficiency. In particular, the average total travel time of users may decrease despite additional walking times \cite{Lotze.2022}.

However, ride sharing is challenged by demand fluctuations over the day \cite{Wang.2019,Wilkes.2021} -- as demand data for ride hailing services in  Manhattan (New York City, USA) exemplify \cite{CityofNewYork.2016,Liu.2019}. Higher demand often provokes higher travel times for users due to additional detours service vehicles need to make.
Ride sharing providers might respond 
by adapting their system to maintain roughly constant travel times. One example response may be to adapt the fleet size  \cite{Altshuler.2017, Vazifeh.2018,Wilkes.2021}, but an increase of the fleet size requires additional vehicles and drivers to be available, often not an economically viable option.
Instead, we here propose to adaptively pool stops to reduce travel time fluctuations. The potential of stop pooling to reduce the travel time is known for steady state operation with constant demand \cite{Lotze.2022}. This potential is higher at higher demand where it is easier to combine close-by stops \cite{Lotze.2023}.
However, the effects of stop pooling on the collective dynamics of ride sharing systems under varying demand has yet to be understood. 

In this article, we demonstrate that stop pooling may reduce travel time fluctuations at a constant fleet size. 
Typically, the travel time increases with the demand. Stop pooling absorbs parts of the increase when users walk further at higher demand. For this purpose, we suggest two simple procedures to adapt the maximum walking distance, i.e.~the maximum distance a user may be asked to walk, to the temporally and spatially varying demand. Both procedures significantly reduce the fluctuations of the travel time without any adaption of the fleet size.

\section{Methods}

To analyze the qualitative effects of stop pooling on the collective dynamics of ride sharing and in particular how stop pooling changes fluctuations in the travel time, we introduce an event-based model (details in Supplementary Note 1 and 2 and in \cite{Lotze.2023}) with three different events: (i) users request trips from an origin to a destination, (ii) ride sharing buses pickup users and (iii) deliver them. 
New users request trips while buses serve other users. Finding the bus routes is thus an online-optimization problem \cite{Agatz.2012,Muhle.2023}. A simple ride sharing algorithm assigns the users to buses (details in Supplementary Note 1.C). The algorithm minimizes the total distance driven by all buses while distributing users over all buses. The algorithm includes rebalancing \cite{Fielbaum.2021b, AlonsoMora.2017,Lu.2021,Wen.2017}, i.e.~sending back idling buses towards a central location to avoid that empty buses get stuck in regions of low demand (details in Supplementary Note 1.C.4).

With stop pooling, users might walk at most a maximum walking distance $r$ per stop with user walk velocity $\vp$. When a user could walk to other user stops within time $\rt=r/\vp$ and (if walking from origin to pickup) arrives at the stop before the bus, both stops are pooled. In this way, the algorithm finds locations of the pooled stops dynamically based on the current demand. 
If users request a very short trip with trip length $\ell\le2r$ they walk their complete trip  (details in Supplementary Note 1.B, cf.~\cite{Wang.2022}). All in all, the system saves at least one stop compared to standard ride sharing services if a user walks. 

We include street network and request data from an example city into the model for interpretable results (details in Supplementary Note 2). 
We take origins, destinations and request time from a data set of taxi cabs in Manhattan (New York, USA) as in 2016 \cite{CityofNewYork.2016} (Supplementary Note 2.E). On a typical day, these requests are served by 7000 to 8000 taxis and even the minimum required taxi fleet would contain almost 6000 taxis \cite{Vazifeh.2018}.
We simulate a ride sharing service with a much smaller fleet size $B=1500$.
Buses drive along a directed street network of Manhattan \cite{OpenStreetMapcontributors.2023, Boeing.2017} that is fine-grained analogously to \cite{Manik.2020} (details in Supplementary Note 2.D). 
Buses drive with a mean-field velocity $\vv=\SI{12}{\kilo\metre\per\hour}$ - a typical average velocity for driving in Manhattan \cite{NewYorkCityDepartmentofTransportation.June2018}. Decelerating, serving users and accelerating is represented by a constant penalty of $\SI{10}{\second}$ per stop  (common for public transport \cite{Dueker.2004}), independent of the number of users entering or exiting the bus at that stop (details in Supplementary Note 1.C.3).
Users walk on an undirected user network that contains the same nodes as the bus street network, but allows users to walk into both directions and to cross the street with a penalty of $\SI{10}{\metre}$ at additional nodes (see Supplementary Note 2.D). Users walk with user velocity $\vp=\SI{4}{\kilo\metre\per\hour}$. The maximum walk distance $r$ is thus equivalent to a walk time limit $\rt=r/\vp$.

We conduct simulations with steady state dynamics by randomly sampling requests from all data and choosing request times from a Poisson distribution with constant request rate $\lambda$. 
We conduct simulations with the actually served taxi trips with varying request rates $\lambda(\tau)$ as resulting from the data of individual requests averaged across ten-minute intervals (see Fig.~\ref{fig:no_stop_pooling}a) on one example day between 6:00 - 24:00 (details in Supplementary Note 2.E). Besides, the spatial demand pattern deviates in the morning from the evening \cite{Lotze.2023,  Liu.2019}.
We observe the travel time $t$ that consists of wait time $\twait$, drive time $\tdriv$ and walk time $\twalk$
(details in Supplementary Note 3.F.1). For steady state simulations, we average all observables over all users. For fluctuating demand, we average all observables within intervals of one hour. Users contribute to that interval in which they pose their request. In all figures, the averages per time interval are represented by one data point in the center of the interval.
We evaluate only times after one hour of simulation time, because simulations start with empty buses randomly distributed over all nodes. In the first hour of simulation time, buses accumulate a planned job list and distribute according to the requests. 
We calculate the request rate $\lambda$ for each interval from the number of requests divided by the length of the time interval. Except for the varying demand, input parameters (e.g.~fleet size, velocities) are constant over the simulation.

\section{Results}

\subsection{No Stop Pooling}
\label{sec:unreliable}

\begin{figure}[t]
	\begin{subfigure}[t]{5cm}
		\begin{center}
			\sfl{a} {\includegraphics{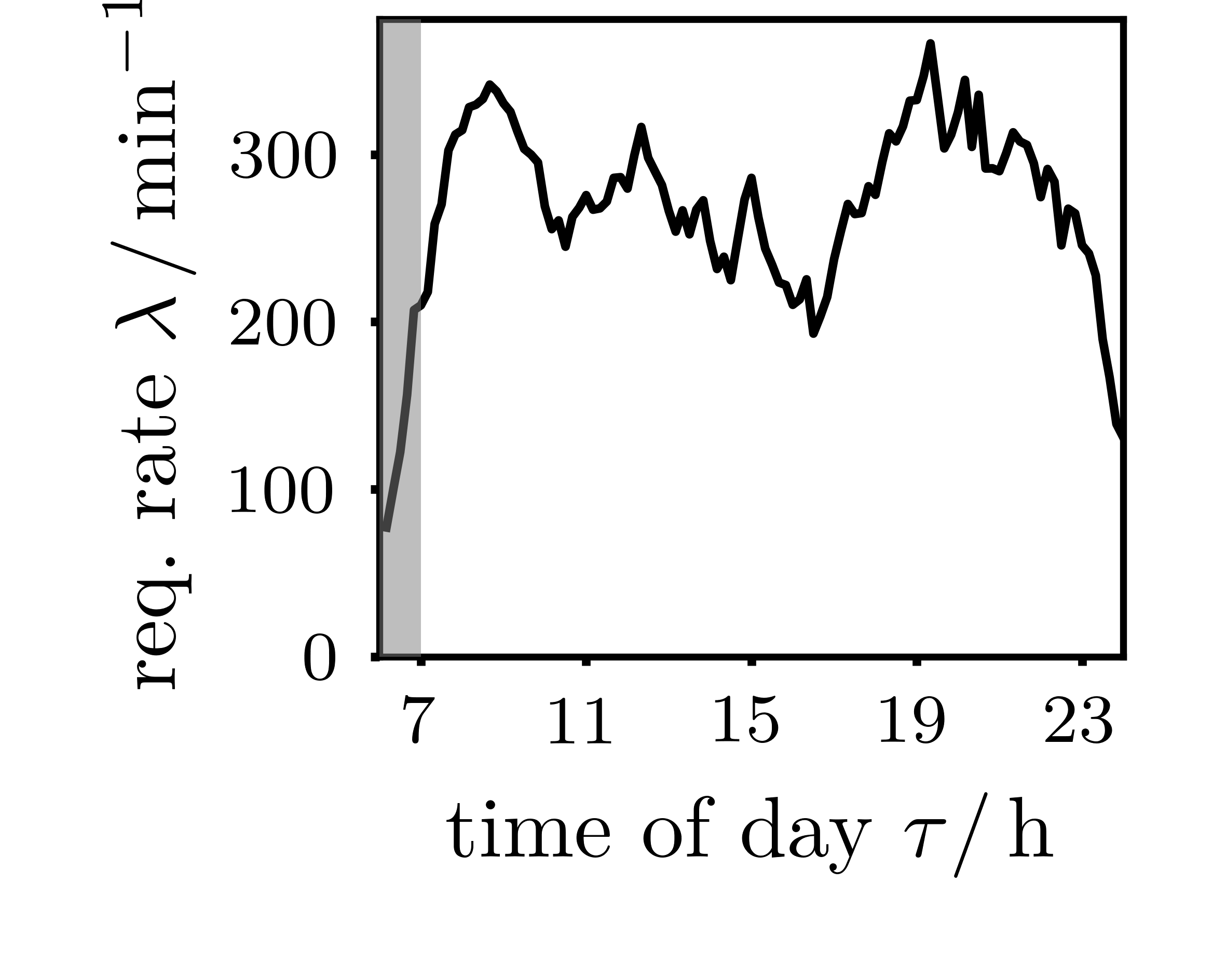}}
		\end{center}
	\end{subfigure}	\hfill
	\begin{subfigure}[t]{5cm}
		\begin{center}
			\sfl{b} {\includegraphics{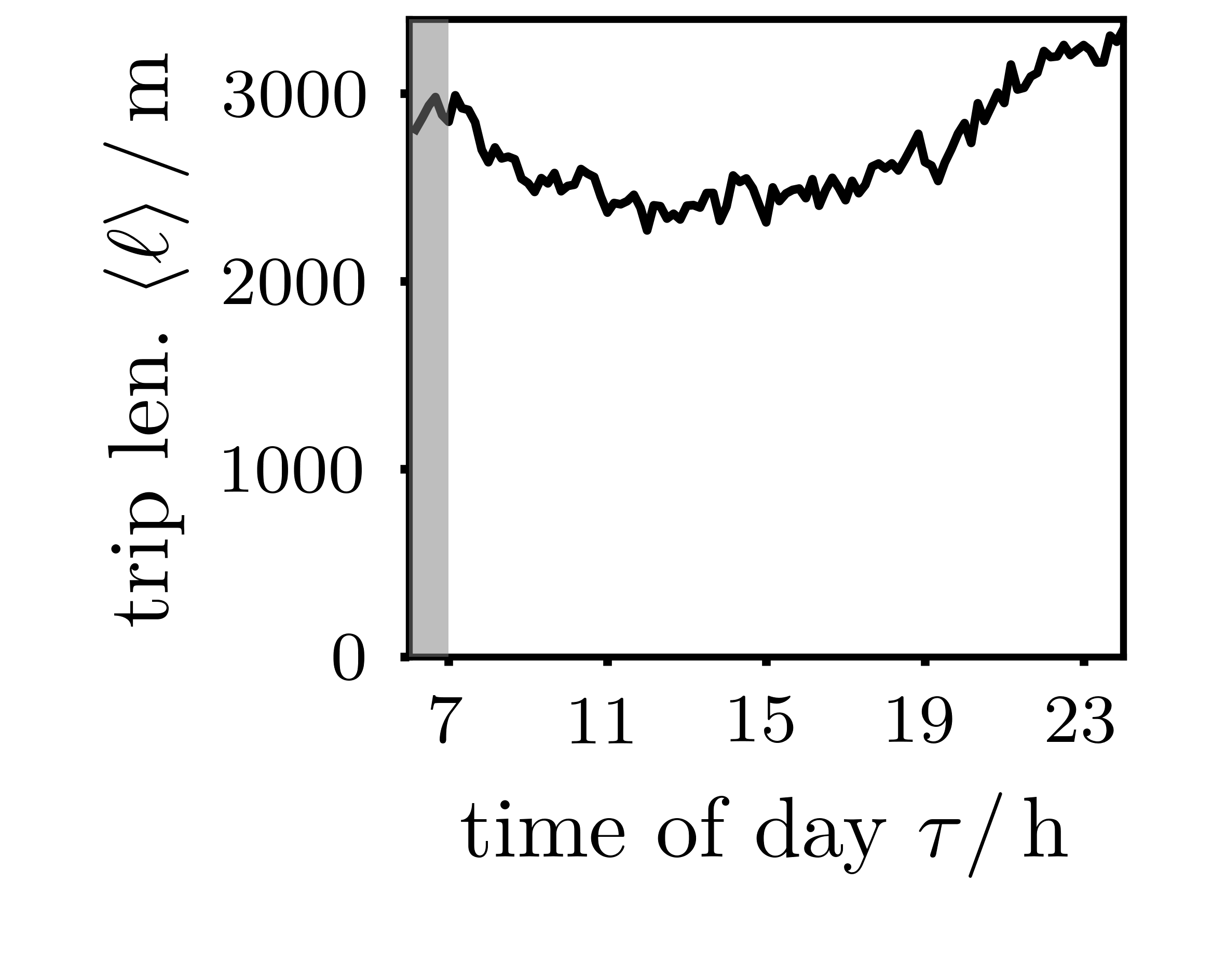}}
		\end{center}
	\end{subfigure}	\hfill
	\begin{subfigure}[t]{5cm}
		\sfl{c}{
			\includegraphics{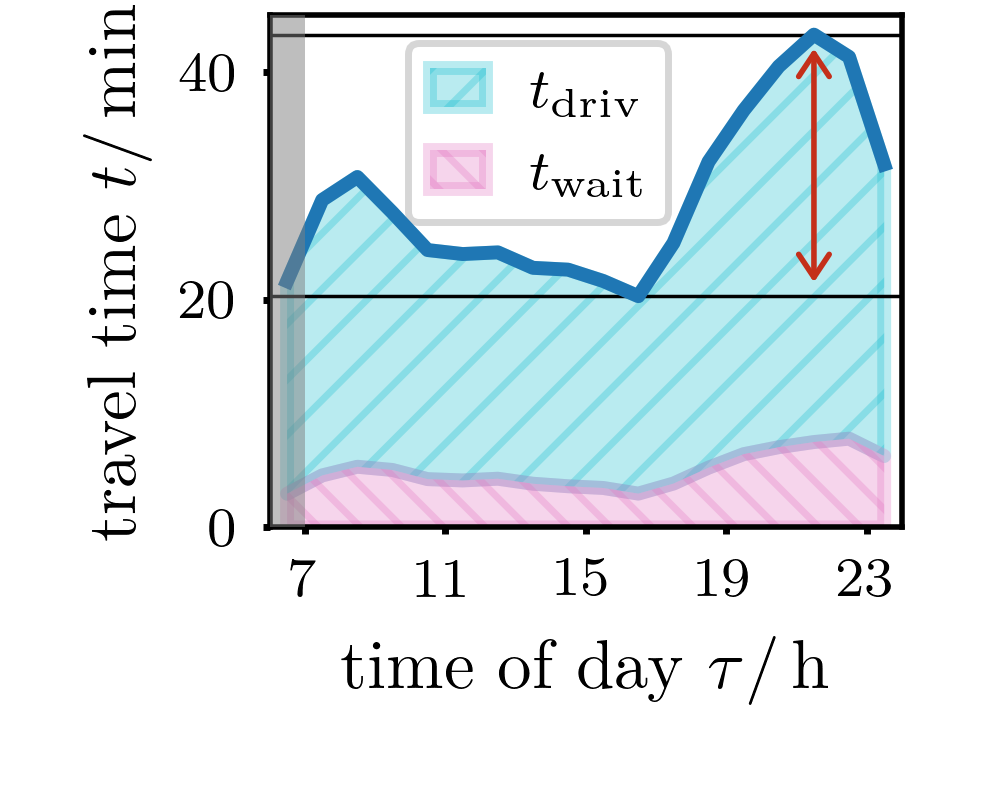}}
	\end{subfigure}
	\vspace{-0.6 cm}
	\caption{\small  \textbf{The travel time in standard ride sharing strongly varies with demand.}
	\small 
	(a,b) Trip demand, i.e. the total trip length requested jointly by all users, varies strongly with the time of day $\tau$. Panel (a) illustrates fluctuating request rate $\lambda$ and (b) varying average trip lengths $\lav$ for one day of requested trips in Manhattan, New York City (based on taxi requests, see Methods for details). (c) Consequently, the travel time $t$ for users fluctuates as well, more than doubling from the minimum at 16:30 to the maximum at 21:30 (red arrow). This variability originates mainly from fluctuations in the average user drive time $\tdriv$ (blue shaded) that is much larger than the wait time $\twait$ (pink shaded).
    $\lambda$ and $\lav$ averaged across ten-minute intervals; travel times $t$ averaged across one-hour intervals, trips requested in the interval. 
    }
\label{fig:no_stop_pooling}
\end{figure}

First, let us consider the collective dynamics and fluctuations in standard ride sharing without stop pooling, $\rt=0$. 
As the demand fluctuates over the day, the user travel time fluctuates as well (Fig.~\ref{fig:no_stop_pooling}). 
For the example shown, the request rate $\lambda$ varies 
with mean $\SI{275}{\per\minute}$ and standard deviation $\SI{44}{\per\minute}$ ($16\%$ of mean) between 7:00 and 24:00 (Fig.~\ref{fig:no_stop_pooling}a).  At the same time, the average trip length $\lav$ of the users varies with mean $\SI{2675}{\metre}$ and standard deviation $\SI{289}{\metre}$ ($11\%$ of mean)  (Fig.~\ref{fig:no_stop_pooling}b).
We consider demand as requested trip length characterized by request rate $\lambda$ and average trip length $\ell$.
The demand is particularly small around 16:00 (neglecting the boundaries) with minimum request rate $\lambda=\SI{193}{\per\minute}$ and more than 50\% higher around 20:00 with maximum request rate $\lambda=\SI{366}{\per\minute}$.

The resulting travel time $t$ with standard ride sharing fluctuates even more strongly, with mean $\SI{29.3}{\minute}$ and standard deviation $\SI{7.3}{\minute}$ ($25\%$ of mean). In the example,  users who request a ride between  21:00 and 22:00 travel on average twice as long as users who requests a ride between 16:00 and 17:00 (Tab.~\ref{tab:var}, Fig.~\ref{fig:no_stop_pooling}c, comparable distributions of individual travel times in both intervals, see Supplementary Note 3.G, Fig.~S10). 
Such high fluctuations make the travel time unreliable for ride sharing users.

\subsection{Static Stop Pooling}
\label{sec:more_reliable}

\begin{figure}[b!]  
	\begin{subfigure}[t]{4.8cm}
		\begin{center}
		\sfl{a}{
			\includegraphics{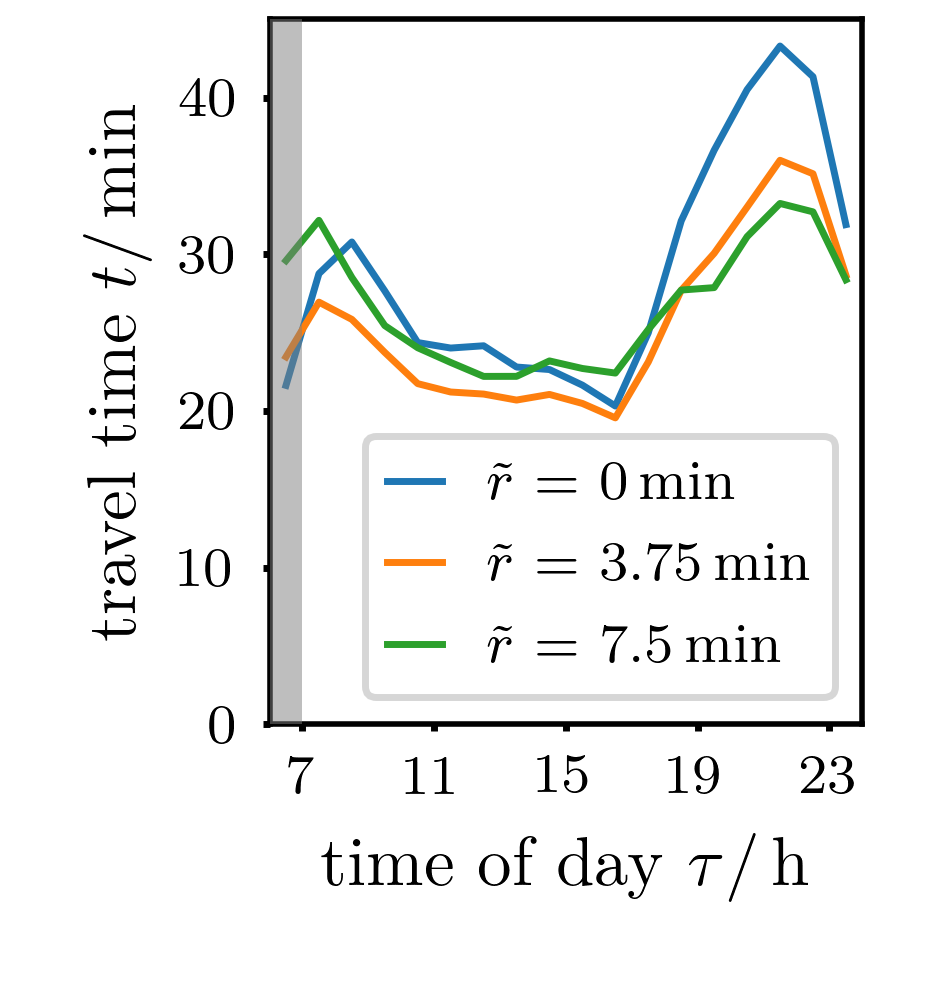}}
		\end{center}
	\end{subfigure}	
	\begin{subfigure}[t]{4.8cm}
		\sfl{b}{
			\includegraphics{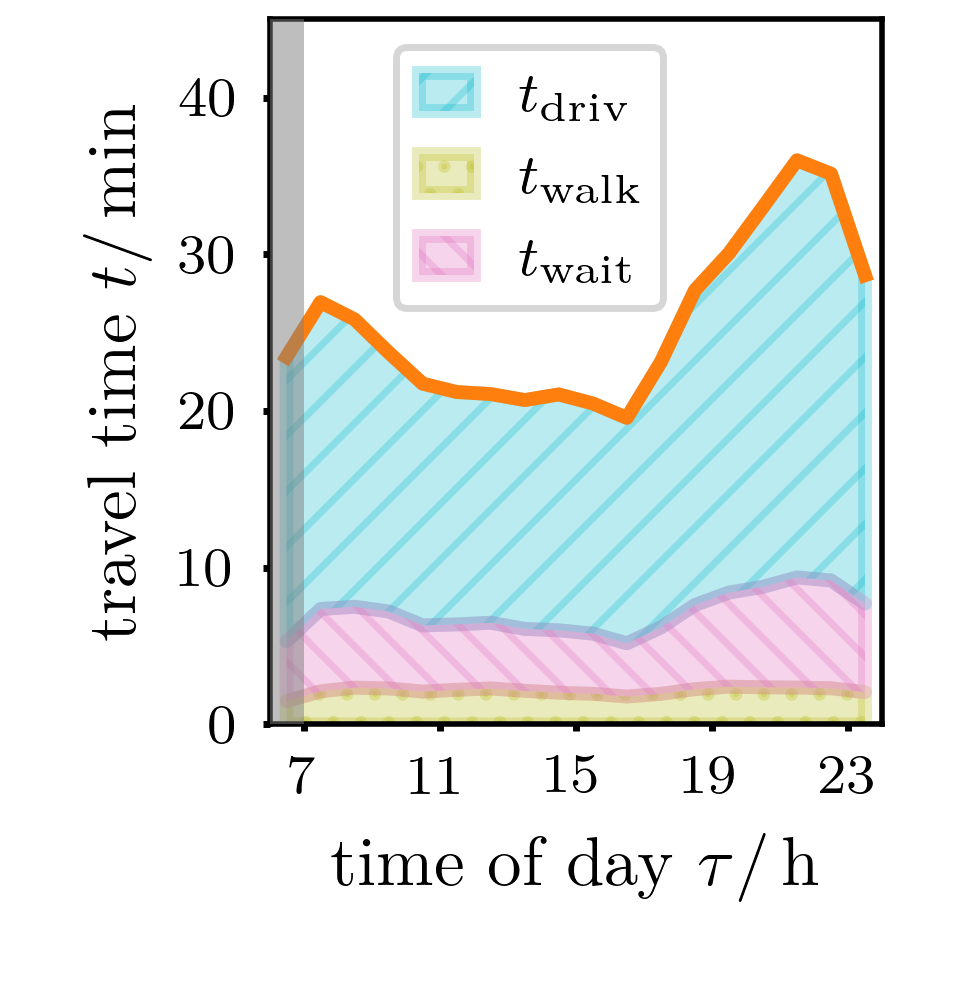}}
	\end{subfigure}
	\begin{subfigure}[t]{5.5cm}
		\sfl{c}{
			\includegraphics{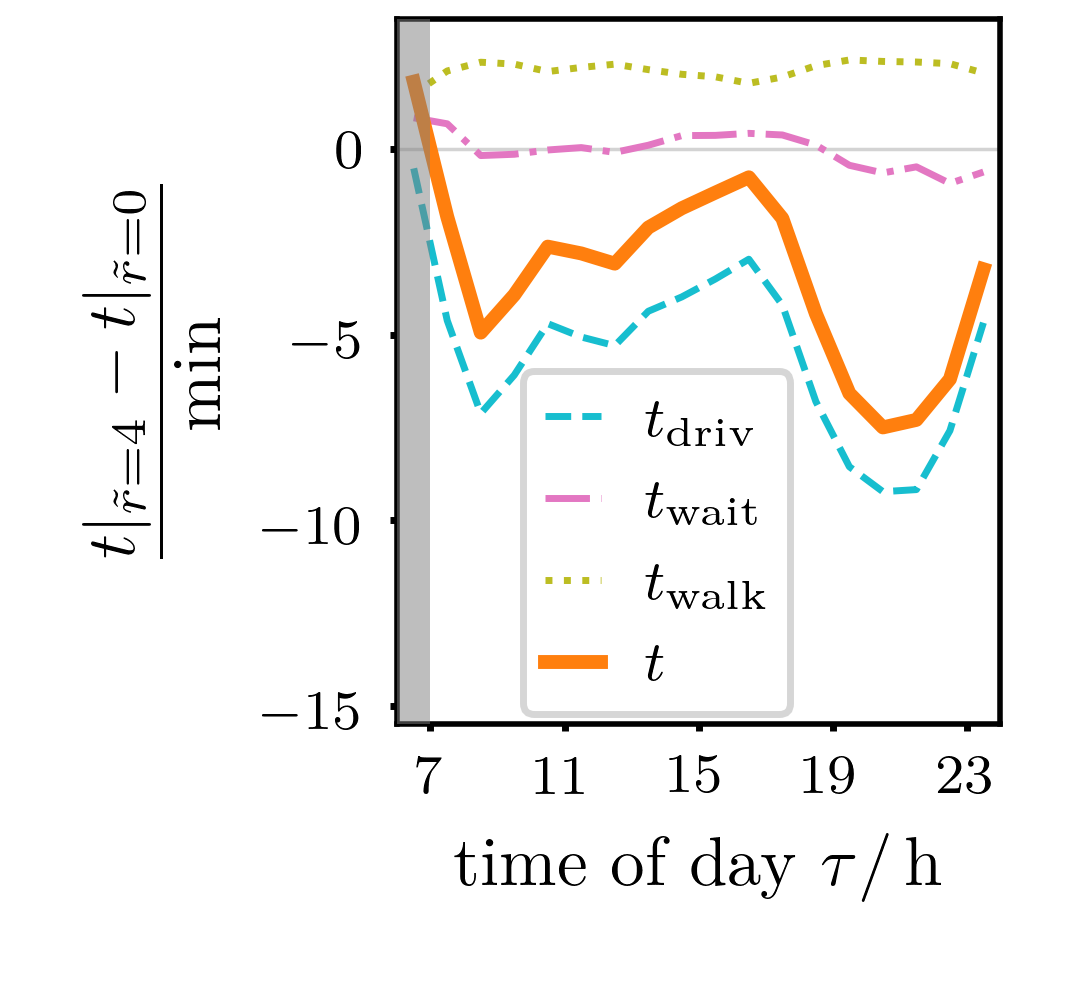}}
	\end{subfigure}
	\vspace{-0.6cm}
	\caption[A fixed maximum walking distance does not use full potential at fluctuating demand.]{\small  \textbf{A fixed maximum walking distance does not use full potential at fluctuating demand.}	
	\small 
	 (a) The travel time $t$ reduces with stop pooling compared to standard ride sharing (blue line), but which walk limit $\rt$ yields the shortest travel time changes across the day? Until 18:00, an intermediate walk limit $\rt=\SI{3.75}{\minute}$ yields best travel time $t$ (orange). During the evening peak, a high walk limit $\rt=\SI{7.5}{\minute}$ yields the best travel time $t$ (green).
    (b) A fixed walk limit $\rt=\SI{3.75}{\minute}$ adds an almost constant walk time $\twalk$ throughout the day (light green dotted). 
    (c) The drive time $\tdriv$ that fluctuates strongly with the demand (cf.~Fig.~\ref{fig:no_stop_pooling}) reduces with stop pooling [while the wait time $\twait$ is roughly constant] such that the overall travel time $t$ reduces (orange solid line) and fluctuates less compared to standard ride sharing (compare panel a, $\rt=\SI{3.75}{\minute}$).}
	\label{fig:static_stop_pooling}
\end{figure}

Static stop pooling, i.e.~stop pooling with fixed walk time limits, already influences the collective dynamics of ride sharing. We thus compare results for different maximum walking distances $\rt>0$ with those for standard ride sharing, $\rt=0$ (no stop pooling).
We find that with stop pooling, the travel time $t$ may fluctuate less (Fig.~\ref{fig:static_stop_pooling}), because (i) stop pooling may reduce the travel time $t$ in general, compare also \cite{Lotze.2022}, and (ii) the reduction is typically higher the higher $t$ at $\rt=0$. 

When users walk at most $\rt=\SI{3.75}{\minute}$ per stop (intermediate walk limit), the travel time $t$ reduces compared to standard ride sharing at most times $\tau$ of the day (Fig.~\ref{fig:static_stop_pooling}a) and also on average (Tab.~\ref{tab:var}). This reduction might seem counterintuitive, because stop pooling requires additional times $\twalk$ for walking (Fig.~\ref{fig:static_stop_pooling}b). However, the reduction is explained by a trade-off of additional walk time and reduced drive time: With a fixed walk limit $\rt$, the walk time $\twalk$ is roughly constant despite fluctuating demand (Fig.~\ref{fig:static_stop_pooling}c). Indeed, users walk on average less than the maximum $2\rt$ ($\rt$ at origin and destination) and even less than $\rt$ (Tab.~\ref{tab:var}). 
When some users walk to pooled stops, ride sharing buses drive to fewer stops and reduce some detours such that the bus routes become more strongly directional. Users profit from such more direct bus routes due to shorter average drive times $\tdriv$ (Fig.~\ref{fig:static_stop_pooling}b). A sufficiently large reduction of $\tdriv$ overcompensates additional walk times (Fig.~\ref{fig:static_stop_pooling}c). 
The reduction is higher at high demand, because many users share one bus. With many users, bus routes in standard ride sharing contain many small detours that stop pooling might save.  There is a high potential to reduce $\tdriv$. In the example, the travel time reduction is particularly high in the demand peak in the evening and rather small in the demand minimum around 16:00 (Fig.~\ref{fig:static_stop_pooling}c). In consequence, the travel time $t$ varies less with stop pooling (standard deviation $\SI{5.4}{\minute}$, $21\%$ of mean at $\rt=\SI{3.75}{\minute}$) than with standard ride sharing, $\rt=0$ (Tab.~\ref{tab:var}). 

Moreover, the results demonstrate that a constant maximum walking distance does not use the full potential of stop pooling at fluctuating demand, because different maximum walking distances yield the shortest travel time $t$ at different times of day $\tau$. At low demand, small reductions in $\tdriv$ buffer only a short walk time $\twalk$. At high demand, a high reduction in $\tdriv$ buffers much longer walk times $\twalk$. Longer walks save more stops and are thus more efficient in reducing $t$.
In the example, an intermediate walk limit $\rt=\SI{3.75}{\minute}$ yields the shortest travel time $t$ before 18:00 while a high walk limit $\SI{7.5}{\minute}$ yields the shortest travel time $t$ after 18:00 (Fig.~\ref{fig:static_stop_pooling}a).
Can we adapt the maximum walking distance to the instantaneous demand to increase service efficiency?

\begin{table}[t]
    \caption{Comparison of four different settings in walk time limit $\rt$, travel time $t$ and walk time $\twalk$ averaged over all time intervals (second column) and for two example time intervals (third and fourth column).}
    \label{tab:var}
    \centering
    \begin{tabular}{l|rrr|rrr|rrr}
        &\multicolumn{3}{c|}{Overall Average}&\multicolumn{3}{c|}{16:00 - 17:00}&\multicolumn{3}{c}{21:00 - 22:00}     \\[3pt]
        &  $\dfrac{\rt}{\min}$&$\dfrac{t}{\min}$ &   $\dfrac{\twalk}{\min}$
        &  $\dfrac{\rt}{\min}$&$\dfrac{t}{\min}$ &   $\dfrac{\twalk}{\min}$
        &  $\dfrac{\rt}{\min}$&$\dfrac{t}{\min}$ &   $\dfrac{\twalk}{\min}$
       \\[7pt]\hline
         No stop pooling ($\rt=\SI{0}{\minute}$)&                                           
        $\SI{0.0}{}$ &$\SI{29.3}{}$&$\SI{0.0}{}$&
        $\SI{0.0}{}$ &   $\SI{20.3}{}$&$\SI{0.0}{}$ &          
       $\SI{0.0}{}$ &$\SI{45.3}{}$&$\SI{0.0}{}$
       \\
         Static stop pooling ($\rt=\SI{3.75}{\minute}$)  &
         $\SI{3.8}{}$& $\SI{25.7}{}$&$2.2$
        &$\SI{3.8}{}$&  $\SI{19.6}{}$&$1.8$ &
        $\SI{3.8}{}$& $\SI{36.0}{}$&$2.4$ 
        \\
         Time-adaptive stop pooling ($\rglob$)  & 
         $\SI{5.3}{}$& $\SI{25.2}{}$  & $\SI{3.9}{}$&  
         $\SI{3.0}{}$&$\SI{19.4}{}$ &   $\SI{1.3}{}$& 
         $\SI{7.4}{}$& $\SI{34.1}{}$  & $\SI{6.4}{}$ 
         \\
         Spatio-time-adaptive stop pooling ($\rloc$) & 
         $4.6$& $\SI{24.8}{}$  & $\SI{5.0}{}$ &   
         $2.0$&$\SI{18.6}{}$ &   $\SI{2.3}{}$& 
         $5.3$& $\SI{33.0}{}$  & $\SI{6.3}{}$
         \\
    \end{tabular}
\end{table}
\subsection{Adapting the maximum walking distances in time}

\begin{figure}[t!]  
 	\begin{subfigure}[t]{7.6cm}
 		\begin{center}\vspace{-0cm}\sfl{a}{\hspace{0.29cm}
 			\includegraphics{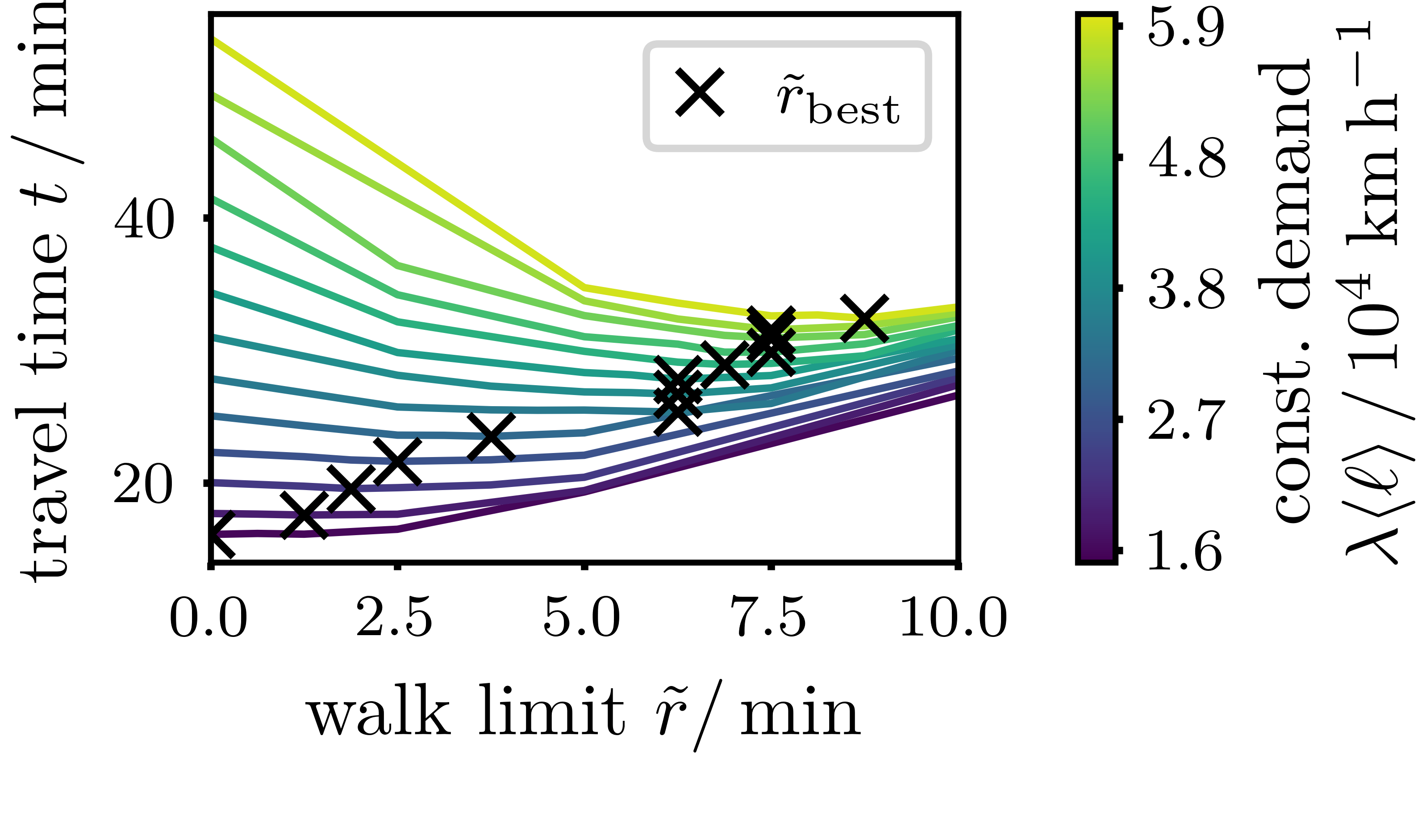}}
 		\end{center}
 	\end{subfigure}
 \begin{subfigure}[t]{7.5cm}
 \begin{center}\vspace{-0cm}\sfl{b}{\hspace{0.2cm}
 	\includegraphics{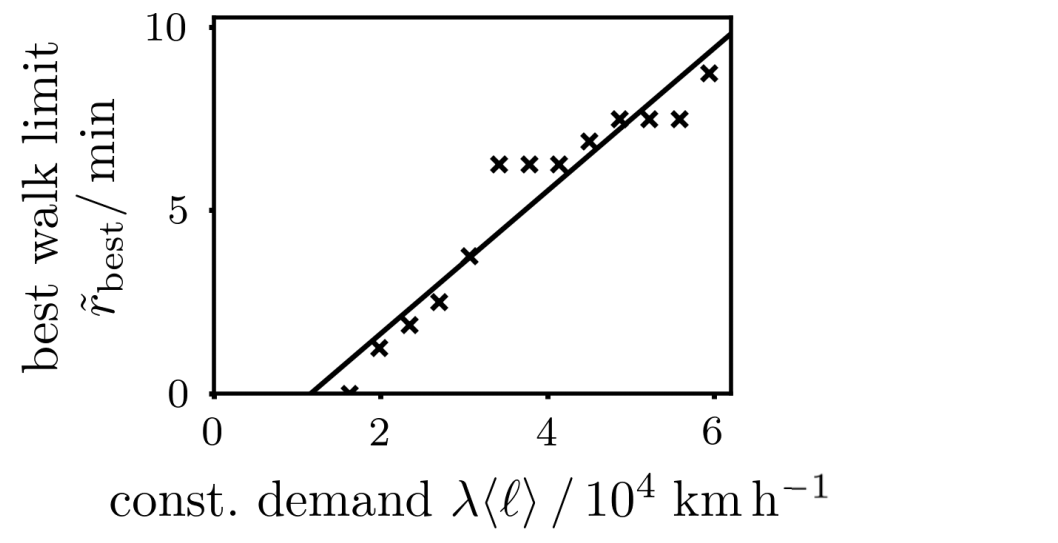}}
 \end{center}
\end{subfigure}
    	\begin{subfigure}[t]{7.6cm}
    		\begin{center}\vspace{-0cm}
    			\sfl{c}{\hspace{0.2cm}
    			\includegraphics{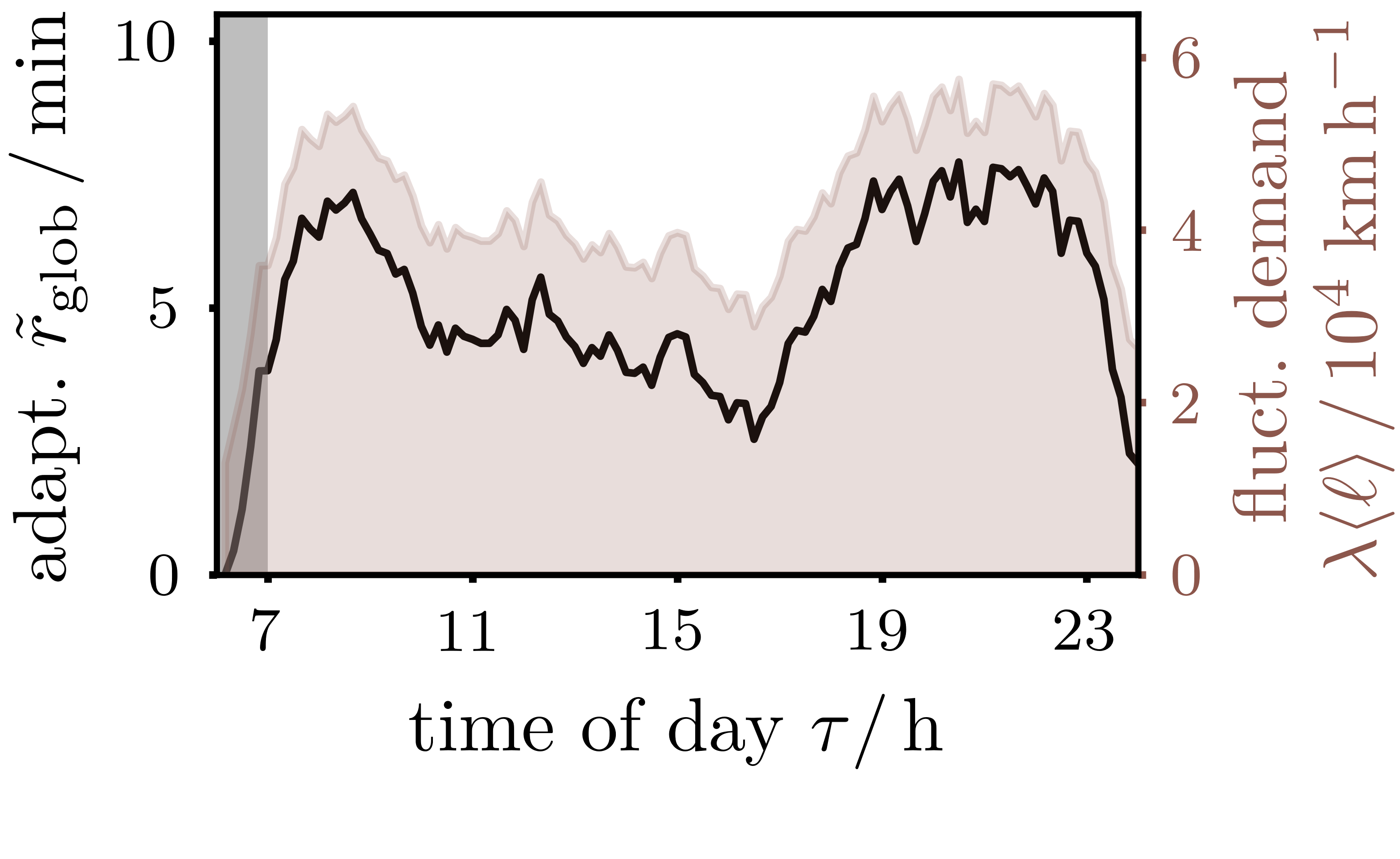}}
    		\end{center}
    	\end{subfigure}
    	\begin{subfigure}[t]{7.5cm}
    		\begin{center}
    			\vspace{-0cm}\sfl{d}{\hspace{0.2cm}
    			\includegraphics{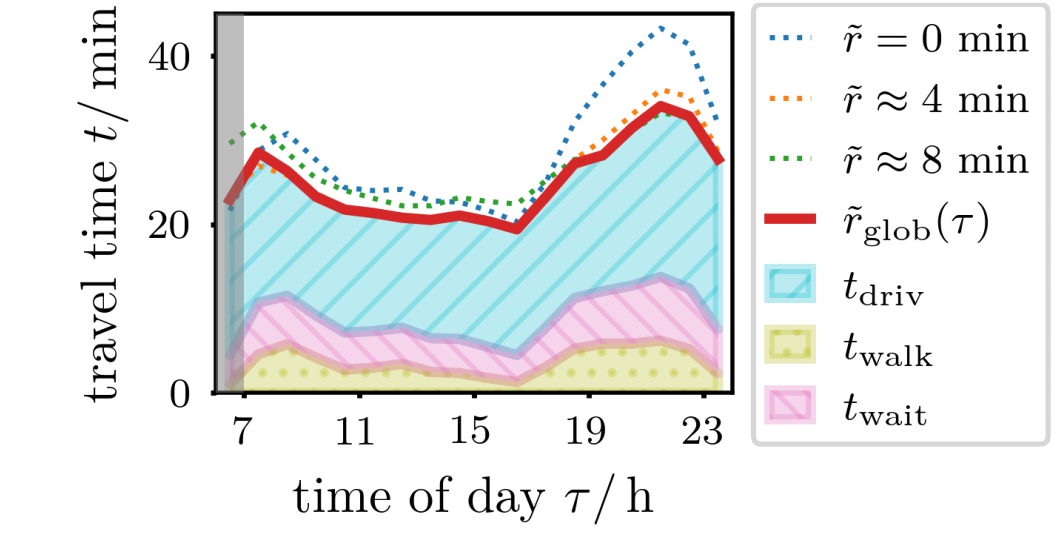}}
    		\end{center}
    	\end{subfigure}
    \vspace{-0.5cm} 
    	\caption[Temporally adapted maximum walk distance achieves consistent travel time reduction.]{\small              \textbf{Temporally adapted walk limit achieves consistent travel time reduction.}
            (a) With constant demand $\lambda\lav$, stop pooling reduces the travel time $t$ up to a best walk limit $\rbt$ (black crosses). 
            (b) This best walk limit $\rbt$ increases approximately linearly with the demand. 
            (c) Applying the linear fit to the fluctuating demand data (brown shaded area) yields a variable walk limit $\rglob(\tau)$ between $\SI{3.1}{\minute}$ and $\SI{7.4}{\minute}$ walk per stop (black line).
            (d) This time-adaptive stop pooling consistently achieves the shortest travel time $t$ at almost all times of the day (thick orange line).
    	}
    	\label{fig:adapt_r_in_time}
    \end{figure}

To study how temporally adapting the maximum walking distance to the instantaneous global demand changes the travel times, we first perform an analysis for steady states that reveals the best suitable maximum walking distance for any given, temporally fixed demand. 
We realize constant demand for the steady state analysis by sampling requests from the example data set. In our analysis, the demand is represented by the product $\lambda\lav$ of request rate $\lambda$ and average trip length $\lav$, i.e.~the total travel distance requested per unit time.
In the model, stop pooling reduces the travel time $t$ at constant demand $\lambda\lav$ up to some best walk limit $\rbt$ (Fig.~\ref{fig:adapt_r_in_time}a). A bisection method finds $\rbt$ for different settings with reduced computation effort (details in Supplementary Note 3.F.2). $\rbt$ increases with the demand $\lambda\lav$ (Fig.~\ref{fig:adapt_r_in_time}b), because more users per bus yield more small detours that determine the potential of reduced $\twait$ and $\tdriv$ to buffer additional walk time $\twalk$ (cf.~previous section). 
For the example setting, this increase roughly fits to a linear function,
\begin{equation}
	\rbt(\lambda\lav)=a\,\lambda\lav+b\,,
	\label{eq:rbest}
\end{equation}
with 
  ${a=[1.95\pm0.18]\,\times\,\si{\minute\hour\per\kilo\metre}}$, ${b=[-2.26\pm0.71]\,\times\,10^4\,\si{\minute}}$ and coefficient of determination \cite{Glantz.2016} $R^2=0.92$.

Using this steady-state analysis, we suggest a simple procedure to adapt the walk limit $\rt$ to the instantaneous demand: 
When a user requests a trip, the global demand at the request time defines their maximum walking distance. In the example, this global walk limit $\rglob$ reads
\begin{equation}
	\rglob(\tau_\mathrm{request})=a\,\lambda(\tau_\mathrm{request}) \,\lav(\tau_\mathrm{request})+b\,.
	\label{eq:rglob}
\end{equation}
for a user with request time $\tau_\mathrm{request}$.
This global walk limit $\rglob(\tau)$ varies over the day following the fluctuations of $\lambda\lav$ (Fig.~\ref{fig:adapt_r_in_time}c).
Again, users walk less than the walk limit $\rglob(\tau)$ on average (Tab.~\ref{tab:var}, Fig.~\ref{fig:adapt_r_in_time}c,d). 

The global walk limit $\rglob(\tau)$ yields the shortest travel times $t$ at almost all times of day $\tau$. Small deviations from the shortest travel time $t$ might result from the fluctuating spatial demand patterns, because the steady state analysis uses a constant mean-field demand pattern. In the example, the travel time has a smaller mean (Tab.~\ref{tab:var}) and fluctuates less (standard deviation $\SI{4.7}{\minute}$, $19\%$ of mean) than with standard ride sharing ($6\%$ less) or stop pooling with an intermediate fixed walk limit ($2\%$ less). 

Adaptive stop pooling efficiently reduces the travel time at fluctuating demand while simultaneously reducing the travel time fluctuations. For this result it is sufficient to adapt the maximum walking distance in time.
However, stop pooling is a local interaction compared to the size of a typical service region, because stops might only be pooled with nearby users. Thus, only the local demand around a user influences their efficient stop pooling setting. Typically, the local mobility demand does not only vary in time but also in space (like the taxi demand, cf.~Fig.~\ref{fig:adapt_r_in_spacetime}a and \cite{Liu.2019}). 
For this reason, let us consider spatio-temporal demand fluctuations for adapting the maximum walking distance instead of using the global time varying demand alone.

\subsection{Adapting the maximum walking distance in time and space}
\label{sec:ASP_local}
\begin{figure}[t!] 
	\begin{subfigure}[t]{4.05cm}
		\begin{center}\vspace{0cm}
			\sfl{a}{
			\includegraphics[width=4.2cm]{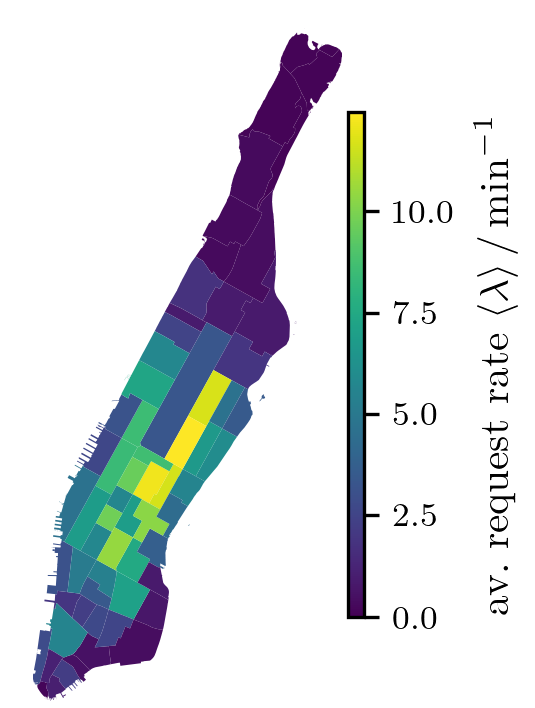}}
		\end{center}
	\end{subfigure}\hfill
	\begin{subfigure}[t]{4.05cm}
		\begin{center}\vspace{0cm}
			\sfl{b}{
			\includegraphics[width=4.2cm]{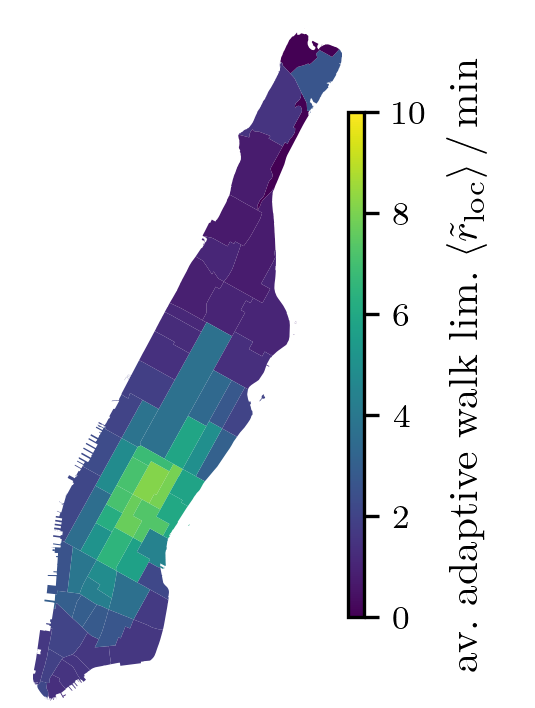}}
		\end{center}
	\end{subfigure}\hfill
	\begin{subfigure}[t]{7cm}
		\begin{center}	
		\vspace{0cm}
			\sfl{c}{
			\includegraphics{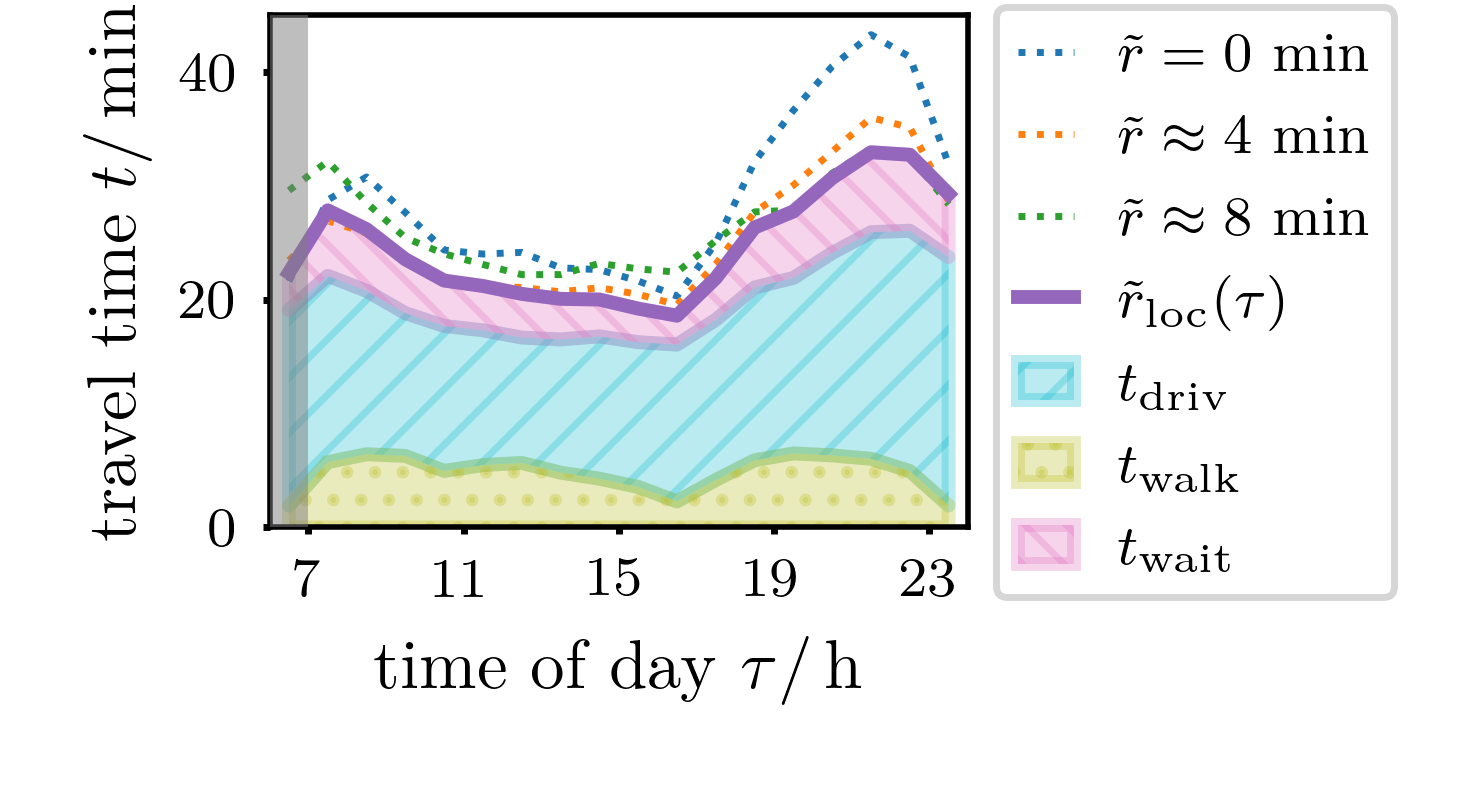}}
			\sfl{d}{	
			\includegraphics{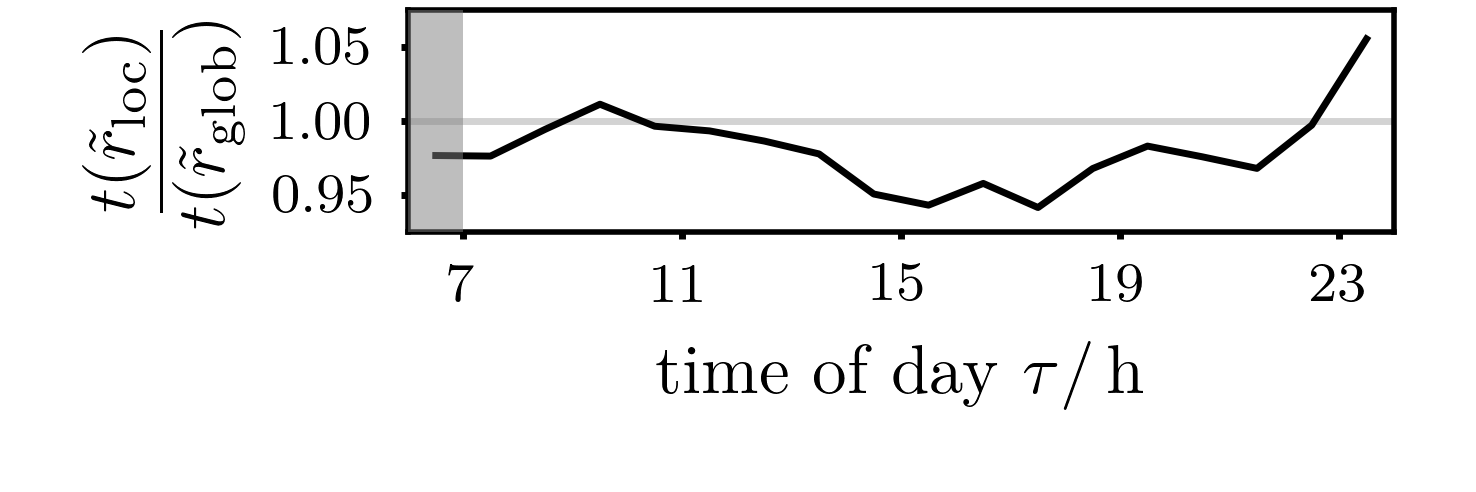}}
		\end{center}
	\end{subfigure}
	\vspace{-0.7cm}
	\caption{\small 
	\textbf{Spatio-temporally adapted walk limit further reduces travel times.}
        (a) The demand is heterogeneous in space as well as in time, illustrated by the daily average request rate per taxi zone.
        (b) Spatially localized adaptations of the walk limit $\rloc(\tau)$ only require users to walk in regions with high demand. 
        (c,d) With this spatio-temporally adapted walk limit, the travel time is further reduced compared to a fixed walk limit by up to $5.8\%$ ($\hat{=}\SI{1.5}{\minute}$ difference). The resulting additional walk time depends on the  time of day $\tau$ (green dotted area in panel c).
        }
	\label{fig:adapt_r_in_spacetime}
\end{figure}

\begin{figure}[b!]  
	\begin{subfigure}[t]{7.7cm}
		\begin{center}
			\sfl{a}{
		\includegraphics{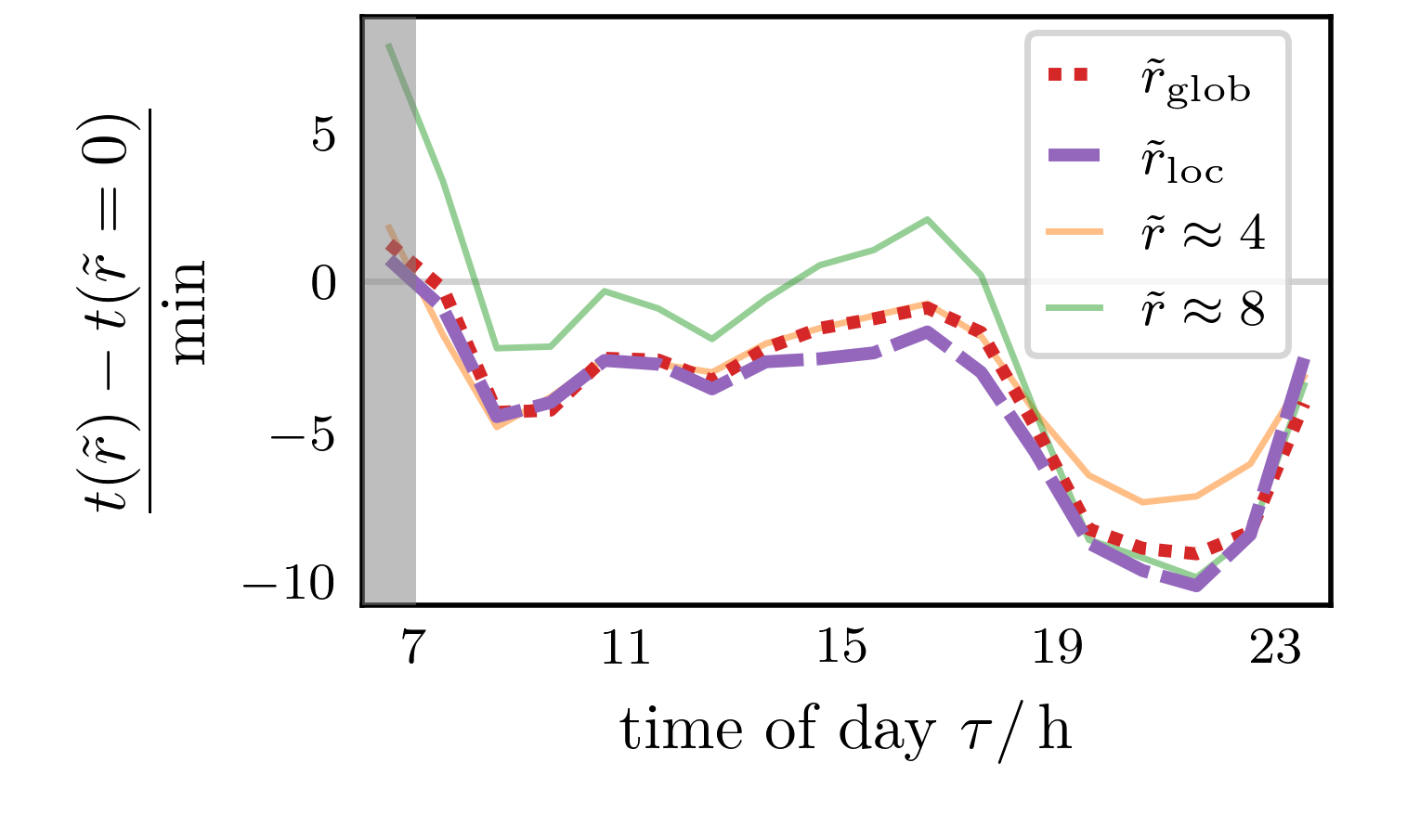}}
		\end{center}
	\end{subfigure}
	\begin{subfigure}[t]{7.5cm}
		\begin{center}
			\sfl{b}{
				\includegraphics{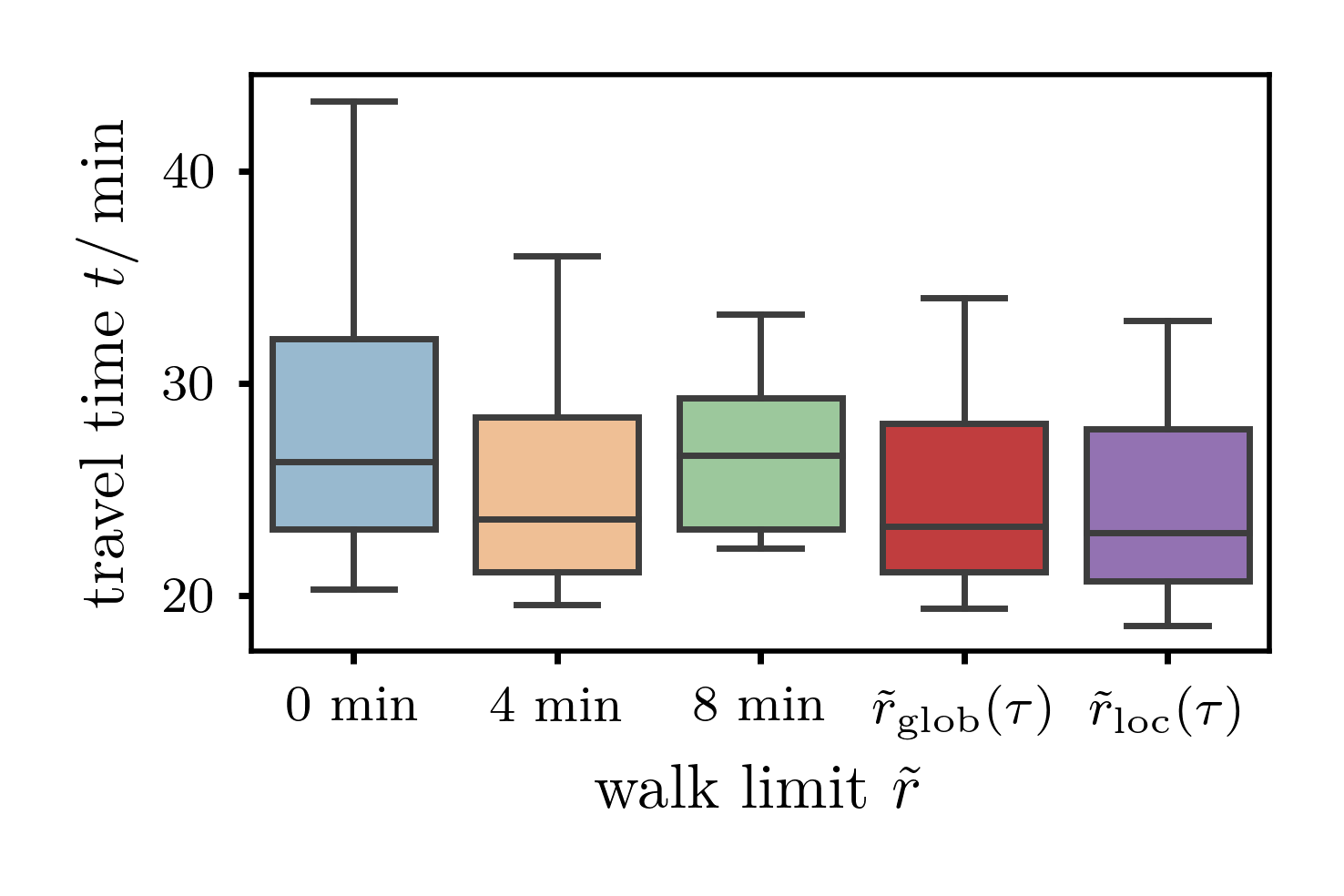}}
		\end{center}
	\end{subfigure}\vspace{-0.5cm}
	\caption{\small  \textbf{Stop pooling reduces travel time and travel time fluctuations.
		}
		\small 
		(a) Adaptive stop pooling significantly reduces the travel time compared to standard ride sharing.	The effect is larger when travel times are high, e.g. during high demand in the evening. Local adaptation of the  walk limit $\rloc$ further improved travel time saving.  
		(b) With stop pooling, the travel time fluctuates less.
		The mean travel time $t$ (horizontal bar) is smaller with adapted $\rglob$ and $\rloc$ than with a fixed walk limit $\rt$. The local walk limit $\rloc$ yields the smallest mean $t$. Furthermore, the averages of $t$ in one-hour-intervals spread in a smaller range with adapted $\rglob$ and $\rloc$ than with standard ride sharing $\rt=0$, making ride sharing travel times more reliable.}
	\label{fig:reduced_fluctuations}
\end{figure}

In principle, the local demand adaption in time \textit{and} space could work analogous to the global adaption in time introduced above: (i) find the best walk limit for steady states in each local region (e.g.~taxi zones) and (ii) adapt the maximum walking distance accordingly. However, the effort of pre-processing for each local region is very high as the region size should reflect the maximum overall acceptable walk time of a few minutes walk. 
We thus suggest a dynamic online adaptation according to the number of users that requested their ride in a local region of the service area.
When a user $i$ requests a trip, we count all $N$ users $j$ 
\begin{itemize}
	\item
with origin $\mathbf{o}_j$ or destination $\mathbf{d}_j$  within a walk distance $\rmaxt/\vp$ around $i$'s origin $\mathbf{o}_i$ and destination $\mathbf{d}_i$ and
\item whose request is at most a maximum overall acceptable walk time $\rmaxt$ ago:
\\${\tau_\mathrm{request}(i)-\tau_\mathrm{request}(j)<\rmaxt}$. 
\end{itemize}
The walk limit $\rloc$ for user $i$ depends on $N$ and a threshold $N_c$:
\begin{equation}
	\rloc(\tau_\mathrm{request}(i))=\begin{cases}
		\rmaxt & N(i)\ge N_c\\
		0 & \textnormal{otherwise}\,.
	\end{cases}
\label{eq:rloc}
\end{equation}
When $i$ has sufficiently many  neighboring users, the walk limit is set to the maximum overall acceptable walk time $\rmaxt$. If not, the walk limit is set to zero. The user $i$ might only walk with sufficiently many neighboring users to potentially pool a stop with, but almost never needs to walk the full distance since a suitable stop is likely to be closer. 
For the example setting, we define $\rmaxt=\SI{10}{\minute}$ (also used for the iterative optimization of $\rbt\in[0,\rmaxt]$). A threshold $N_c=1000$ yields best results. 

With this simple adaption scheme, the users walk only in the regions and during times of high demand (Fig.~\ref{fig:adapt_r_in_spacetime}a,b, cf.~Supplementary Note 3.G, Fig.~S11).  
The adaption scheme avoids that users walk in sparse demand regions while all other users are allowed to walk a maximum acceptable walk limit $\rmaxt$. In practice, the stop pooling algorithm determines how far each user walks, often much shorter than $\rmaxt$ (Tab.~\ref{tab:var}, Fig.~\ref{fig:adapt_r_in_spacetime}c). Users walk on average longer than $\rt$ (maximum is $2\rt$ - once at origin and once at destination) and longer
than with exclusive temporal adaption (Tab.~\ref{tab:var}). 
Nonetheless, spatio-temporal adaptive stop pooling reduces the travel time $t$ even more than stop pooling with exclusive temporal adaption (Tab.~\ref{tab:var}, Fig.~\ref{fig:adapt_r_in_spacetime}c,d). 
In the example, it yields on average $1.5\%$ and at most $5.8\%$ smaller travel times $t$ (cf.~Fig.~\ref{fig:adapt_r_in_spacetime}c,d). The mean travel time is smaller than with a global walk limit $\rglob$, with standard ride sharing, $\rt=0$, or with intermediate fixed walk limit, $\rt=\SI{3.75}{\minute}$, 
\begin{equation}
	\langle{t}\rangle(\rloc)<\langle{t}\rangle(\rglob)<\langle{t}\rangle(\rt=\SI{3.75}{\minute})<\langle{t}\rangle(\rt=0)\,.
\end{equation}
In the example, the spatio-temporal adaption yields on average $13.4\%$ and at most $24.7\%$ smaller $t$ compared to standard ride sharing. 
Moreover, the travel time fluctuates less (standard deviation $\SI{4.8}{\minute}$, $19.5\%$ of mean) than with standard ride sharing or stop pooling with intermediate fixed walk limit, $\rt=\SI{3.75}{\minute}$. A user who requests a ride between 21:00 and 22:00  travels on average 
less than twice as much than users who requests a ride between 16:00 and 17:00 (Tab.~\ref{tab:var}). The fluctuations are slightly higher than with a global walk limit $\rglob$.
Still, adaptive stop pooling efficiently reduces the travel time at fluctuating demand while simultaneously reducing the travel time fluctuations.

All in all, stop pooling reduces fluctuations of the average travel time of ride sharing users at a constant fleet size. In the given example, this is true no matter if the walk limit is adapted in time or in time and space or not at all (Fig.~\ref{fig:reduced_fluctuations}b). In addition, stop pooling consistently reduces the travel time when the maximum walking distance is adapted to the instantaneous demand. In the example, a spatio-temporal adaption yields a slightly higher reduction of the travel time than a purely temporal adaption (Fig.~\ref{fig:reduced_fluctuations}a).

\section{Discussion}

Ride sharing users might experience unreliable, highly fluctuating travel times over the day induced by fluctuating demand. 
We here propose to reduce these travel time fluctuations by adaptive stop pooling, requiring some users to walk a short distance and pool their stops with other users.
Interestingly, some ride sharing services already include short walks from exact locations to close-by virtual bus stops \cite{Esch.2023,Kostorz.2021,Stiglic.2015,Oh.2020,Goel.2017}. 
Instead of sending users to the closest virtual bus stop, the proposed stop pooling scheme combines user stops flexibly, which might be particularly efficient when adapting the maximum walking distance to the current demand. 

In this article, we study the qualitative collective influence of stop pooling in basic models of ride sharing fleets operating at fluctuating demand using event-based simulations. We find that stop pooling may reduce the user travel time, because buses drive along more direct routes. The reduction is larger the higher the  travel time initially. Consequently, stop pooling also reduces travel time fluctuations.
The optimal maximum walking distance depends on the demand: Users should walk further for higher demand, where higher time savings buffer higher walk times. We demonstrate this with two example procedures of adaptive stop pooling adjusting the maximum walking distance to the current demand: (i) setting the maximum walking distance to the best suitable distance for the current global demand or (ii) deciding if a user walks or not based on the local demand around the user. With both procedures, the travel time is on average smaller and fluctuates less than with standard ride sharing.

In general, ride sharing operators have different options to suitably design their service to adapt to fluctuating demand. 
A common strategy is to adapt the fleet size \cite{Vazifeh.2018,Wilkes.2021} requiring to provide additional vehicles and drivers which increases the overall carbon emissions and costs of the ride sharing system. Alternatively, providers may adapt their dispatcher which requires less effort and allows finer and faster adaptation. For instance, dynamic pricing \cite{Banerjee.2015, Cachon.2017} might increase the user incentive for sharing trips \cite{Ruijter.2020} if necessary, but dynamic pricing has already been misused by drivers to artificially increase the cost for a ride \cite{Schroder.2020}. 
As demonstrated above, adaptive stop pooling  requires only a straightforward adaption of the dispatcher, without requiring additional or higher-capacity vehicles (details in Supplementary Note 3.G, Fig.~S12). 
Adaptive stop pooling may thus contribute to cheap and sustainable ride sharing with reliable travel times.

Our analysis has focused on qualitative effects of adaptive stop pooling. 
The quantitative results depend on parameters like fleet size or vehicle velocity 
(Supplementary Note 4.H) and should be seen as examples. 
For instance, the model uses a constant mean-field velocity. Typically, vehicle velocities reduce during times of high demand (rush hour), further contributing to high travel times. However, stop pooling reduces travel times more strongly compared to standard ride sharing when the difference between driving and walking velocity is smaller, since longer walk times are possible (see Supplementary Note 4.H). Consequently, we expect the potential of stop pooling to reduce variability of travel times across the day to remain robust.
Moreover, the model uses a simple assignment algorithm, but we show that the result is robust for a more complex algorithm (Supplementary Note 4.I): When limiting the user travel time by a maximal delay, providers have to reject users if the demand exceeds the supply. Then, the rate of rejected users fluctuates instead of the travel time and adaptive stop pooling reduces the fluctuations of the rejection rate. 
Besides, the model does not include that short user walks might reduce the perceived service quality.
In general, walk time is typically valued less than waiting for or driving in the vehicle \cite{Wardman.2004}, walking might provoke safety risks (especially at night) and walking might not even be possible for some users. 
However, stop pooling requires only some users walk while others are served from door to door.
Further research questions result from the suggested methods to adapt the maximum walking distance to current demand. For example one might avoid repeated pre-processing when discovering universal scaling laws either for the best maximum walking distance or for the optimal minimum number of neighbors to pool stops with. We found this threshold by trial and error, but results are robust for slight deviations $N_c\in[750,1250]$. 
In addition, one may further develop the adaption methods themselves. For example, the spatio-temporal adaption might improve by differentiating between the local demand around the origin and that around the destination of a user, or by taking the age of stops into account.

The results in this article contribute to a fundamental understanding of the collective dynamics of ride sharing systems under conditions of fluctuating demand.
Moreover, the results might motivate (i) ride sharing providers to include stop pooling into their service, because the service may become more efficient, and (ii) users to participate in a stop pooling service, because their total travel time may -- counterintuitively -- decrease.
The presented basic stop pooling algorithm that includes two procedures to adapt the maximum walking distance to the current demand might serve as a basis for future adaptive dispatchers. Such ride sharing services including adaptive stop pooling may contribute to sustainable and reliable human mobility.

\bibliographystyle{unsrt}
\bibliographystyle{naturemag}
\bibliography{references.bib}

\vspace{1.55cm}
\subsection*{Acknowledgments}
The authors thank Verena Krall, Christoph Steinacker, Kush Mohan Mittal, Felix Jung and all members of Chair for Network Dynamics for valuable discussions. This work was partially supported by the Volkswagen Foundation under grant no. 99 720 and the German Federal Ministry for Education and Research (BMBF). CL acknowledges support from the German Federal Environmental Foundation (Deutsche Bundesstiftung Umwelt DBU). The authors are grateful to the Centre for Information Services and High Performance Computing (ZIH) TU Dresden for providing facilities for high throughput calculations.

\vspace{0.5cm}
\subsection*{Author contributions}
CL: Conceptualization, literature review, modeling,  simulation code, simulations and data analysis, manuscript writing, and editing. 
PM: Modeling and simulation code.
MS: Conceptualization, modeling, supervising simulations, data interpretation, manuscript writing, and editing. 
MT: Conceptualization, data analysis, manuscript writing, and editing. 

\null
\newpage
\renewcommand{\thefigure}{S\arabic{figure}}
\renewcommand{\thetable}{S\arabic{table}}
\renewcommand{\theequation}{S\arabic{equation}}

\setcounter{figure}{0}
\setcounter{table}{0}
\setcounter{equation}{0}
\setcounter{page}{1}

\section*{Supplementary Note 1: Model Details}
    	\label{ch:model}
    	This supplementary note provides a detailed description of the model. The model grounds on a basic ride sharing model and adds stop pooling with flexible stop locations.  Next, we explain how the original street network is translated into a bus street network and a separate user walk network. Moreover, we describe how we implement taxi trip data as requests into the model and how a ride sharing algorithm assigns users to buses.

    	\subsection{Basic ride sharing model}
    Let us consider an event-based model where users want to travel certain trips at certain times. The only option to travel for all users is a ride sharing service that provides a bus fleet of fixed fleet size $B$ to transport the users. The interaction of users and buses defines the collective dynamics of the system. Agent based simulations reflect these interactions explicitly. 
    	The simulations are based on three events:  
    	\begin{enumerate}
    		\item Request: a user $i$ requests a trip, that is they want to travel from their origin $\mathbf{o}(i)$ to their destination $\mathbf{d}(i)$ as soon as possible after their request time $t_{\mathrm{request}}(i)$. 
    		\item Job: a bus stops at a stop location (origin or destination) to serve a user $i$, either
    		\begin{enumerate}
    		\item Pickup: bus stops at the origin $\mathbf{o}(i)$ of user $i$. The user $i$ enters the bus at pickup time $\tau_{\mathrm{pickup}}(i)$.
    		\item Delivery: bus stops at the destination $\mathbf{d}(i)$ and the user $i$ leaves the bus. The user $i$ leaves the bus at delivery time $\tau_{\mathrm{delivery}}(i)$.
    		\end{enumerate}
    	\end{enumerate}
    	When a request $i$ appears, the    	
    	service assigns  user $i$ to a bus using an assignment algorithm (Sec.~C) that plans a feasible pickup $\mathrm{P}_i$ at the user's origin $\mathbf{o}(i)$ and a delivery $\mathbf{d}_i$ at the user's destination $\mathbf{d}(i)$. The algorithm assigns the user in a way that the bus detour is minimized, even if this means long travel times for the user. Simulations do not include any alternative service
    	and users always accept the ride suggested. Thus, no decision criteria such as trip prices are necessary.
       	The assignment algorithm generates a list of planned jobs for each bus, the \textit{job list}. The bus visits all stop locations in the job list one after another, driving on the shortest route between each two stop locations with a constant bus velocity $\vv$. As soon as at least one stop is planned for a bus, the bus immediately starts executing its job list. A bus is busy (i.e.~driving with constant bus velocity $\vv$) between the time when the first request is assigned to the bus and the time the bus delivers the last assigned user such that the job list runs dry. Then, the bus becomes idle until a new request is assigned to the bus.

    	\subsection{Stop Pooling with Dynamic Stop Locations Maintains Flexibility}
    	\label{ch:stop_pooling}
    	\begin{figure}[b]
    		\begin{center}
    		\includegraphics[ width=0.5 \textwidth]{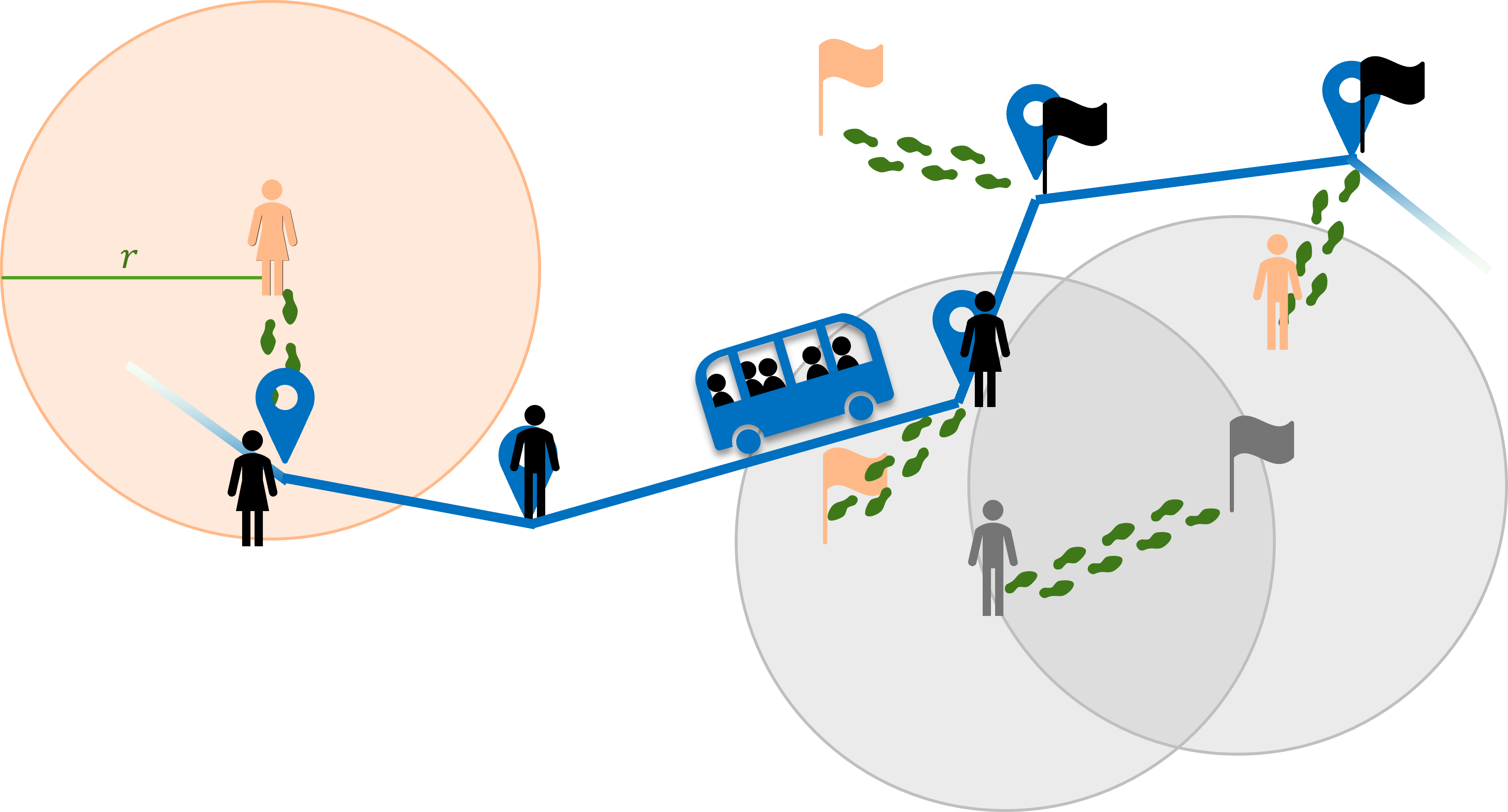}
    	\end{center}
    \vspace{-0.5cm}
    		\caption[Dynamic stop locations keep the ride sharing system flexible.]{\small\textbf{Dynamic stop locations keep the ride sharing system flexible.} 	\small 
    			In continuous space, the walk limit $r$ spans a disk with radius $r$ around each stop.  When a neighbored stop is within a walk distance $r$ a stop is served indirectly (rose) and the user walks to a pooled stop (blue pin). Otherwise, users are served directly at the door (black).  Some users with a short  trip length $\ell<2r$ are  rejected (gray). In continuous space their disks around origin and destination overlap (gray circles). The users walk their complete trip, but never more than $2\,r$. }
    		\label{fig:stop_pooling_sketch}    
    	\end{figure}
    	
    	With stop pooling, users might walk a short distance with user velocity $\vp$ to a pooled stop, a nearby stop location where they are served together with at least one other user. Buses do not serve all users exactly from door to door and save small detours (Fig.~\ref{fig:stop_pooling_sketch}). 
    	When a user requests a trip, the assignment algorithm searches in all bus job lists for planned stops within a walk distance limit $r$ around the request's origin $\mathbf{o}$ and destination $\mathbf{d}$. 
    	If there is a stop $\mathbf{s}_1$ within a distance $r$ around the origin $\mathbf{o}$ that the user reaches by foot before the bus arrives, the user may walk from $\mathbf{o}$ to $\mathbf{s}_1$ to be picked up at $\mathbf{s}_1$. If there is a stop $\mathbf{s}_2$ within a distance $r$ around the destination $\mathbf{d}$, the user may be delivered to $\mathbf{s}_2$ and walk from $\mathbf{s}_2$ to their destination $\mathbf{d}$. 
    	Once a user has been assigned to a bus and received a pick up and delivery location, these locations are fixed. All stop locations where users have been promised to be served without walking must thus be visited by a bus and might become pooled stops if other users request a trip with origin or destination within a distance $r$. In principle, any requested stop location (origin or destination) might become a pooled stop. The locations of the pooled stops are selected flexibly from requested origins and destinations. Users only walk if their stop is pooled with at least one other user. By contrast, at fixed stations as in public transportation multiple users might be served at once, but users also have to walk to a station even if they will be the only user that is served there.

    	\label{ch:cpwalk}	
    	In the model, users might not only walk to a nearby pooled stop but even their complete trip if the trip length $\ell$ is below the total walk limit - twice the walk limit $r$ per stop, 
    	\begin{equation}
    		\ell\le2\,r \Leftrightarrow \textnormal{user walks their complete trip}\,.
    		\label{eq:cpwalk}
    	\end{equation}
		The requested stop locations (origin and destination) of these users are rejected. This procedure is a simple solution for the problem of users with short trip length \cite[Sec.~4.1]{Wang.2022}. An alternative is to filter out those requests because they are unlikely in real demand patterns \cite{Fielbaum.2021b}. 
    In summary stop pooling treats requested stop locations (origin and destination) in three different ways (see Fig.~\ref{fig:stop_pooling_sketch}):
    	\begin{enumerate}
    		\item A stop is served directly. When the bus stops exactly at the requested location to pick up or deliver a user. When more than one user is served at the stop location, a directly served stops is a pooled stop.
    		\item A stop is served indirectly. The bus stops within a distance $r$ around the requested location at a pooled stop. The user walks from/to the indirectly served stop location to/from the pooled stop.
    		\item A stop is rejected if $\ell\le2\,r$ and the user walks their complete trip. The user is not served by a bus.
    	\end{enumerate}
    	A user's walk distance is the sum of the spatial distance between the user's origin and pickup point $\mathbf{s}_1$ and the spatial distance between the user's delivery point $\mathbf{s}_2$ and their destination
	    \begin{equation}
	    	\dwalk = d(\mathbf{o},\mathbf{s}_1)+d(\mathbf{s}_2,\mathbf{d})\,\,.
	    \end{equation}
	    If a user is picked up at their origin and delivered at their destination, the walk distance equals zero.
    	These stop types yield three different user types:
    	\begin{enumerate}
    		\item A user does not walk, when the bus stops exactly at their origin and destination (${\dwalk=0}$).
    		\item A user walks partially, when origin and/or destination are served indirectly. They walk a short distance, but still use the ride sharing service ($\dwalk>0$).
    		\item A user is rejected, if $\ell\le2\,r$. They walk their complete trip and do not use the ride sharing service ($\dwalk=\ell$).
    	\end{enumerate}

      	\subsection{Ride Sharing Algorithm Assigns Users by Reducing Bus Detour}
    	\label{ch:algorithm}
\begin{figure}[t!]
    		\vspace{-0.1cm}
    			\begin{center}
    				\includegraphics{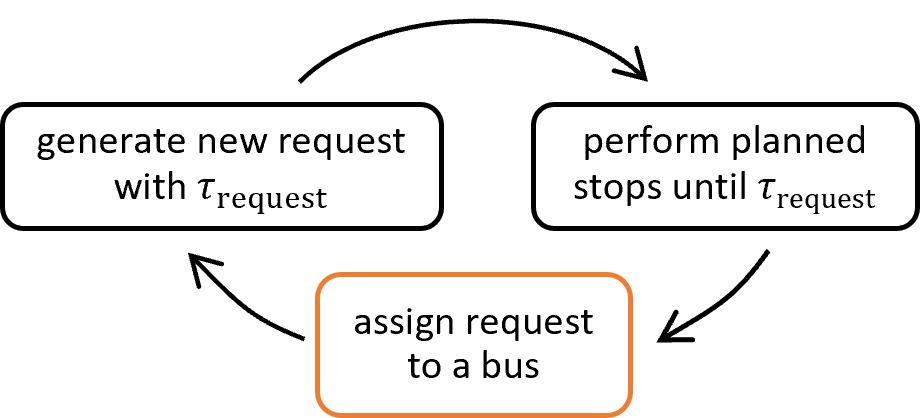}
    			\end{center}
    		\vspace{-0.6cm}
    			\caption[The ride sharing simulations are event based.]{\small  \textbf{The ride sharing simulations are event based.} \small 
			The algorithm repeats a loop that (i) generates the next requests, (ii) performs all events until the next request time and (iii) assigns the new request.}
    			\label{fig:simulation_procedure}
    			\vspace{-0.1cm}
    	\end{figure}

    	To keep the number of parameters in the system as small as possible, the basic assignment algorithm is very simple. For example, the algorithm does not collect a batch of requests before assigning users as in \cite{AlonsoMora.2017, Qin.2021, Fielbaum.2021, Fielbaum.2021b,Engelhardt.2020,Azar.06192017,Ke.2020} to avoid the influence of the batching time \cite{Qin.2021}. The algorithm assigns requests immediately at their request time $\tau_\mathrm{request}$. From this results a simple simulation procedure (Fig.~\ref{fig:simulation_procedure}): The simulation assigns a request and then performs all scheduled bus jobs (picking up and delivering users) that are planned before the next request time. Then, the next request is assigned.
    	
    	\subsubsection{Standard Ride Sharing Algorithm}
    	
    	\begin{figure}[b!]
    		\begin{center}
    			\includegraphics{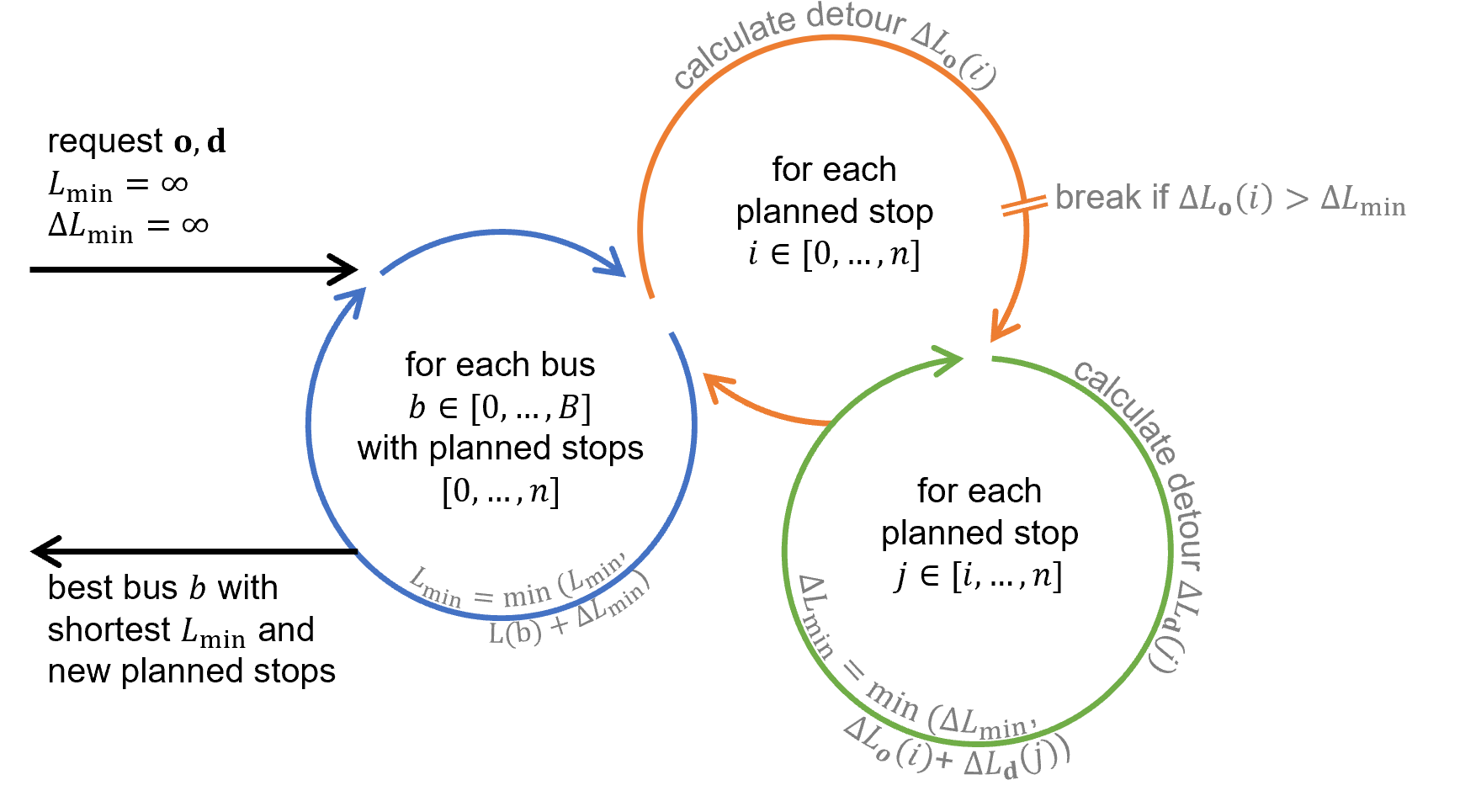}
    		\end{center}
    		\vspace{-0.9cm}
    		\caption[A simple ride sharing algorithm optimizes  bus routes in three loops.]{\small  \textbf{A simple ride sharing algorithm optimizes  bus routes in three loops.} \small 
    			The assignment algorithm contains three loops: one loop over all buses (blue), a second loop over all planned stops (orange), and a third loop over the rest of planned stops (green). The first  loop chooses the bus with shortest distance driven  $L+\Delta L_\mathrm{min}$ after inserting the request's origin $\mathbf{o}$ and destination $\mathbf{d}$. The second  loop finds the best position to insert the  origin $\mathbf{o}$ into the list of planned stops of bus $b$. The third  loop finds the best position to insert the  destination $\mathbf{d}$ into the list of planned stops of bus $b$ after the origin $\mathbf{o}$.}
    		\label{fig:alg_RS}
    	\end{figure}
    	The simple algorithm checks all possible insertion options using three loops  (Fig.~\ref{fig:alg_RS}): 
    	\vspace{-0.2cm}
    	\begin{itemize}
    		\item A loop over all buses $b\in[1,2,\ldots,B]$ finds the best suitable bus $b$ for the request.
    		\item For each bus $b$, a loop over all planned stops  $i\in[1,2,\ldots,n(b)]$ calculates the detour $\lo$ to insert the origin $\mathbf{o}$ as a pickup point. $\lo$ is the difference in the planned distance driven $L(b)$ of the bus route of $b$ before and after insertion of $\mathbf{o}$.
    		\item For each stop $i$, a loop over the rest of the planned stops  $j\in[i,i+1,\ldots,n(b)]$ calculates the detour $\lde(j)$ to insert the destination $\mathbf{d}$ as a delivery point after the pickup. 
    		$\lde$ is the difference in the planned distance driven of the bus route of $b$ with only the origin $\mathbf{o}$ and with both origin $\mathbf{o}$ and destination $\mathbf{d}$.
    		Together, both loops find the best combination of pickup position $i_\mathrm{b}$ and delivery position $j_\mathrm{b}$ with minimal detour $\Delta L_\mathrm{min}= \lo(i_\mathrm{b})+\lde(j_\mathrm{b})$ for  bus $b$. 
    	\end{itemize}
    The algorithm chooses the bus $b$  with shortest planned distance driven after insertion of the new request
    	\begin{equation}
    		L_\mathrm{min}=\min_b(L(b)+\Delta L_\mathrm{min}(b))\,\,,
    		\label{eq:alg_min_L}
    	\end{equation} 
    	given as the sum of of the planned distance driven $L$ without the new request and the shortest detour $\Delta L_\mathrm{min}$ when assigning the new request to that bus.
    	Indeed, a simpler definition for the best bus by the shortest detour $\Delta L_\mathrm{min}$ alone risks that one bus serves all users, because a bus with a long list of planned stops has a higher chance to pass a new stop than a bus without planned stops. 
    	Instead, minimizing the new total distance driven $L+\Delta L_\mathrm{min}$ favors buses with fewer planned stops and thus balances the number of planned stops between all buses. If the bus distance driven $L$ was similar for all buses, the assignment algorithm would exclusively minimizes the detour $\Delta L_\mathrm{min}$ by the new request. 
    	All in all, the assignment algorithm minimizes the detour induced by a new user while simultaneously spreading the requests over all buses. The algorithm does not consider user travel times at all. 
    	With a simple trick, the computing procedure becomes more efficient:
    	If the current minimal detour $L_\mathrm{min}(b)$ in the loop is smaller than the detour for inserting the origin  $\lo(i)$ would be, the algorithm skips the third loop for inserting the destination for this $i$ and jumps to $i+1$.  
    	This standard ride sharing algorithm was designed and implemented by Philip Marszal. 
    	
    	\begin{figure}[t!] 
    		\begin{subfigure}[t]{4.95cm}
    			\begin{center}
    				\sfl{a}{
    					\includegraphics[width = \textwidth]{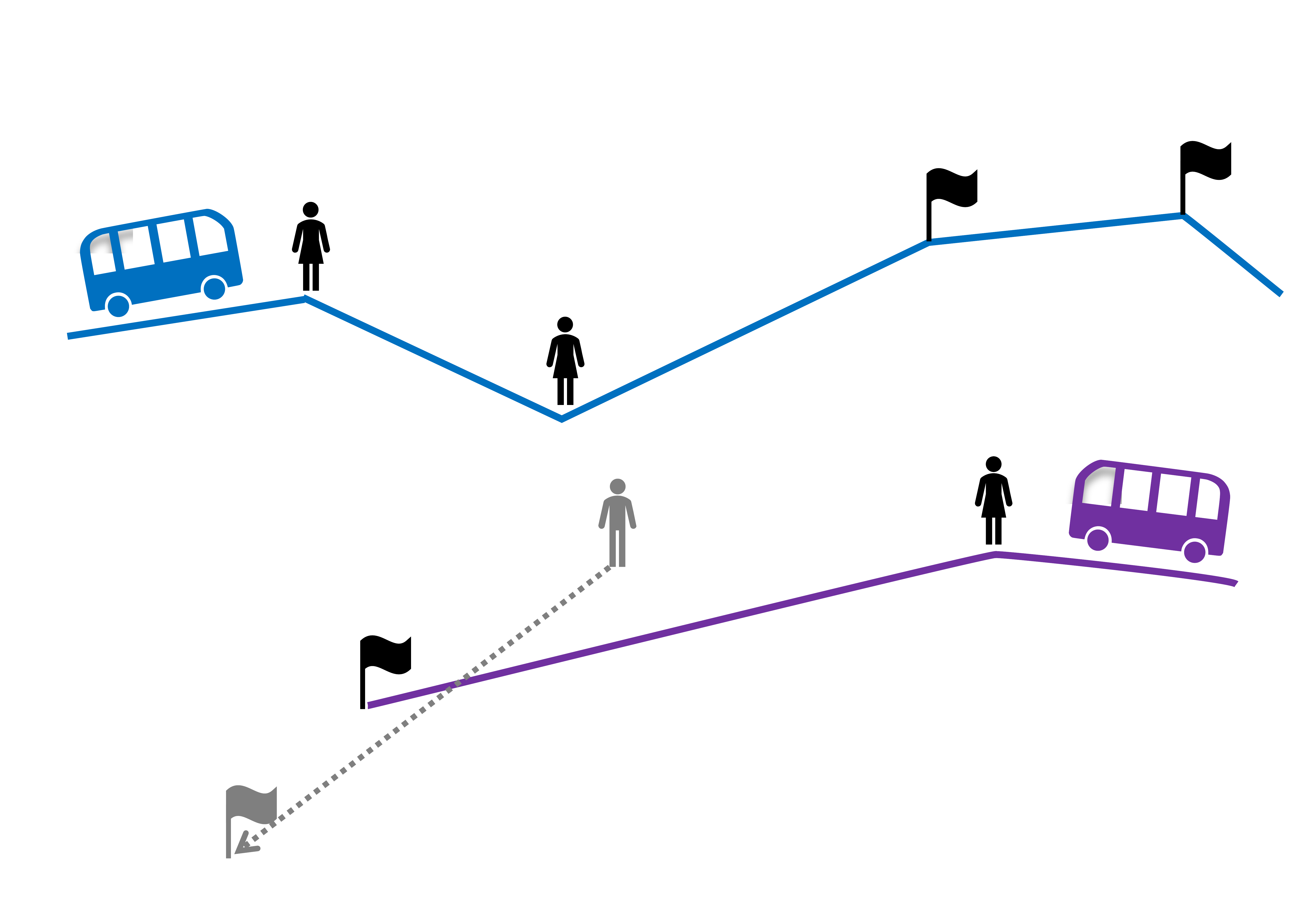}}
    			\end{center}
    		\end{subfigure}\hfill 	
    		\begin{subfigure}[t]{4.95cm}
    			\begin{center}\sfl{b}{
    					\includegraphics[width = \textwidth]{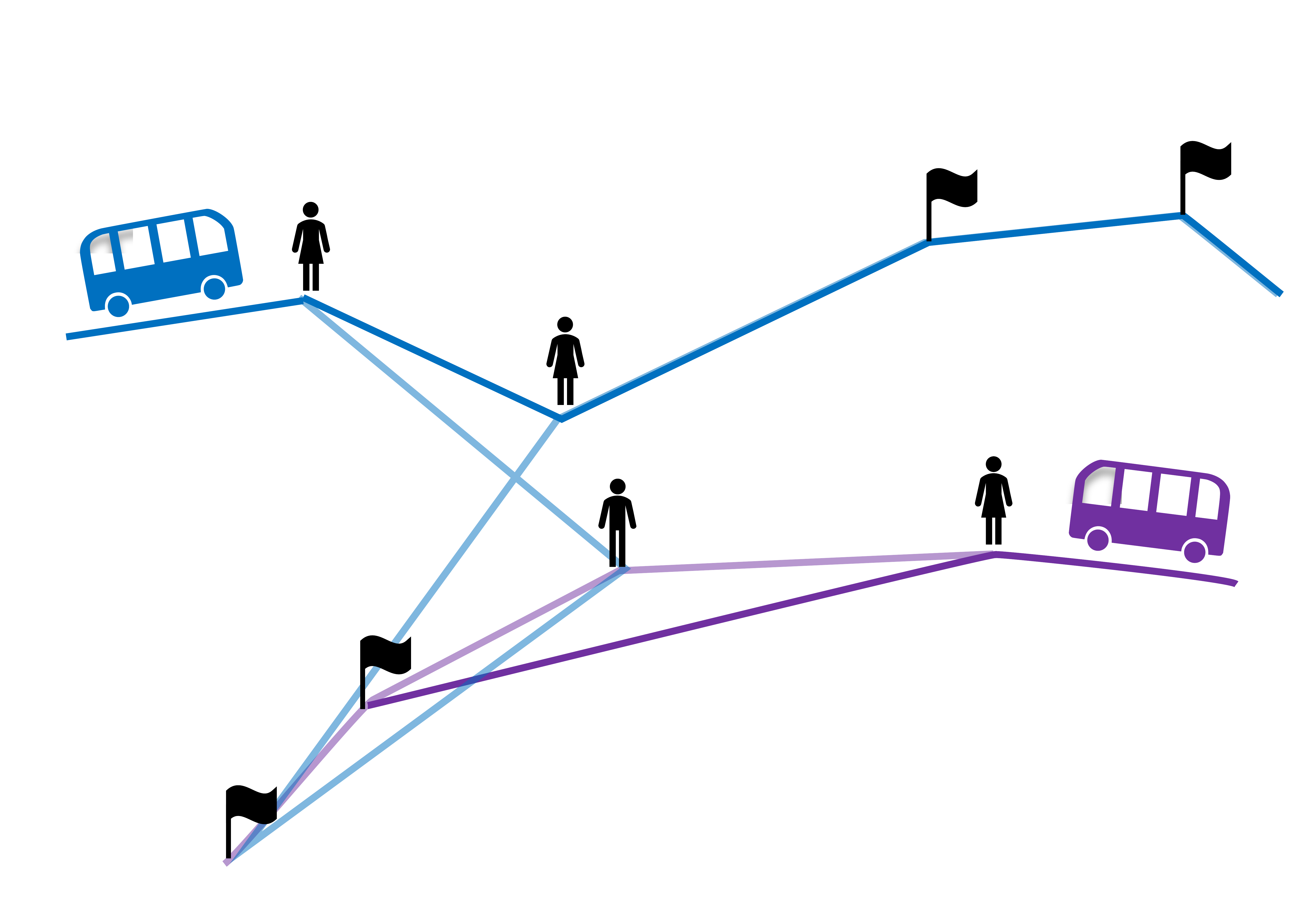}}
    			\end{center}
    		\end{subfigure}\hfill 	
    		\begin{subfigure}[t]{4.95cm}
    			\begin{center}\sfl{c}{
    					\includegraphics[width = \textwidth]{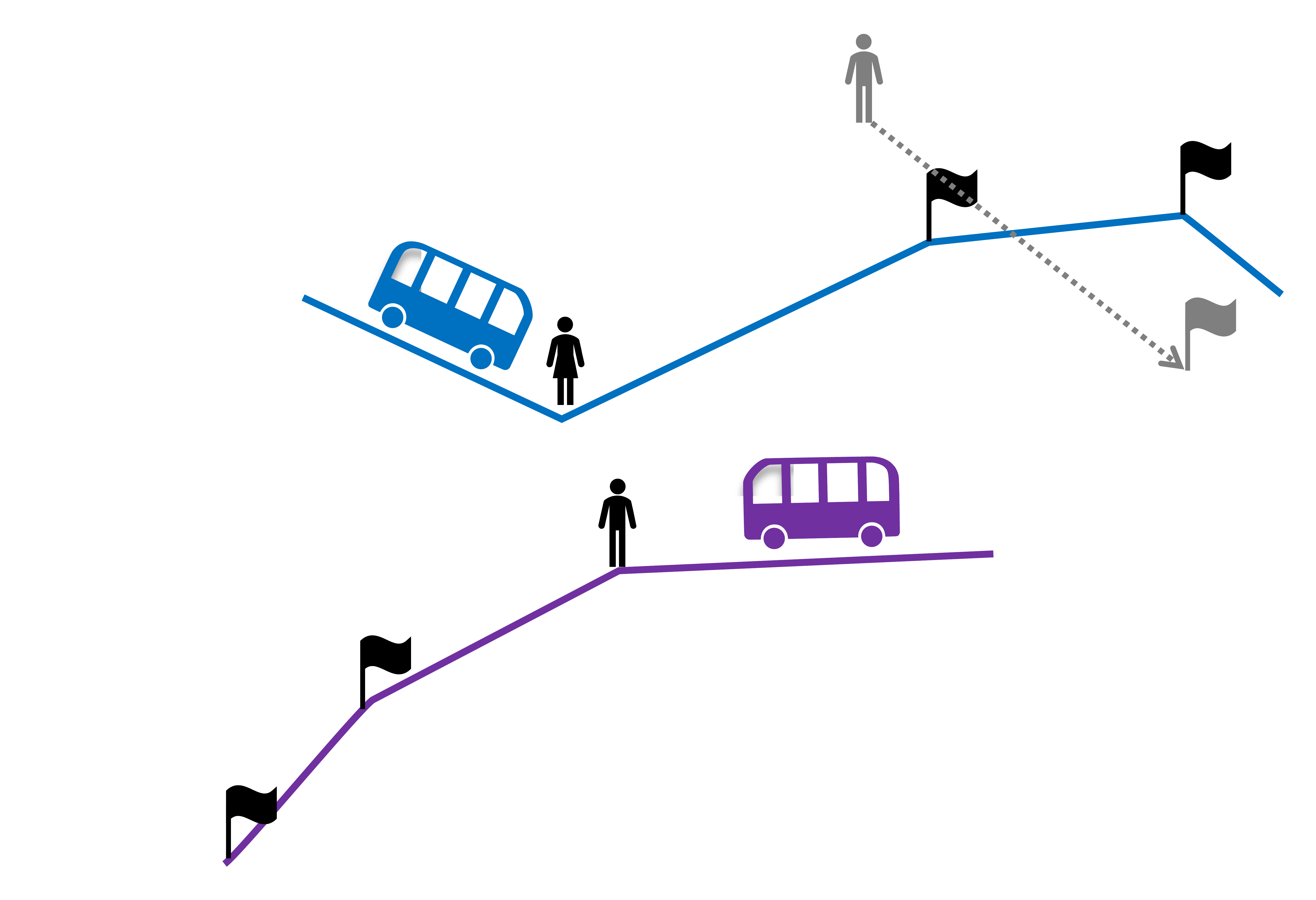}}
    			\end{center}
    		\end{subfigure}
    		\vspace{-0.5cm}
    		\caption[Ride sharing is an online optimization problem]{\small  \textbf{Ride sharing is an online optimization problem} \cite{Muhle.2023}.
    			\small 
    			While the ride sharing buses serve their users, new requests appear. Consider two example buses, each with a planned route (line) (a) while buses visit their planned stops, a new request appears (gray). (b) The algorithm checks for each bus, where an insertion of the new user stops induces the smallest detour, first for the origin and second for the destination. The best new route option of each bus is marked by a transparent line. The algorithm assigns the new request to the bus with shortest planned route with the new user and adds the stops to the planned bus route  (here purple bus). (c) While the buses continue driving along their planned route, new requests appear.}
    		\label{fig:iterative_bus_route}
    	\end{figure}
    
	    	Together, the simulation procedure and the assignment algorithm work as follows (Fig.~\ref{fig:iterative_bus_route}): buses visit their scheduled stops one after another. From time to time, a new user requests a ride. The assignment algorithm finds the best option to insert the origin and the destination to the planned route of each bus with minimal detour $\Delta L_\mathrm{min}$. Then, the algorithm decides for the bus with shortest distance driven after insertion of the new stops $L+\Delta L_\mathrm{min}$, which is often also the bus with shortest detour $\Delta L_\mathrm{min}$.

		\subsubsection{Stop Pooling Algorithm}
    	
    	\begin{figure}[t!]
    		\begin{center}
    			\includegraphics{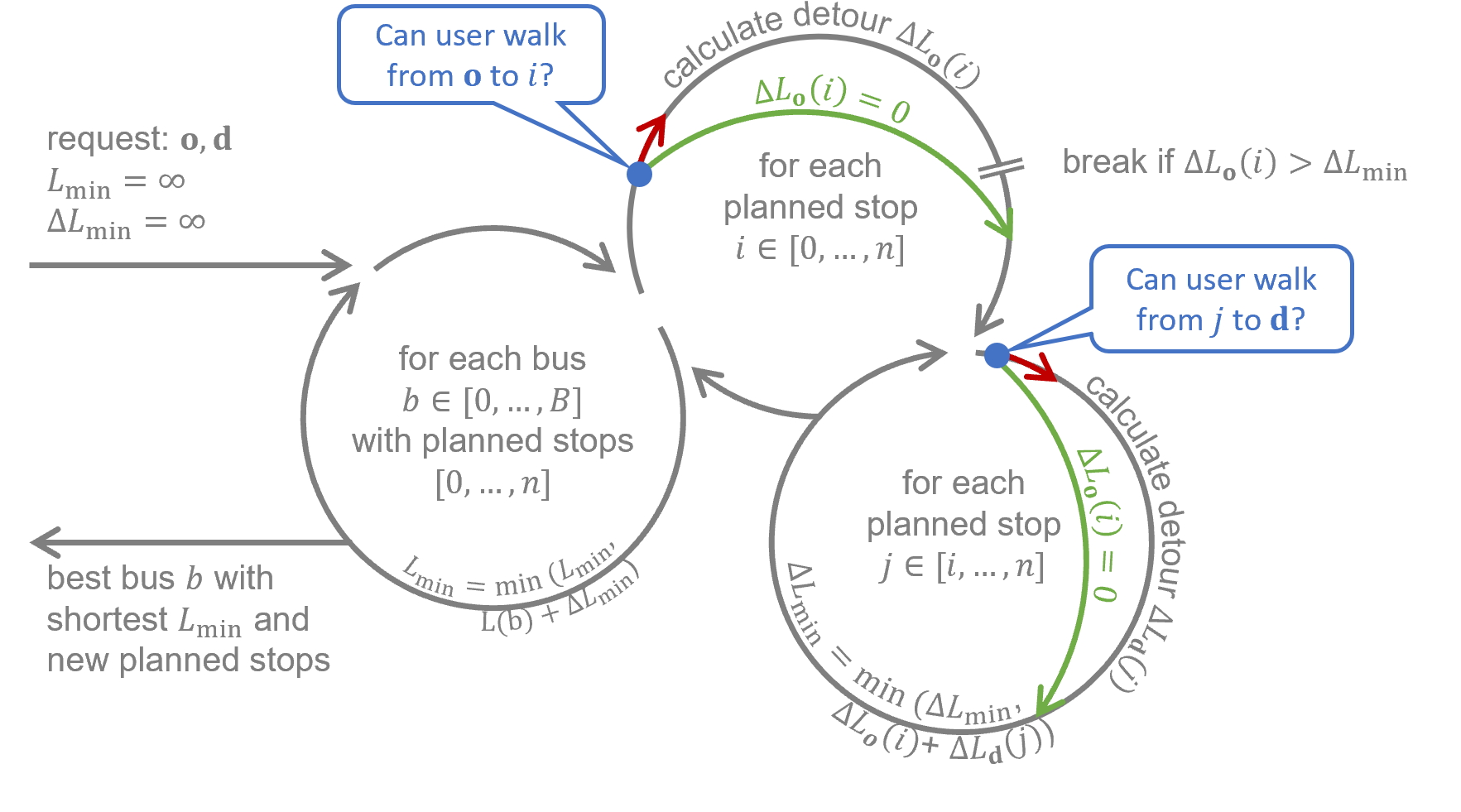}
    		\end{center}
    	\vspace{-1cm}
    	\caption[Stop pooling algorithm still contains only the three loops.]{\small  \textbf{Stop pooling algorithm still contains only the three loops.} \small 
    		With stop pooling, the assignment algorithm has the same structure (gray). Two additional if clauses (blue) check if stops ($\mathbf{o}$ and $\mathbf{d}$) might be pooled: If poolable (green arrow), the stop is inserted without any detour. If not (red arrow), the algorithm proceeds as for standard ride sharing.}
    		\label{fig:alg_SP}
    	\end{figure}
    	To derive a feasible assignment algorithm including stop pooling requires some simplifications, 
    	because the problem of merging stops in stop pooling is NP hard \cite{Aliari.2022,Wang.2022}. In principle, a pooled stop location should allow as many users as possible with a short walk length for all users coupled with temporal constraints because users have to reach a pooled stop before the bus that picks them up. This model allows only stop locations that are already planned in a bus stop list to become pooled stops. This is based on the assumption that once a user is informed where they will be picked up, this pickup point should not change anymore. For this reason, only the users that request their trip later need to walk to the stops of users that have been assigned earlier. This reduces the amount of possible meeting points drastically compared to algorithms where users meet between their originally requested locations. 
    	Thanks to this assumption, the assignment algorithm for stop pooling does not require additional loops which keeps the computing time comparable to the algorithm without stop pooling (Fig.~\ref{fig:alg_SP}). 
    	
    	The assignment algorithm requires only one simple preprocessing that filters all requests with $\ell\le2r$ and lets these users walk their complete trip. The algorithm only assigns requests with  $\ell>2r$ to a bus. For these requests, the assignment algorithm for standard ride sharing is extended by two if clauses: 
    	\begin{itemize}
    	\item For each stop $i$, the algorithm checks if the user could walk from the origin to the stop 
    	\begin{equation}
    		d(\mathbf{o},i)\,\le r\,.
    	\end{equation}
    	If so, the algorithms checks whether the user would arrive at stop $i$ before the bus $b$ does
    	\begin{equation}
    		\tau_\mathrm{request}+\dfrac{1}{\vp} d(\mathbf{o},i)< \tau_\mathrm{arrival}(b,i)\,\,.
    		\label{eq:cond_o}
    	\end{equation}
	    If so, the origin $\mathbf{o}$ is \textit{poolable} with $i$ and does not require insertion of a new stop $\lo(i)=0$. If not, the algorithm calculates the detour $\lo(i)$ as without stop pooling.
    	\item For each stop $j$, the algorithm checks if the user could walk from the  stop $j$ to the destination
    	\begin{equation}
    		d(j,\mathbf{d})\,\le r\,.
    		\label{eq:cond_d}
    	\end{equation}
    	If so, the destination $\mathbf{d}$ is \textit{poolable} with $j$ and does not require insertion of a new stop $\lde(j)=0$. If not, the algorithm calculates the detour $\lde(j)$ as without stop pooling.
    	\end{itemize}
    	In this way, stop pooling is always preferred over insertion of new stops as it yields zero detour for the pooled stop. However, if pooling the origin $\mathbf{o}$ yields a high detour $\lde(j)$ for the destination it is possible that the algorithm finds an option where no stop is pooled with shorter detour $\lo(i)+\lde(j)$.
    	Evidently,  options where both origin $\mathbf{o}$ and destination $\mathbf{d}$ are pooled yield shortest possible detour $\Delta L_\mathrm{min}=0$. If there are multiple options to pool both stops, the algorithm chooses the option with shortest user walk distance $d(\mathbf{o},i)+d(j,\mathbf{d})$. Accordingly, a second trick helps to reduce the computation effort: If the shortest detour $\Delta L_\mathrm{min}=0$, the algorithm jumps from any stop $i$ that is not poolable with the origin $\mathbf{o}$ to $i+1$ without even calculating $\lo(i)$ (for simplicity not illustrated in Fig.~\ref{fig:alg_SP}).

    \subsubsection{Explicit Stop Times Ensure Penalty For Each Stop}
    \label{ch:stop_times}

    In a real ride sharing system, each additional stop induces a delay. Typically buses take a detour to reach the new stop. Even without any detour, buses at least have  to decelerate, serve the user and accelerate again.
    On a network, buses drive from stop to stop on predefined paths, passing many other nodes. It is possible that a new request has origin and destination exactly on a bus route such that the bus serves the new user without any spatial detour. This is more likely, the more buses there are per node. 
    To ensure a delay for any new stop, we extent the model by the stop time $\ts$ - a constant time that a bus spends at each stop before continuing their driving. This stop time $\ts$ represents the time a bus requires to decelerate, drive to the side of the street, open the doors, let users enter or leave the bus, close the doors and accelerate again, regardless how many people enter or leave the bus at one stop. A fixed $\ts$ per stop assumes that users leave and enter rather quickly compared to deceleration and acceleration.
    In public transportation, stop times typically range between ten and twenty seconds \cite{Dueker.2004}. In the example setting in Manhattan we take $\ts=10\,$s. 
    
    With explicit stop times, the ride sharing algorithm does no longer minimize the distance driven of buses (Eq.~\eqref{eq:alg_min_L}) but the route finishing time. The route finishing time is the planned time for the last delivery of a bus. It contains the planned distance driven plus once the stop time $\ts$ for each planned stop. The algorithm
    chooses the best bus as that bus $b$ with shortest planned route finishing time after insertion of the new request.

    \subsubsection{Imbalanced Demand Requires Rebalancing of Buses}
    \label{ch:rebalancing}

    Due to imbalanced requests in the Manhattan taxi data (details in Supplementary Note 2 and \cite[Ch.~3,A]{Lotze.2023}), buses have a tendency to drive to the boundaries of the bus street network and get stuck there. 
    Different algorithms exist to move empty buses actively towards regions with high demand \cite{Fielbaum.2021b, AlonsoMora.2017,Lu.2021,Wen.2017}. We implement a simpler algorithm that always sends back buses towards a central node with high average request rate without looking for current regions of high demand.
    The rebalancing algorithm sends any bus towards a center node - the Time Square - when idling. 
    An empty bus with empty job lists drives into direction of the Time Square either until the bus reaches it or until a new request is assigned to the bus. A new request becomes more likely when the bus is more central and not at the boundary. Buses rarely reach the rebalancing node, but only drive into its direction. 
    With the simple rebalancing procedure, buses distribute according to the requests and more buses are in the center than in the north of Manhattan (details in \cite[Ch.~3]{Lotze.2023}). One consequence of rebalancing are shorter user travel times, because users are better distributed over all buses  (details in \cite[Ch.~A]{Lotze.2023}).

\null
\newpage
\section*{Supplementary Note 2: Manhattan Street Network and Taxi Data}  
The model uses data from Manhattan (New York, USA) as example city to yield interpretable results. This note introduces the data and describes how they contribute to the model. 
For further details we refer the reader to \cite[Ch.~3,A]{Lotze.2023}.

  \subsection{Fine-Grained Street Network Enables Short Walk Distances}
    \label{ch:street_data}

    \begin{figure}[b!]   
    	\begin{center}
    		\includegraphics[height=6cm]{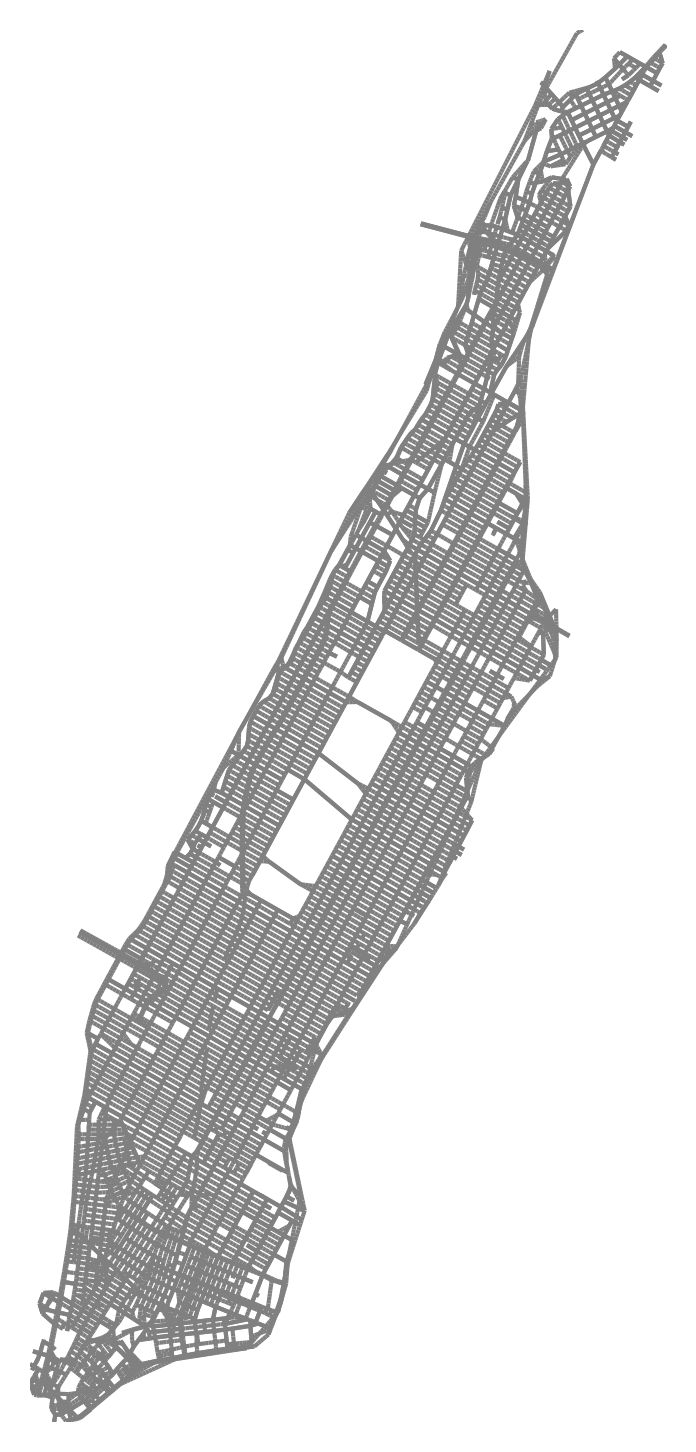}
    	\end{center}
    	\vspace{-0.6cm}
    	\caption[An example setting uses the street network of Manhattan (New York, USA).]{\small  \textbf{An example setting uses the street network of Manhattan (New York, USA)} \small 
    		with more than $\num{4000}$ corners. }
    	\label{fig:manhattan_map}
    \end{figure}

\begin{figure}[t!]   	
	\begin{subfigure}[t]{7.4cm}
		\textbf{a}
		\vspace{-0.5cm}
		\begin{center}
			\includegraphics[width=7cm]{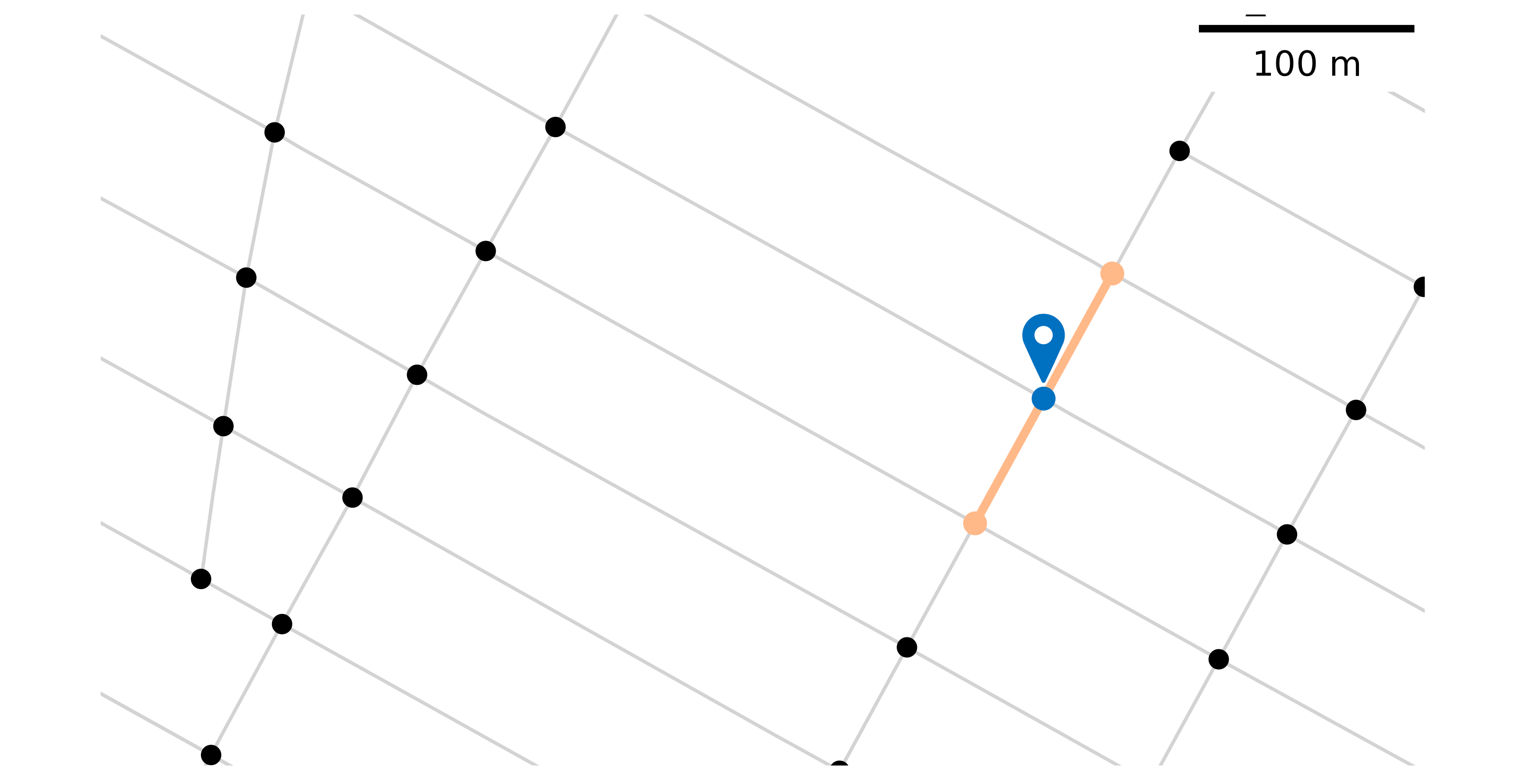}
		\end{center}
	\end{subfigure}\hfill
	\begin{subfigure}[t]{7.4cm}
		\textbf{b}
		\vspace{-0.5cm}
		\begin{center}
			\includegraphics[width=7cm]{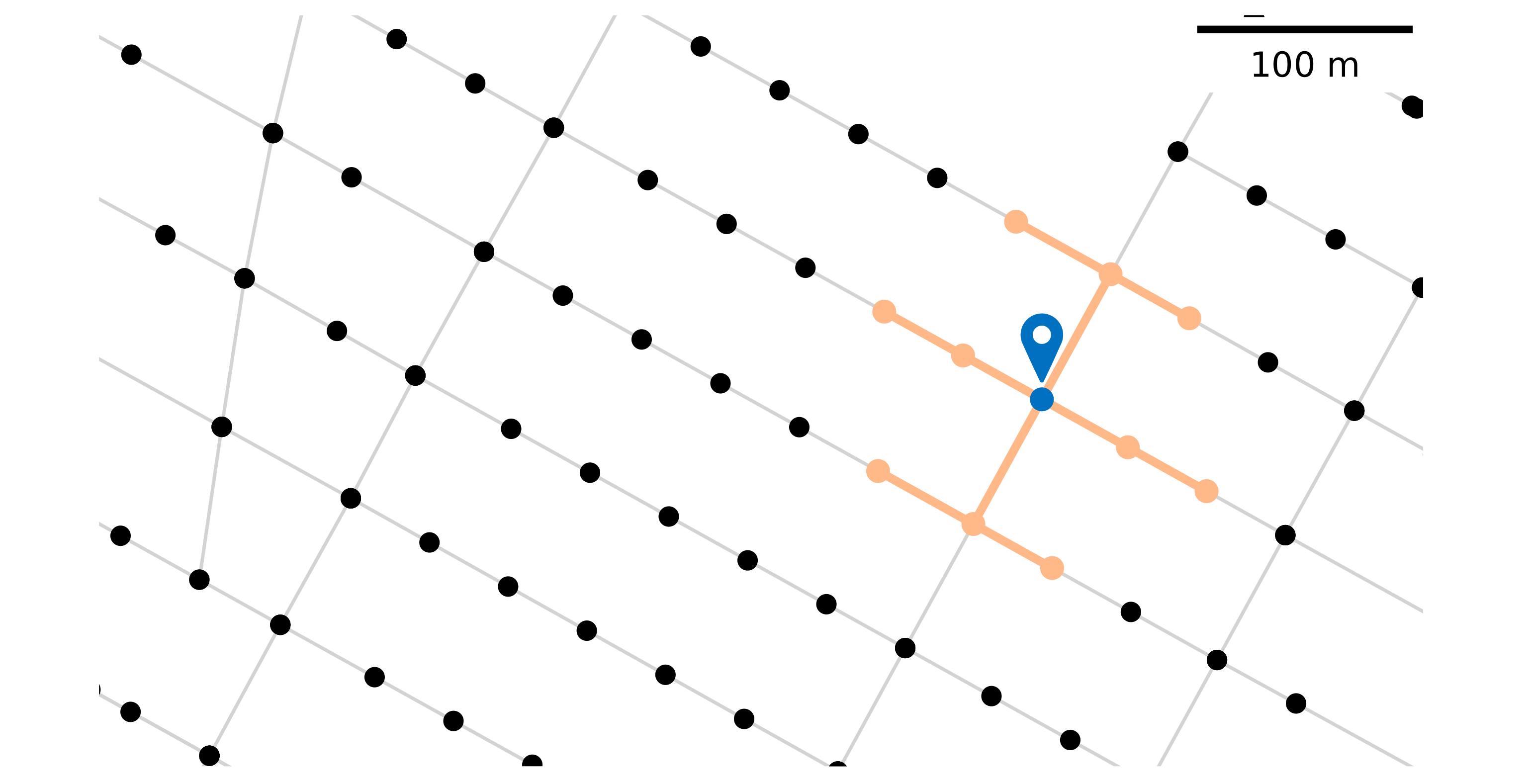}
		\end{center}
	\end{subfigure}
\vspace{-0.25cm}
	\caption[Modeling short user walks requires sufficiently fine street network.]{\small  \textbf{Modeling short user walks requires sufficiently fine street network. } \small  (a) When nodes are only situated at corners of the street network users hardly reach any stop with a short walk limit $r=\SI{100}{\metre}$ (e.g.~blue pin only reachable from orange nodes along orange paths). Depending on the local network structure, users might only walk into some directions but not into others. (b) With a finer network with a maximal edge length of $\dmax=\SI{50}{\metre}$, users reach the example stop (blue pin) already from multiple neighbored stops at a short walk limit of $\SI{100}{\metre}$ and the walk options are comparable  into  all directions (e.g.~blue pin reachable from all orange nodes along orange paths).}
	\label{fig:r_in_Manhattan}
\end{figure}

    In the model, users and buses move along a network that represents a real street graph of an example city: Manhattan (New York, USA).
    Street network data is publicly available on OpenStreetMap \cite{OpenStreetMapcontributors.2023}. In Python, it can be download easily using the package \textit{osmnx} \cite{Boeing.2017}. The provided street network consists of nodes at corners connected by edges that represent streets. The network is directed and contains one way streets. The model weights each edge with the respective length of each street provided by OpenStreetMap \cite{OpenStreetMapcontributors.2023}, but has no information on the priority or average speed along a street. Again, buses and users move with constant velocity 
    that is the same on all streets.
    In this network, all nodes are possible stop locations (origins, destinations, pickup points or delivery points). 
    Thus, users only walk between two nodes in stop pooling. To enable an analysis with different short walk lengths, a finer network is required than only the street corners provided by OpenStreetMap. Additional in-between nodes along the edges divide each edge into $N_\mathrm{ib}$ equidistant edges of maximal edge length $d_\mathrm{e,max}$ (Fig.~\ref{fig:r_in_Manhattan}).
    The procedure goes as follows (adaption of code by Debsankha Manik \cite{Manik.2020}):
    The number of in-between $N_\mathrm{ib}$ nodes  is the integer quotient of the edge length $d$ and the maximal edge length $d_\mathrm{e,max}$
    \begin{equation}
    	N_\mathrm{ib}=\mathrm{int}\left(\dfrac{d_\mathrm{e}}{d_\mathrm{e,max}}\right)\,\,.
    \end{equation}
    If the edge length $d_\mathrm{e}$ is already smaller than $d_\mathrm{e,max}$, no new stop is inserted, $N_\mathrm{ib}=0$. If the edge length $d$ is larger than $d_\mathrm{e,max}$, $N_\mathrm{ib}$ in-between stops are inserted, each connected by an edge of length
    \begin{equation}
    	d_\mathrm{ib}=\dfrac{d_\mathrm{e}}{N_\mathrm{ib}+1}\le d_\mathrm{e,max}\,,
    \end{equation}
    that is at most $d_\mathrm{e,max}$.
    In this way, the network edges are all of length $d_\mathrm{e}\le d_\mathrm{e,max}$ and the street length between to corners 
    is maintained.

    The positions of the in-between nodes require additional treatment, because the original street network is a directed graph.  If there are two equivalent edges between two nodes $i$ and $j$ in opposite direction $e_{ij}$ and $e_{ji}$ the in-between nodes on both edges have the same locations. A pair of these nodes represents one spatial location on opposite sides of the street. In the bus street network, these in-between node pairs remain unconnected, because buses are not allowed to turn within a street.    	
    By contrast, a separate, undirected user walkway network represents that users walk along streets in both directions, independent of the side of the street or one way streets. This user walkway network is mainly equivalent to the directed bus street network, but in addition connects in-between nodes with similar location by a short edge of length $d_0=\SI{10}{\metre}$. These short edges model the option for users to cross a street. 
    
	The average speed in Manhattan ranges between $6.6$ and $9.1$ miles per hour  (approximately $\SI{10}{\kilo\metre\per\hour}$ to $ \SI{15}{\kilo\metre\per\hour}$) \cite{NewYorkCityDepartmentofTransportation.June2018}. 
	Consequently, buses drive on a directed bus network with average velocity  $\vv= \SI{12}{\kilo\metre\per\hour}=5.\bar{5}\,\si{\metre\per\second}$ and users walk on a separate undirected user walk network with similar nodes with a constant velocity $\vp= \SI{4}{\kilo\metre\per\hour}=1.\bar{1}\,\si{\metre\per\second}$ if not stated differently.
     
    The example street network in Manhattan has $\num{4321}$ corners. The fine grained street network with maximal edge length $\dmax=\SI{50}{\metre}$ contains $\num{22483}$
    nodes with an 
     average shortest path length of $\SI{7200}{\metre}$ and a 
     diameter of $\max(d)=\SI{22462}{\metre}$. 
    The undirected user network contains paths that do not exist in the directed bus network. Consequently, it has a shorter average shortest path length of $\SI{3991 }{\metre}$ and a shorter diameter of $\max(d)= \SI{12538}{\metre}$.
    The original Manhattan street network resembles a grid with an average degree $k=4.4$ (see \cite[A.1]{Lotze.2023}).  A grid has   not the best topology for ride sharing but it is also not  the worst  \cite{Manik.2020} and is hence a feasible example setting to analyze the dynamics of ride sharing.

    \subsection{Data-Driven Demand is Heterogeneous}
    \label{ch:demand_data}

    The New York Yellow Cab taxi data \cite{CityofNewYork.2016} is widely used to estimate the demand for ride sharing services  \cite{Aliari.2022,AlonsoMora.2017, Barann.2017,Santi.2014}.
    This model draws requests from  the taxi trip data Manhattan in 2016, using the positions where taxi users have been picked up as origins and the positions where taxi users have been delivered as destinations and the pickup times as request times. We only consider trips with start and end within Manhattan, where trips are dense.
    Each taxi ride generates one request, independent of the passenger count (infinite bus capacity), such that per se no trips are shared and no stops are pooled. The stop locations are mapped to the geographically closest network node. If two nodes have similar positions, we shifted their location 1 millimeter along the outgoing directed edge in the bus street network. In this way, the node mapping is unique and repeatable. In the demand data, request rates fluctuate over time. We model different request rates: 
    \begin{enumerate}
    	\item Artificial constant request rate $\lambda=\,$const,
    	\item Real request rates, $\lambda$ from input.
    \end{enumerate}
    For the artificial request rates requests are sampled from all taxi trips and the request time is drawn from a Poisson process with request rate $\lambda$. 
    For the real request rates $\lambda$, a time window is chosen, including the fluctuating request rates and the varying directions in the requests. 
    Liu \textit{et al.}~identify four clusters in the New York Yellow Cab taxi data \cite{Liu.2019}: 00:00-08:00, 08:00-10:00, 10:00-18:00, 20:00-24:00. Different directions also influence the performance of ride sharing \cite{Ruijter.2023}.
    Besides, the demand pattern depends on the daytime and the weekday (details in \cite[Ch.~3]{Lotze.2023}). 
    
As an example we choose
Wednesday, the 24th of February 2016 (no holiday). Typically, ride sharing providers react to high fluctuations in the request rate $\lambda$ by adapting the fleet size $B$ \cite{Oh.2020}. To avoid extreme fluctuations in the request rate $\lambda$
    the time window is restricted to $\tau>\,$6:00.
	In addition, a short analysis of the taxi data supports the chosen bus velocity $\vv={\SI{12}{\kilo\metre\per\hour}}$. The pickup and delivery times in the Manhattan request data yield an average trip duration of $\SI{16.2}{\minute}$ \cite{Kedari.2020}. With an average trip length of $\SI{2804}{\metre}$ 
	a bus velocity $\vv={\SI{12}{\kilo\metre\per\hour}}$ yields a comparable average trip duration of $ \SI{14}{\minute}$ on the shortest path. Thus, the bus velocity  ${\vv=\SI{12}{\kilo\metre\per\hour}}$ is slightly above the real average speed of taxis in Manhattan and overestimates the service quality.

\null
\newpage
\section*{Supplementary Note 3: Analysis and Results}

		\subsection{Analysis}
		Simulations start with empty, randomly distributed buses that have no stop planned. 
    	We evaluate only times after 7:00. Between 6:00 and 7:00, buses accumulate a planned job list and distribute according to the request rate. For fluctuating demand, the simulations do not equilibrate to a steady state because of fluctuating demand pattern. For steady state analyses, we sample requests to achieve constant request rate $\lambda$ and average trip length $\lav$. Then, the results are evaluated in the steady state within a fixed observation time window $\tau_\mathrm{min}\le\tau\le\tau_\mathrm{max}$ after sufficient time for equilibration  (analogue to \cite{Manik.2020}). 
	\subsubsection{Observables}
    	\label{ch:analysis}

    	\begin{figure}[b!]   	
    		\begin{subfigure}[t]{7.4cm}
    			\begin{center}\sfl{a}{
    				\includegraphics[height=3cm]{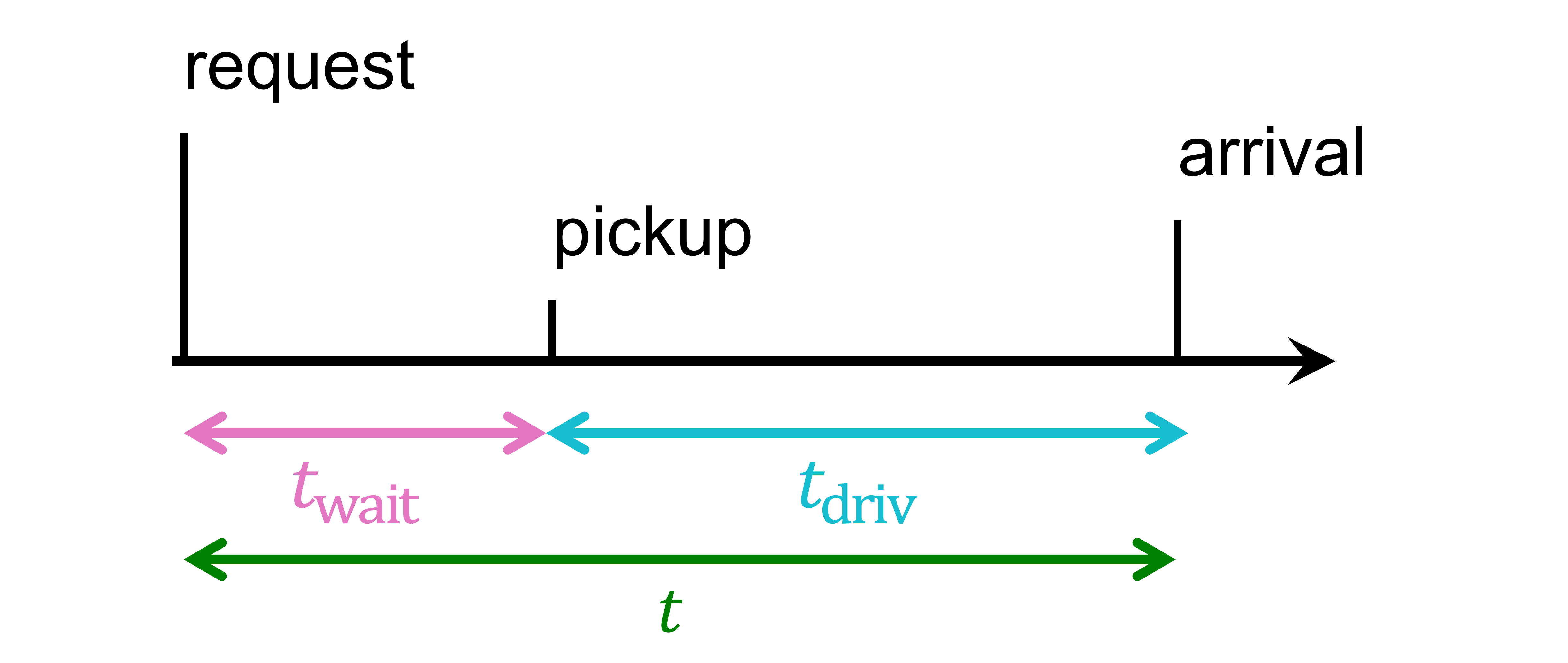}}
    			\end{center}
    		\end{subfigure}\hfill
    		\begin{subfigure}[t]{7.6cm}
    			\begin{center}\sfl{b}{
    				\includegraphics[height=3cm]{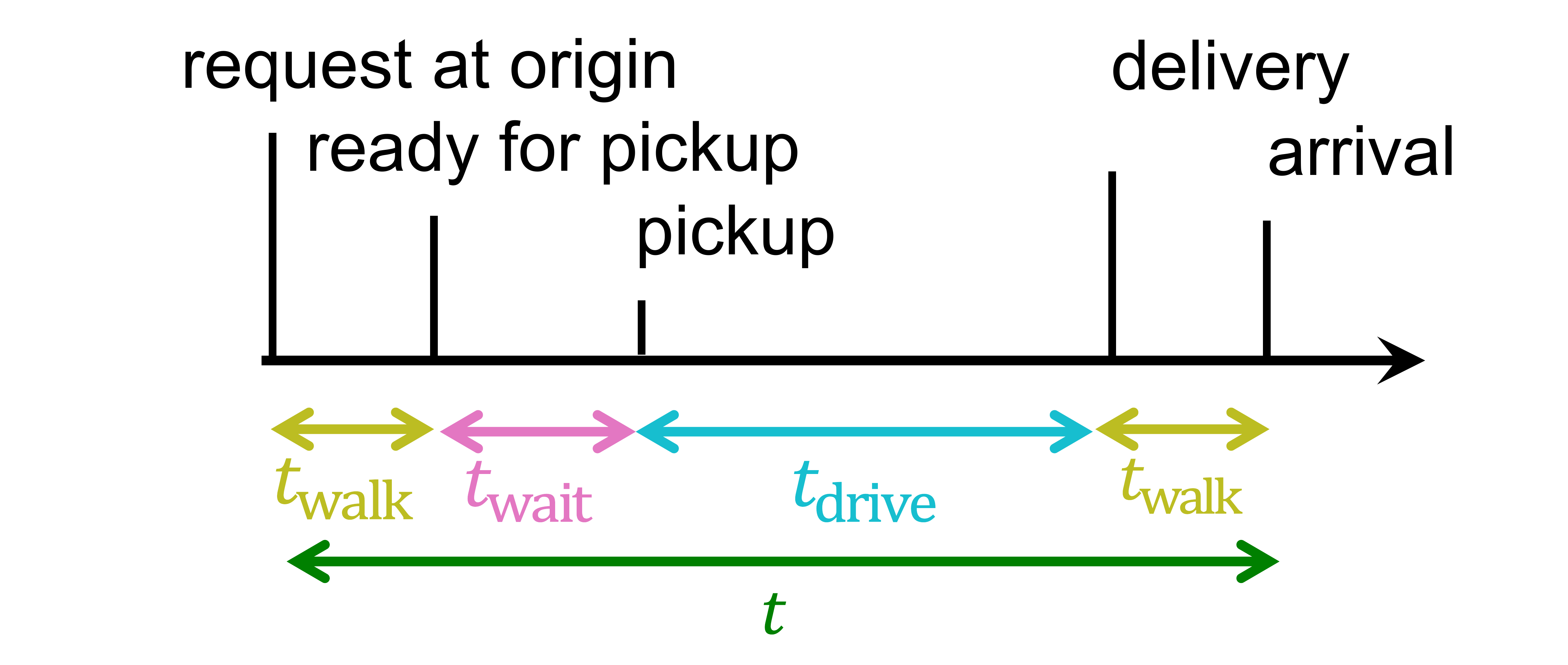}}
    			\end{center}
    		\end{subfigure}
    	\vspace{-0.25cm}
    		\caption[The travel time consists of walk time, wait time and drive time.]{\small  \textbf{The travel time consists of walk time, wait time and drive time.}
    		\small  
    			In general, the travel time $t$ is the total time between a user's request and their arrival at the destination. (a) In standard ride sharing, the travel time consists of the wait time $\twait$ between a user request and pickup and the drive time $\tdriv$ between a user pickup and delivery. (b) With stop pooling, users might walk from their requested origin to a nearby stop where they are picked up and from the stop where they are delivered to their final destination. This might add additional walk time $\twalk$ to the travel time $t$. Although the symbols denote averages over all users in this thesis, this figure illustrates time spans for one individual user.   	}	
    		\label{fig:t_split}
    	\end{figure}
    	
    	The key observable is how long users travel. The travel time $t$ is the average time span between a user's request at time $\tau_{\mathrm{request}}$ and their arrival at their destination at $\tau_{\mathrm{arrival}}$, either by bus or by feet. This travel time 
    	\begin{align}
    		t:=\langle \tau_{\mathrm{arrival}}-\tau_{\mathrm{request}}\rangle=\twalk+\twait+\tdriv\,
    		\label{eq:t}
    	\end{align} 
    	splits into walk time $\twalk$, wait time $\twait$, and drive time $\tdriv$. The walk time of a user is proportional to their walk distance
    	\begin{equation}
			\twalk=\dfrac{\dwalk}{\vp}\,.
    	\end{equation}
    	If a user does not walk (i.e.~the user is picked up at their origin and delivered at their destination) the walk time equals zero.
    	The wait time $\twait$ is the time span between the time the user arrives at their pickup point $\mathbf{s}_1$ - the request time plus the walk time at the origin - until they are picked up. 
    	\begin{align}
    		\twait= \tau_{\mathrm{pickup}}-\left(\tau_{\mathrm{request}}+\dfrac{d(\mathbf{o},\mathbf{s}_1)}{\vp}\right)\,\,.
    	\end{align} 
    	The time a user walks to the pickup point does not contribute to the wait time.
    	If the user is picked up at the origin, the walk distance at the origin $d(\mathbf{o},\mathbf{s}_1)=0$ and the wait time $\twait$ is simply the time span between request and pickup.
    	The drive time is the time span from the pickup of the user until their delivery
    	\begin{align}
    		\tdriv= \tau_{\mathrm{delivery}}-\tau_{\mathrm{pickup}}\,.
    	\end{align} 
    	Fig.~\ref{fig:t_split} illustrates how the travel time $t$ splits into wait time $\twait$, drive time $\tdriv$ and walk time $\twalk$. Typically, users value walk, wait and drive time differently \cite{Wardman.2004}. However, we simply compare the total travel time $t$ as the sum of walk, wait and drive time.
    	The travel time of rejected users is the time they walk their complete trip, which reads
    	\begin{align}
    		t_\mathrm{r}=\dfrac{\ell}{\vp}\,,
    	\end{align}
    	because they walk their trip length $\ell$ with constant user velocity $\vp$.

        In general, we consider demand as requested trip length characterized by request rate $\lambda$ and average trip length $\ell$. This corresponds to the counter of the load \cite{Molkenthin.2020,Manik.2020}
    	\begin{equation}
    		q=\dfrac{\lambda\lav}{B\vv}\,.
    		\label{eq:load}
    	\end{equation}
    	This load $q$
    	is a dimensionless rescaled request rate relative to the number $B$ of buses available. The load $q$ reduces two important simulation inputs: the demand in terms of request rate $\lambda$ and average trip length $\lav$, and the fleet in terms of fleet size $B$ and bus velocity $\vv$. An intuitive interpretation of the load $q$ is the  minimal bus occupancy \cite{Molkenthin.2020}. 
    	Higher loads $q$ yield more efficient ride sharing \cite{Molkenthin.2020}. The volume covered by a planned bus route increases linearly with the load - independent on the network topology \cite{Manik.2020}. 
    	
    \subsubsection{Bisection Method to Find Minimal Travel Time with Small Effort}
    \label{ch:bisection_method}
    
\begin{figure}[b!]   
	\begin{subfigure}[t]{ 0.49\textwidth}
		\begin{center}\sfl{a}{
				\includegraphics{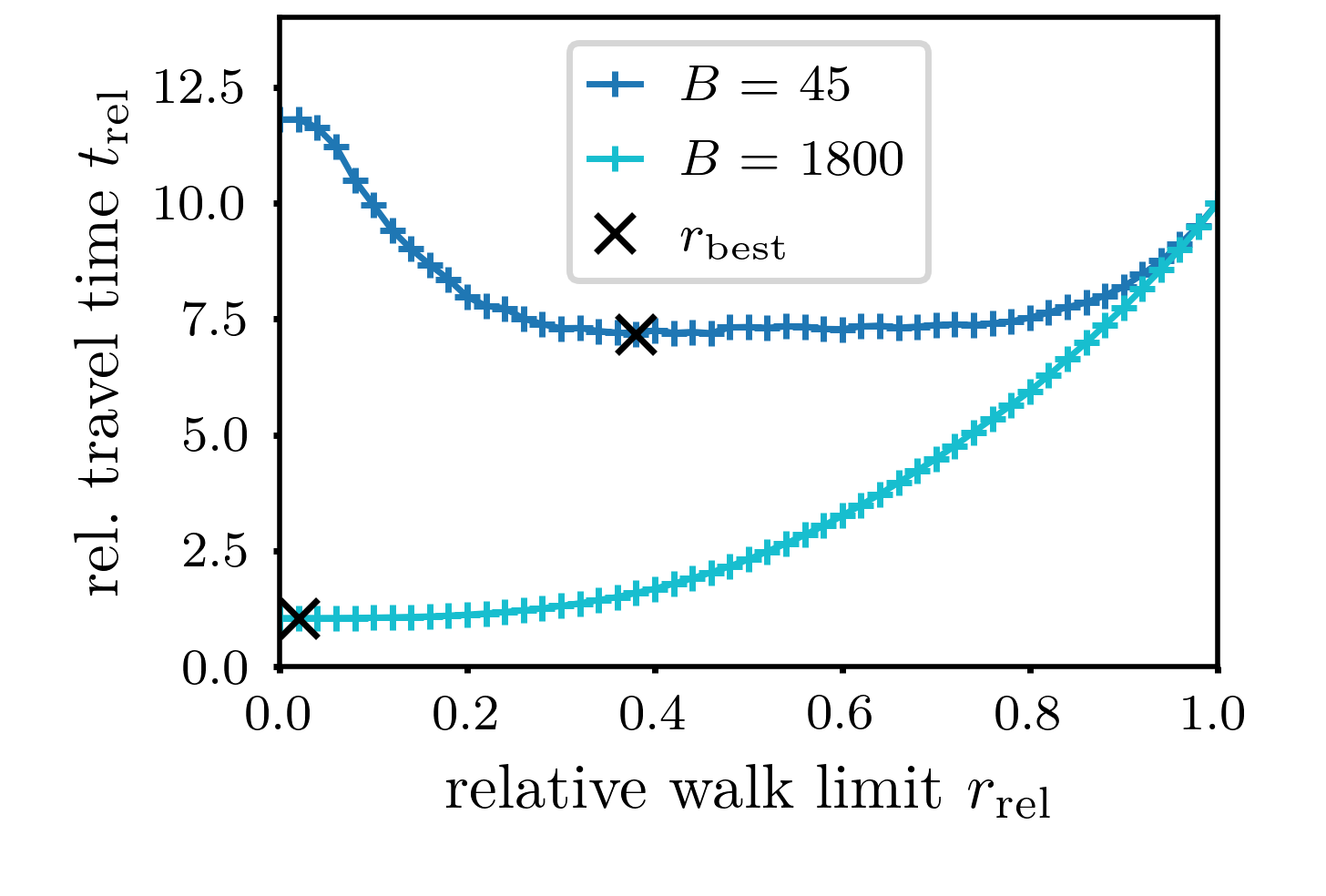}}
		\end{center}
	\end{subfigure} 
	\begin{subfigure}[t]{ 0.49\textwidth}
		\begin{center}\sfl{b}{
				\includegraphics{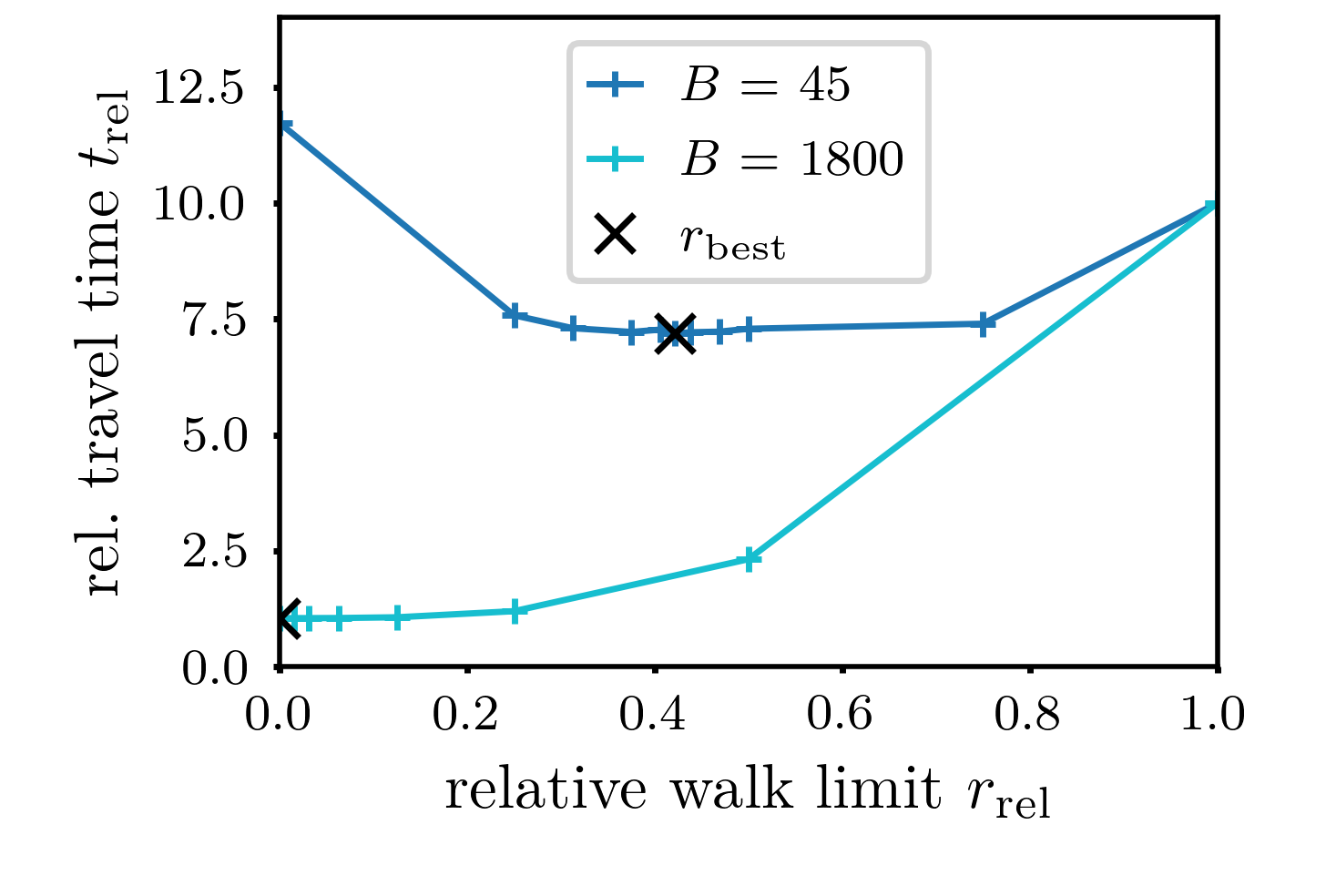}}
		\end{center}
	\end{subfigure} 
	\begin{subfigure}[t]{\textwidth}
		\begin{center}
			\begin{tabular}{l|c|c|c|c|c|c|}
				\textbf{c}& \multicolumn{3}{c|}{$B=45$}&\multicolumn{3}{c|}{$B=1800$}\\
				&$N$& $\rb$& $\min(t)$&$N$& $\rb$& $\min(t)$\\
				\hline
				Equidistant simulations with $\Delta r=0.05$&51&$0.095$&$2.398$&51&$0.005$&$0.352$\\
				Bisection method, $\varepsilon=0.05$&11&$0.105$&$2.398$&8&$0.000$&$0.351$\\
			\end{tabular}
		\end{center}
	\end{subfigure} 
	\caption[The bisection method reduces the computation effort.]{\small  \textbf{The bisection method reduces the computation effort.} \small 
	The bisection method (panel b) finds the minimum travel time $\min(t)$ and best walk limit $\rb$ (black x) with smaller number $N$ of simulations (panel c) but almost the same precision as an equidistant simulation (panel a).
	When stop pooling simulations yield roughly constant the travel times $t$  for many $r$ the results of the bisection method deviates slightly from the result of equidistant simulations ($B=45$). }
	\label{fig:opt_alg}
\end{figure}

    The travel time $t$ is convex as a function of the walk limit $r$ with a minimum in a fixed interval of $r$ (Fig.~\ref{fig:opt_alg}). 
    Let the best walk limit 
    \begin{equation}
    	\rb=r|_{t=\min(t)}\,,
    \end{equation}
	be that walk limit $r$ with minimal travel time $t$.
	A bisection method \cite{Burden.1985} is feasible to find $\rb$ numerically. The bisection method assumes that there is only one minimum for the travel time $t$ in the considered closed interval $r\in[0,\rmax]$. A pseudo code describes the bisection method in detail (\cite[A.2]{Lotze.2023}).
	In general, the bisection method tests the travel time numerically at test points $r_1,r_2\in[a_i,b_i]$ and then chooses a new interval $[a_{i+1},b_{i+1}]$ with half the size $b_{i+1}-a_{i+1}=(b_{i}-a_{i})/2$ depending on the comparison on the values $t(r_1)$ and $t(r_2)$. This procedure is repeated until the interval size $b_{i}-a_{i}$ reaches a pre-set lower bound $\varepsilon$. In this way, the algorithm approaches the desired value $\min(t)$ iteratively. Here,
    the boundaries $r=0$ or $r=\rmax$ might yield the minimal travel time $\min(t)$ as well. Thus, the bisection method also considers the boundaries $r=0$ or $r=\rmax$ as possible solutions.

	The algorithm works for different example functions as long as there is only one minimum which is true for the convex travel time functions observed. When simulation results fluctuate, the minimum found with the bisection method might differ slightly from the minimum of multiple equidistant simulations that scan all $r$ in the interval in fine steps (Fig.~\ref{fig:opt_alg}). In particular when fluctuations lead to multiple minima, the bisection method might yield imprecise results. Clearly, the bisection method requires less steps and hence less simulations and evaluations of the travel time $t$ (Fig.~\ref{fig:opt_alg}c). It is thus computationally more feasible. 
	
\subsection{Additional Results}

\begin{figure}[b!]
	\begin{subfigure}[t]{7.5cm}
		\begin{center}
		  16:00 - 17:00 
			\sfl{a}{\vspace{0.1cm}
			\includegraphics{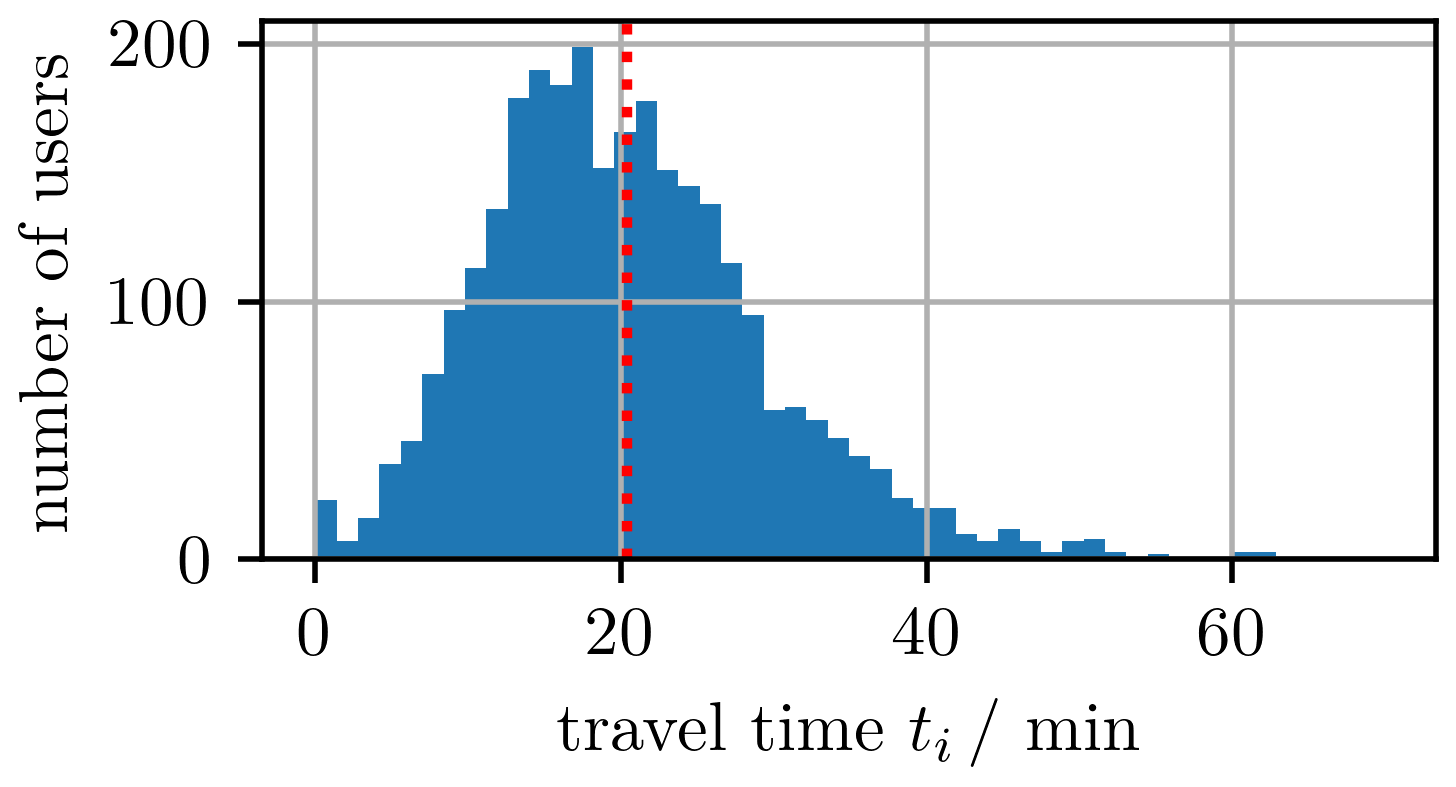}}
		\end{center}
	\end{subfigure}
	\begin{subfigure}[t]{7.5cm}
		\begin{center}
		 21:00 - 22:00 
			\sfl{b}{\vspace{0.1cm}
			\includegraphics{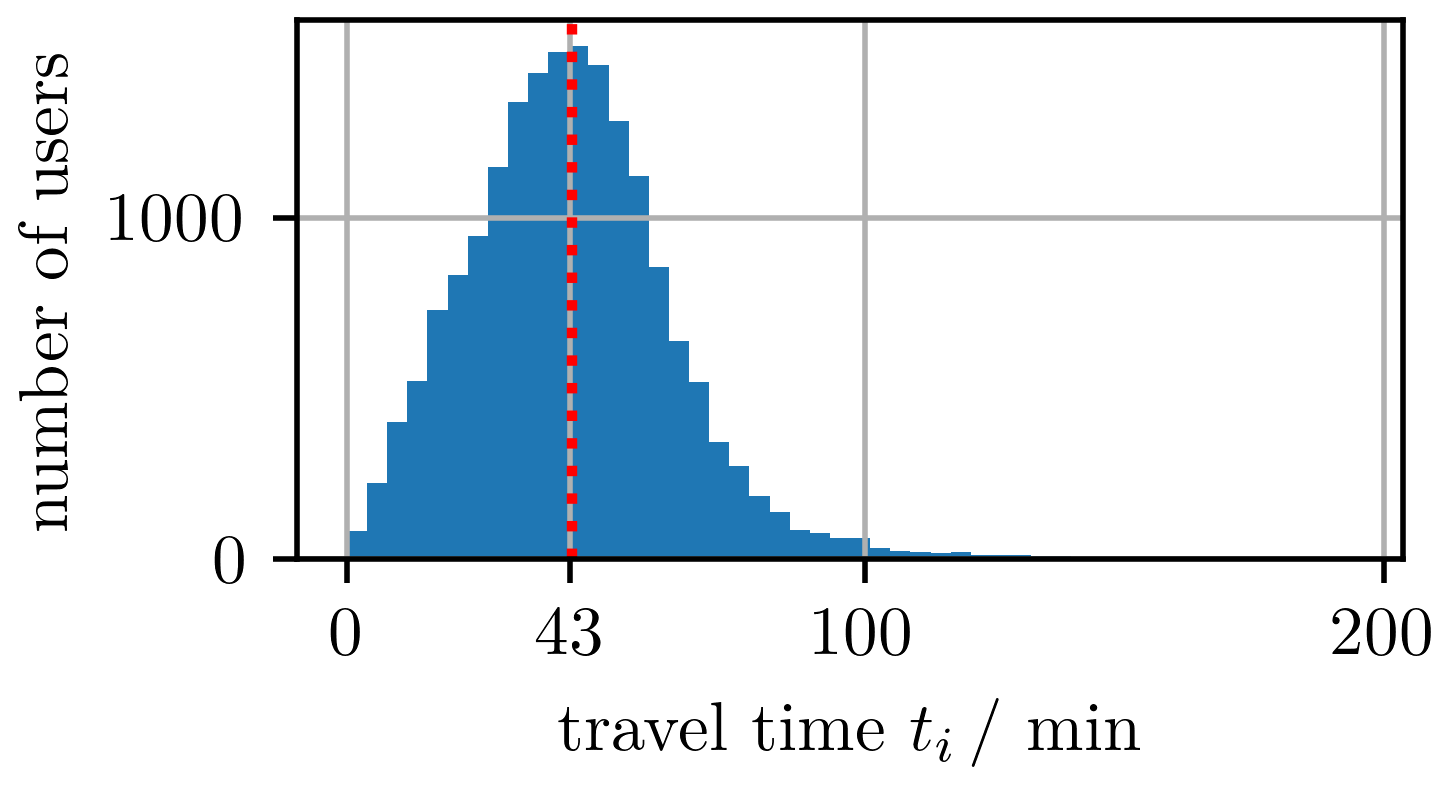}}
		\end{center}
	\end{subfigure}
	\begin{subfigure}[t]{7.5cm}
		\begin{center}
			\sfl{c}{\vspace{0.1cm}
			\includegraphics{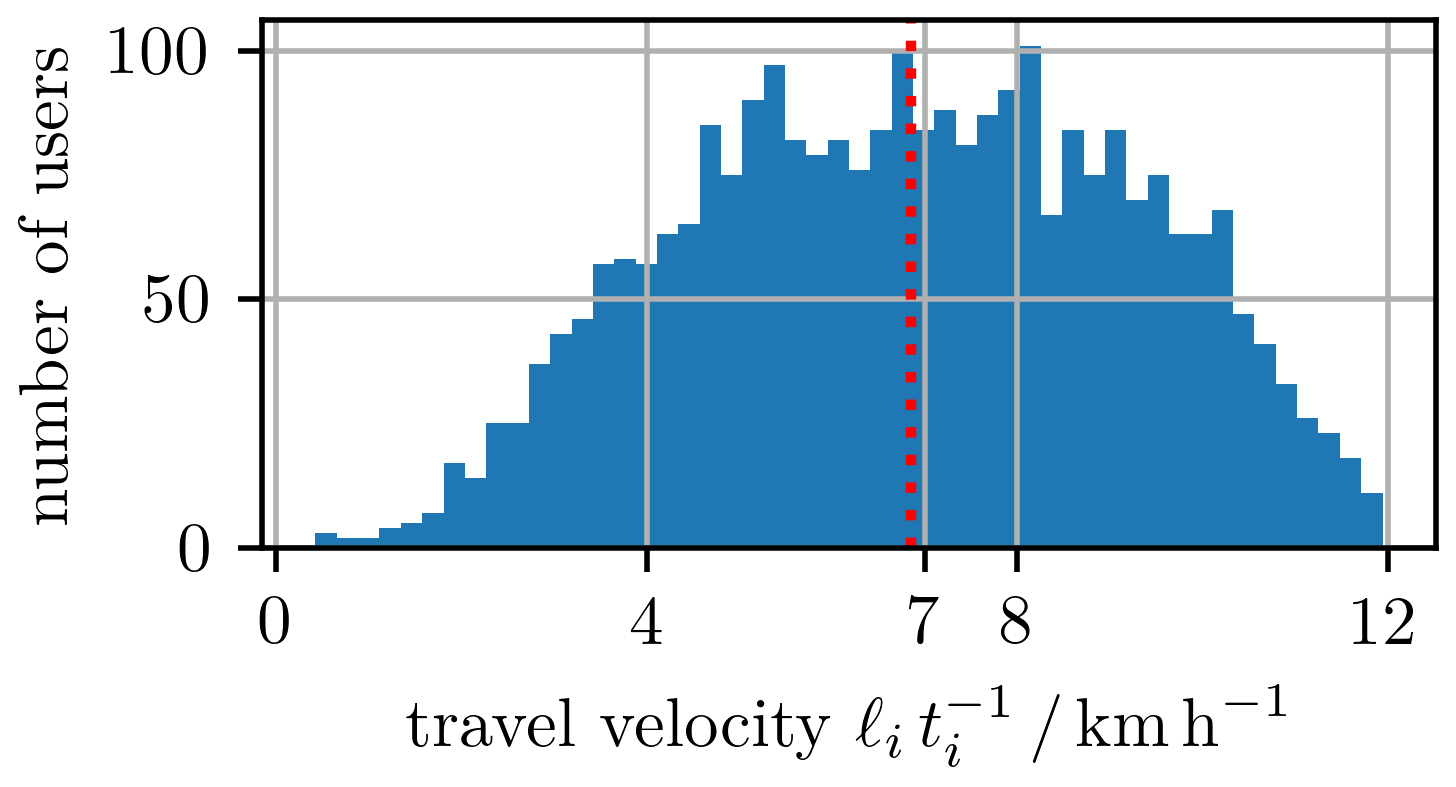}}
		\end{center}
	\end{subfigure}
	\begin{subfigure}[t]{7.5cm}
		\begin{center}
			\sfl{d}{\vspace{0.1cm}
			\includegraphics{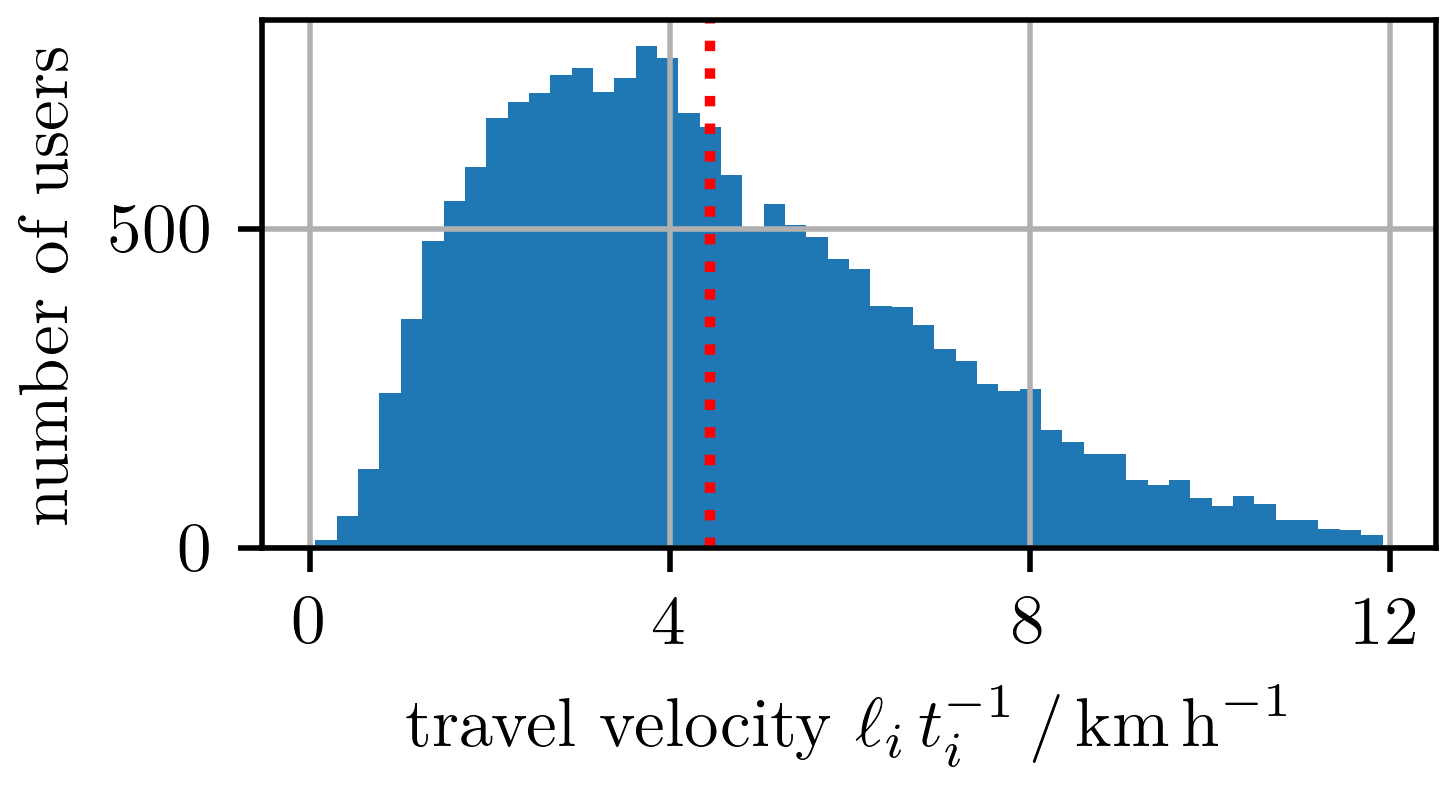}}
		\end{center}
	\end{subfigure}
	\caption{\small\textbf{The travel time distribution in both intervals is comparable.} The individual travel time $t_i$ of users who request a ride between 16:00 and 17:00 (panel a) is less than half of the travel time $t_i$ of users who request a ride between 21:00 and 22:00 (panel b) on average (dotted red line), but the shape of the distributions is similar - demonstrated by histograms with 50 bins. The travel velocity (direct trip length $\ell_i$ divided by travel time $t_i$) of users who request a ride between 16:00 and 17:00 (panel c) is respectively larger than that of users who request a ride between 21:00 and 22:00 (panel d). The travel velocity 	 is smaller than the driving velocity of buses, $\vv=\SI{12}{\kilo\metre\per\hour}$, because of wait times and detours while driving.  
	 }
	\label{fig:travel_time_dist}
\end{figure}

\begin{table}[b!]
    \caption{Key indicators of distribution of individual travel time $t_i$ and travel velocity (direct trip length $\ell_i$ divided by travel time $t_i$) compared for users who request a ride between 16:00 and 17:00 and between 21:00 and 22:00.}
    \label{tab:travel_time_dist}
    \centering
    \begin{tabular}{l|l|r|r}
        observable&indicator&{16:00 - 17:00}&{21:00 - 22:00}     \\[3pt]\hline
        individual travel time $t_i\,/\min$ & mean & $\SI{20.4}{}$&$\SI{43.3}{}$\\
                                    & std & $\SI{9.4}{}$&$\SI{20.2}{}$\\
                                    & min & $\SI{0.0}{}$&$\SI{0.0}{}$\\
                                    & max & $\SI{69.8}{}$&$\SI{194.0}{}$\\[3pt]\hline
        individual travel velocity $\ell_i\,t_i^{-1}/\,\mathrm{km\,h}^{-1}$ & mean & $\SI{6.9}{}$&$\SI{4.4}{}$\\
                                                    & std & $\SI{2.4}{}$&$\SI{2.3}{}$\\
                                                     & min & $\SI{0.4}{}$&$\SI{0.1}{}$\\
                                                    & max & $\SI{11.9}{}$&$\SI{11.9}{}$\\
    \end{tabular}
\end{table}

We here provide details on the distribution of the individual travel time $t_i$  for standard ride sharing, $\rt=0$, in the example time intervals from Tab.~1 in the main manuscript. In general, the shape of the distributions is comparable (Fig.~\ref{fig:travel_time_dist}a,b). The travel time has a smaller mean and smaller standard deviation between 16:00 and 17:00 than between 21:00 and 22:00 (Tab.~\ref{tab:travel_time_dist}). In consequence, the travel velocity (direct trip length $\ell_i$ divided by travel time $t_i$) is higher between 16:00 and 17:00 than between 21:00 and 22:00 (Tab.~\ref{tab:travel_time_dist}). Because the travel velocity has an upper limit in the driving velocity of buses, $\vv=\SI{12}{\kilo\metre\per\hour}$, the shape of the distributions differs. The distribution for users between 21:00 and 22:00 is more weighted to the left than the distribution of users between 16:00 and 17:00 (Fig.~\ref{fig:travel_time_dist}c,d).

\begin{figure}[t!]
	\begin{subfigure}[t]{5cm}
		\textbf{a}\hfill  $\rt=\SI{3.75}{\minute}$\hfill \null
		\begin{center}
			\vspace{-0.2cm}
			\includegraphics[height=6cm]{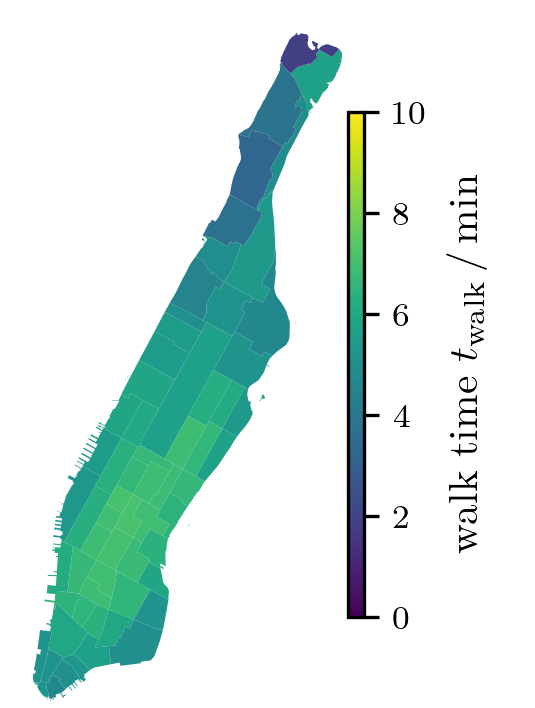}
		\end{center}
	\end{subfigure}
	\begin{subfigure}[t]{5cm}
		\textbf{b}\hfill  $\rglob(\tau)$\hfill \null
		\begin{center}
			\vspace{-0.2cm}
			\includegraphics[height=6cm]{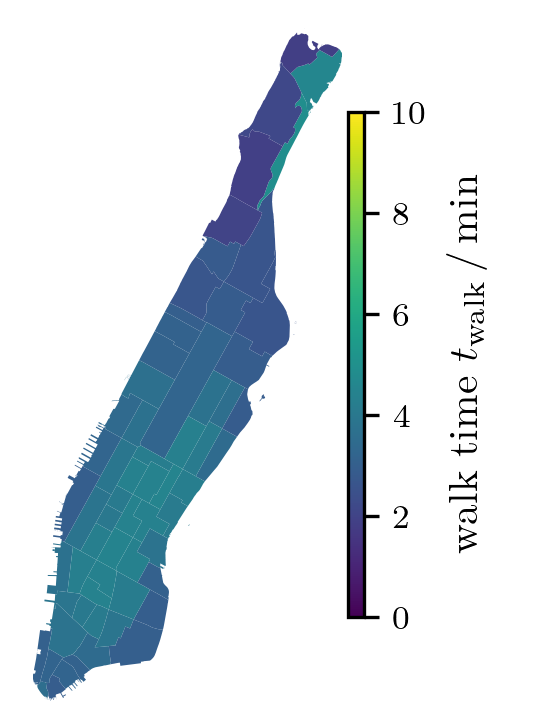}
		\end{center}
	\end{subfigure}
	\begin{subfigure}[t]{5cm}
		\textbf{c}\hfill $\rloc(\tau)$\hfill \null
		\begin{center}
			\vspace{-0.2cm}
			\includegraphics[height=6cm]{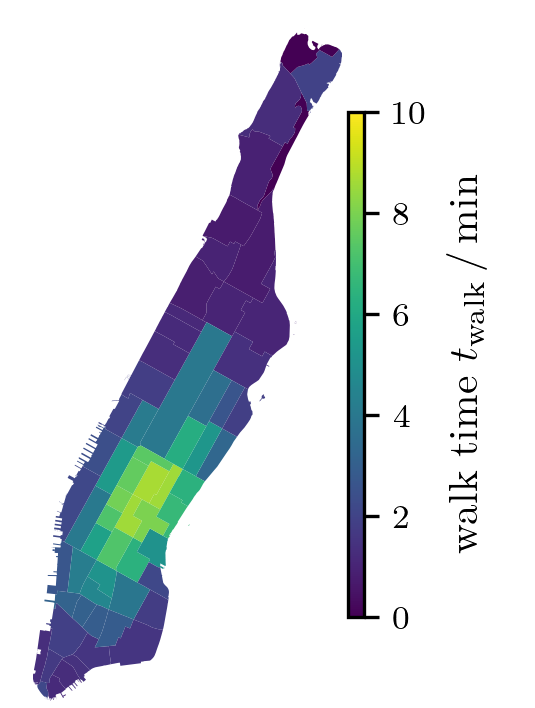}
		\end{center}
	\end{subfigure}
	\caption{\small  \textbf{A local walk limit lets users walk in regions of high demand.}
	\small  
	 With a fixed mean-field walk limit $\rt=\SI{7.5}{\minute}$ (panel a) and a global walk limit $\rglob(\tau)$ (panel b), the average walk time $\twalk$ is almost homogeneous in all taxi zones. Instead, the heterogeneous pattern of the average walk time $\twalk$  with a local walk limit $\rloc(\tau)$ (panel c) is comparable to that of the requests (main manuscript, Fig.~5a). The average walk time $\twalk$ is higher in zones with high request rates (panel c).}
		\label{fig:adapt_r_in_spacetime_supp}
\end{figure}

\begin{figure}[h!]  
	\begin{subfigure}[t]{\textwidth}
		\begin{center}
				\includegraphics{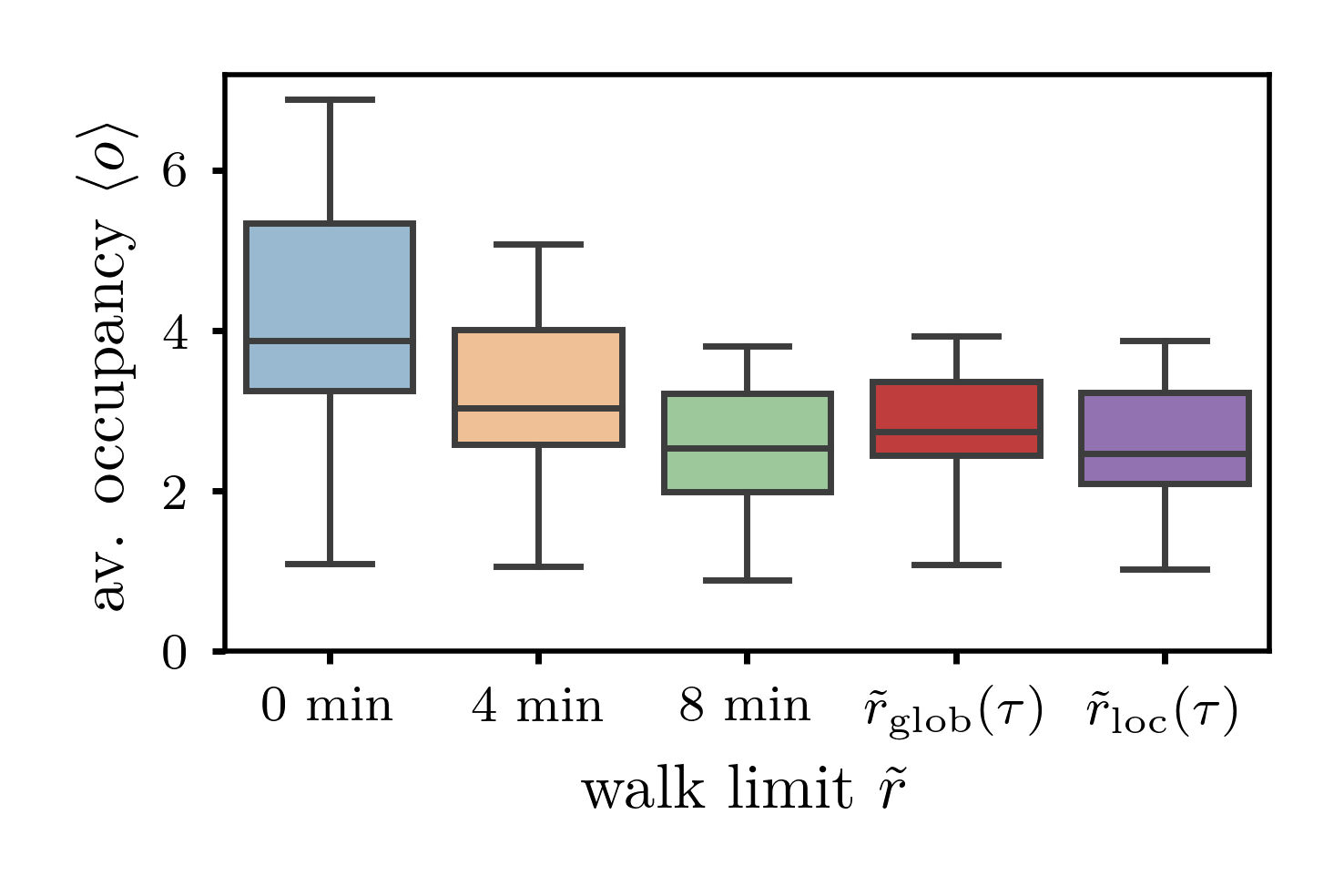}
		\end{center}
	\end{subfigure}\vspace{-0.5cm}
	\caption[Travel time and  occupancy fluctuate less when users accept short walks.]{\small  \textbf{Travel time and  occupancy fluctuate less when users accept short walks.
		}
		\small 
		The average values of travel time $t$ (cf.~main manuscript, Fig.~5, 6b) and the average occupancy $\oav$ have a smaller mean (horizontal bar) with global walk limit $\rglob$ and local walk limit $\rloc$ than with a fixed walk limit $\rt$. Furthermore, $t$ and $\oav$ spread in a smaller range with adapted $\rglob$ and $\rloc$ than with standard ride sharing $\rt=0$.}
	\label{fig:reduced_fluctuations_occ}
\end{figure}

A local adaption of the walk limit lets users walk in those regions with high demand. With an intermediate fixed walk limit, $\rt=\SI{3.75}{\minute}$, and with an globally adapted walk limit $\rglob$ the walk time $\twalk$ is much more homogeneous in space than with a local walk limit $\rloc$ (Fig.~\ref{fig:adapt_r_in_spacetime_supp}). 

An adaption of the walk limit does not only reduce the fluctuations of the travel time (main manuscript, Fig.~6b) but also those of the average occupancy $\oav$ (Fig.~\ref{fig:reduced_fluctuations_occ}). This result suggests that stop pooling enables the usage of smaller vehicles than standard ride sharing.

\null
\newpage

\section*{Supplementary Note 4: Effect of Stop Pooling is Robust}
\label{sec:ASP_robust}
     This note addresses the robustness of the results from the main manuscript. 
     First, users should walk further for higher demand. While this effect holds qualitatively, the best walk limit is quantitatively different for each parameter setting. If one wants to use the scaling of the best walk limit with the demand, one has to analyze it separately for each service. Further research might focus on identifying universal scaling laws.
     Second, we provide an analysis of the fluctuations for a more detailed setting with  constraints. Due to the limited user delay, the travel time is roughly constant. Instead, the rejection rate fluctuates. Stop pooling reduces the fluctuations of the rejection rate.

 \subsection{Robust Result for Different Parameters}
 \label{sec:DSP_Man_rob}

\begin{figure}[b!]   	
	\begin{subfigure}[t]{6.7cm}
		\begin{center}
			\sfl{a}{\hspace{0.25cm}
				\includegraphics{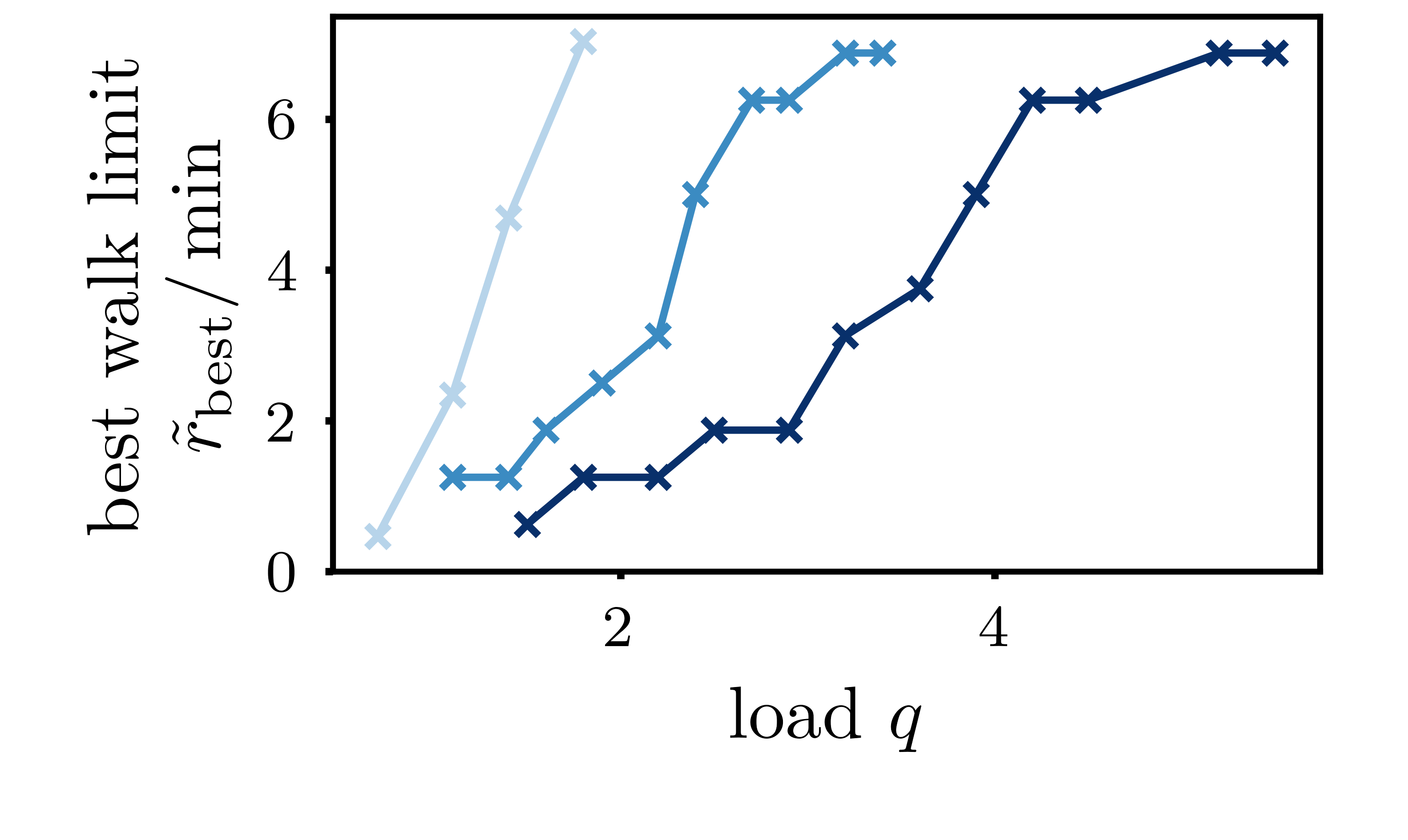}}
		\end{center}
	\end{subfigure}\hfill	
	\begin{subfigure}[t]{8.4cm}
		\begin{center}
			\sfl{b}{\hspace{0.25cm}
				\includegraphics{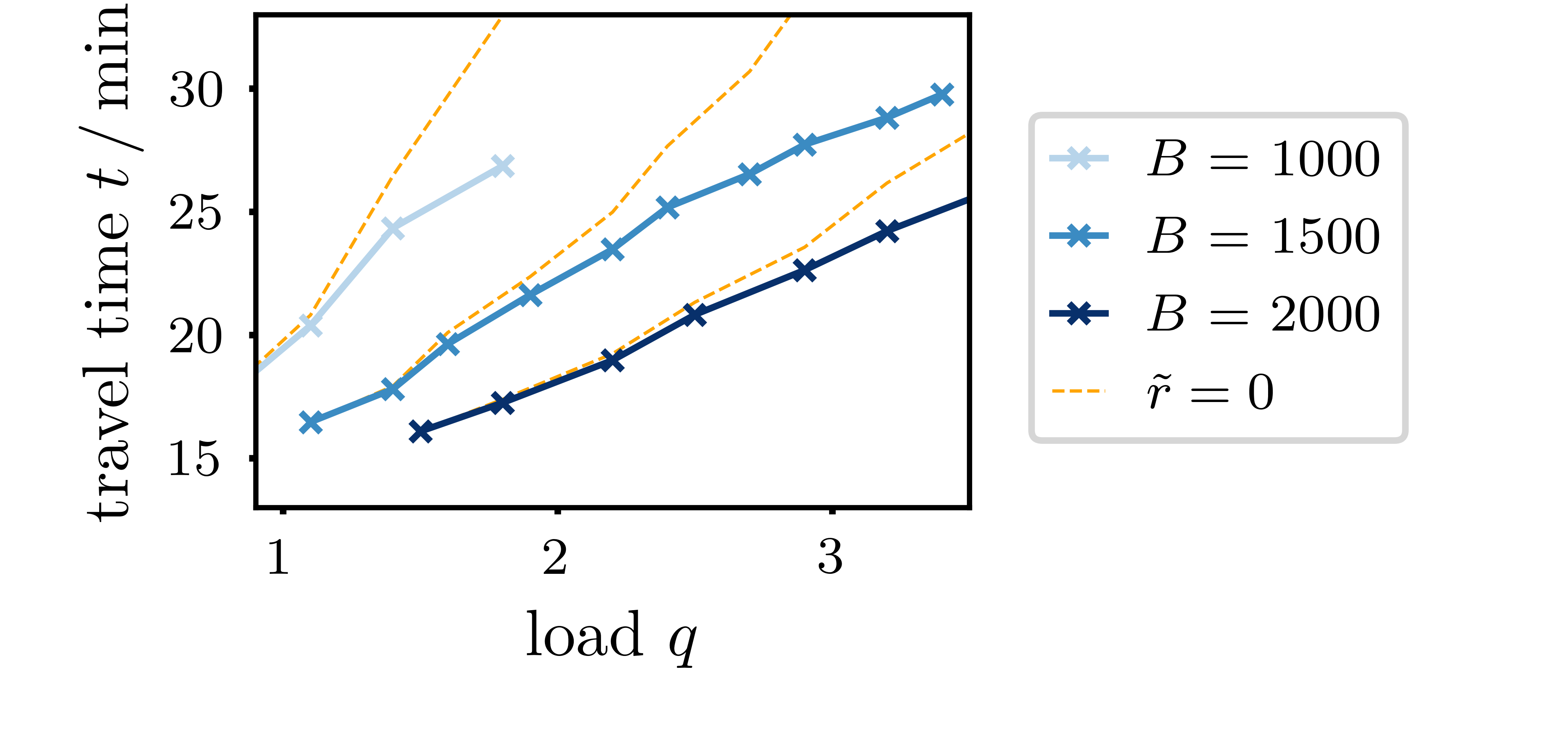}}
		\end{center}
	\end{subfigure}\hfill
	\begin{subfigure}[t]{6.7cm}
		\begin{center}
			\sfl{c}{\hspace{0.25cm}
				\includegraphics{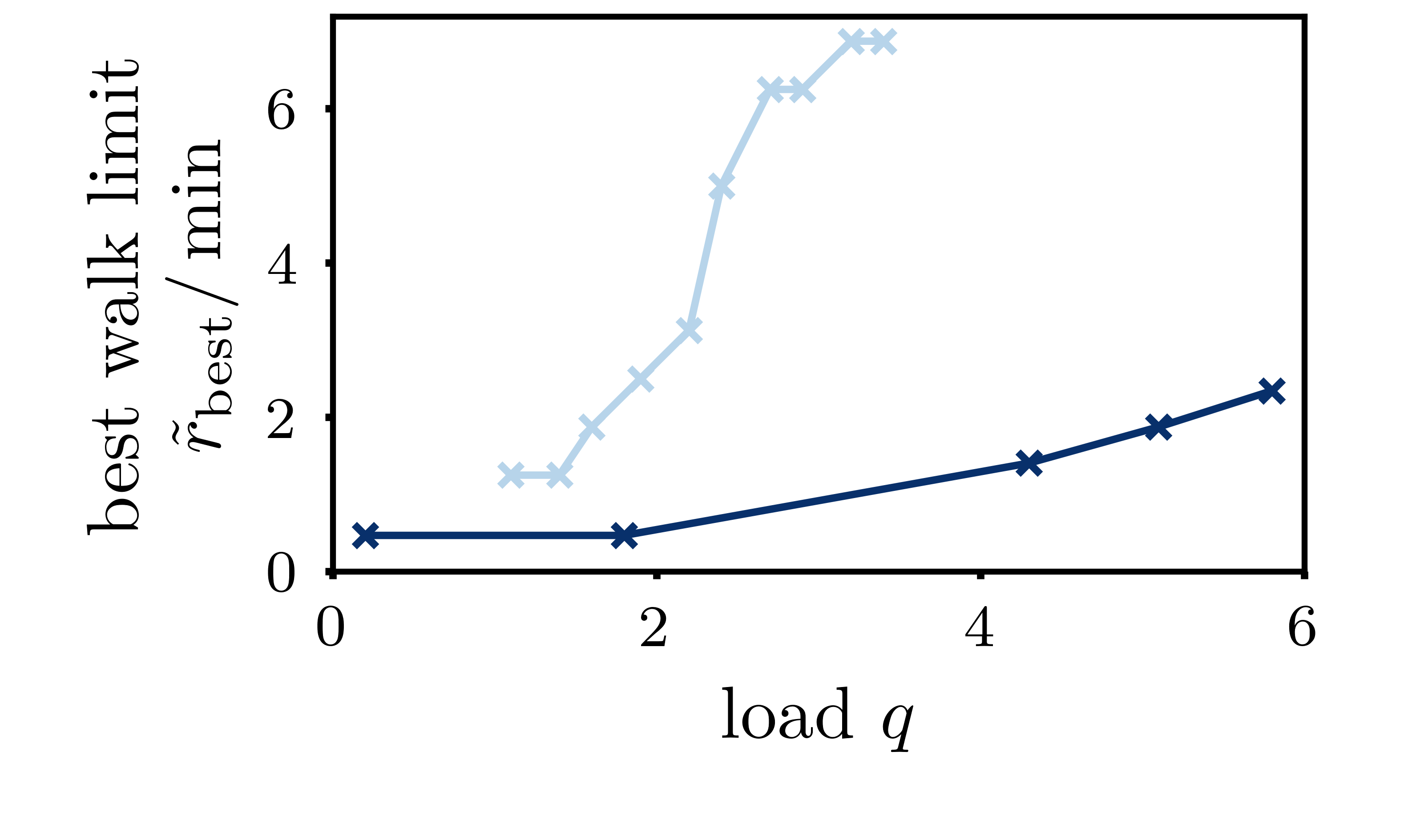}}
		\end{center}
	\end{subfigure}\hfill
	\begin{subfigure}[t]{8.4cm}
		\begin{center}
			\sfl{d}{\hspace{0.25cm}
				\includegraphics{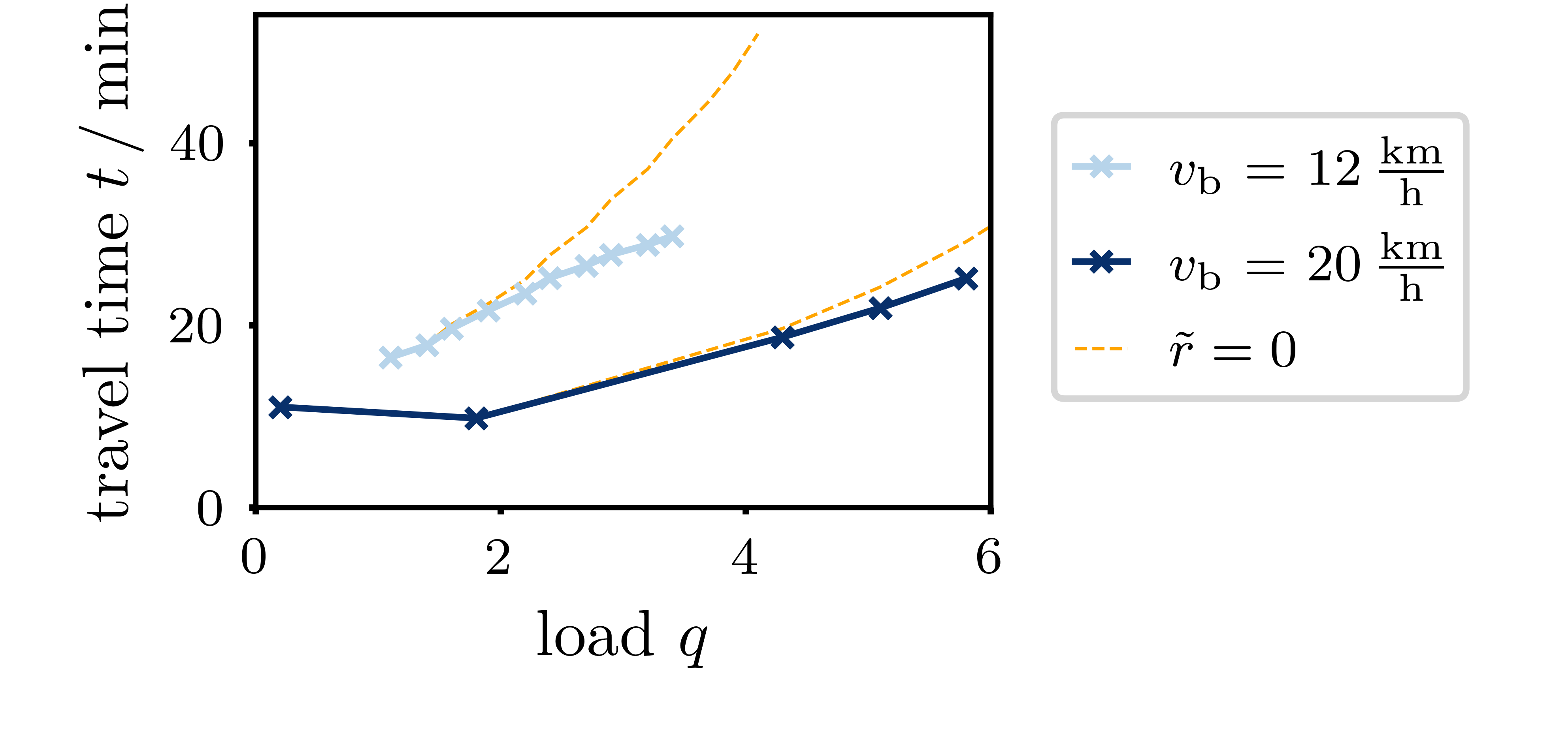}}
		\end{center}
	\end{subfigure}  
	\vspace{-0.5cm} 
	\caption{\small  \textbf{The travel time reduction is robust for different parameters.} \small 
		Stop pooling (solid lines) reduces the travel time $t$ compared to standard ride sharing (orange dashed lines, panels b,d).
		The best walk limit $\rbt$ increases with increasing load $q$ (panels a, c). $\rbt$ increases faster for settings that yield worse travel time $t$  (panels b, d) - for example with smaller fleet sizes $B$ (panels a,b) or bus velocities $\vv$ (panels c,d).}
	\label{fig:dsp_manhattan_robustness}
\end{figure}

Stop pooling reduces the fluctuations of the travel time $t$, because it reduces $t$ the more the higher $t$ without stop pooling  (main manuscript, Fig.~\ref{fig:static_stop_pooling}a). The drive $\tdriv$ and wait time $\twait$ increase with the load $q$.e
For higher $q$, reductions in $\tdriv$ and $\twait$ buffer higher walk times $\twalk$.
The best walk limit $\rbt$ where $t$ is minimal increases with the load $q$. This result is robust with different inputs for  parameters like fleet size $B$ (Fig.~\ref{fig:dsp_manhattan_robustness}a) and velocity (Fig.~\ref{fig:dsp_manhattan_robustness}b). Indeed, the slope of the best walk limit $\rbt$ depends on different parameters.
For smaller fleet sizes $B$ or smaller $\vv$, users should walk further than for larger $B$ or $\vv$ (Fig.~\ref{fig:dsp_manhattan_robustness}). 
The best walk limit $\rbt(q)$ increases faster with smaller $B$ and smaller $\vv$. 
The results are even robust for more complex assignment algorithms with maximal delay for the users and maximal bus capacity (next section).
    \subsection{Robust For Constrained Assignment Algorithm}
    
    This section addresses the robustness of the result with a more complex assignment algorithm that restricts the maximal user delay. So far, the simple assignment algorithm allows arbitrary detour for users which might result in very high delay for some users - much higher than the typical delay that users accept, for example 1.7-2.3 times of direct route in public transportation \cite{vanExel.2010}.
A simple way to avoid unacceptable high detours for users is to restrict the maximal detour $\dtm$. When the assignment algorithm finds no feasible option to serve a user within this time window $\dtm$, the user is rejected. 
     In detail, the assignment algorithm is extended as follows:
    \subsubsection{Limit User Delay}
    \label{ch:constraint}
    The simple assignment algorithms might yield very high travel times for some users.
    A typical procedure to avoid unfeasible long travel times $t$ uses fixed time windows to constrain the user travel time \cite{Mourad.2019, Stiglic.2018,Santos.2015,Li.2018, Muhle.2023}, sometimes even split in a separate wait time window and a drive time window \cite{Ghilas.2016}. 
	To keep it simple, each user is allowed to travel at most their direct trip time plus a fixed maximal delay $\dtm$
	\begin{equation}
		\tau_\mathrm{delivery}-\tau_\mathrm{request}\le \dfrac{\ell}{\vv}+\dtm\,.
	\end{equation}
    To do so, the algorithm checks the currently planned detour of all users 
    that have already been assigned and for the new user for each assignment option and neglects options where at least one user would be delivered to late. This reduces the options to insert new users. 
    If the algorithm does not find any option to serve a user under the delivery constraint, the user is rejected. In the evaluation, rejected users are treated like users that walk their complete trip. This is a simple procedure to include a penalty for rejections.
 	Notably, with a maximal user delay $\dtm$ rebalancing of buses at imbalanced demand is particularly important to avoid high rejection rates (details in \cite[A.10]{Lotze.2023})

    \subsubsection{Fix Bus Capacity}
    \label{ch:capacity}
    A very natural and frequently used constraint is a fixed bus capacity $c$ \cite{Bischoff.2017,Fagnant.2018,Mourad.2019,Santos.2015,AlonsoMora.2017, Wang.2022,Oh.2020,Liu.2019, Muhle.2023} that reflects the limited number of seats in a bus - typically 12 - 14 seats in a minibus and 25 to 60 seats in a bus \cite{Cervero.2000}.
    When a user is assigned to a bus, there must be a seat for the user at any time between pickup and delivery. The extended assignment algorithm with bus capacity $c$ 
    checks the planned bus occupancy between a request's potential pickup and delivery for any assignment option and sorts out all options that exceed the capacity $c$.  Without delivery constraint $\dtm$, the algorithm with capacity constraint does reject users but in the worst case adds them to the end of the bus job queue, which might yield particularly unfeasible travel times. If the system load is higher than the bus capacity, the length of the bus job queue increases continuously with simulation time and simulations never equilibrate. Thus, it makes sense to apply delivery and capacity constraint together. 
    The capacity constraint might not significantly change results if the shareability of the trips lies sufficiently below the constraint (see \cite[A.14]{Lotze.2023}). For example, Ruch \textit{et al.} model ride sharing that is only feasible in small buses with 4 to 6 seats \cite{Ruch.2020}.

\subsubsection{Users Should Walk Further For Higher Demand}
\begin{figure}[b!]   
	\begin{subfigure}[t]{6.7cm}
		\begin{center}
			\sfl{a}{\hspace{0.25cm}
				\includegraphics{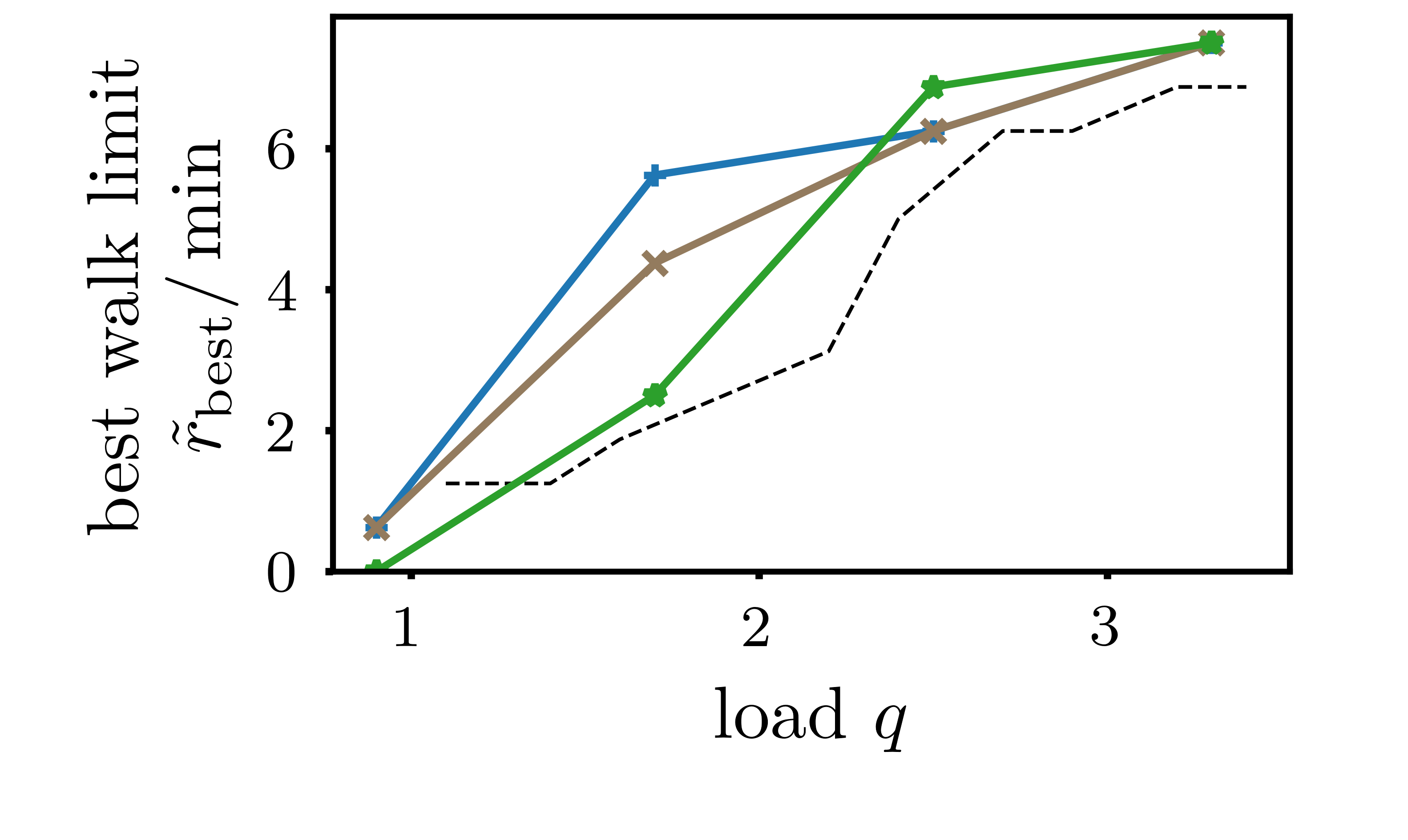}}
		\end{center}
	\end{subfigure}\hfill
	\begin{subfigure}[t]{8.4cm}
		\begin{center}
			\sfl{b}{\hspace{0.25cm}
				\includegraphics{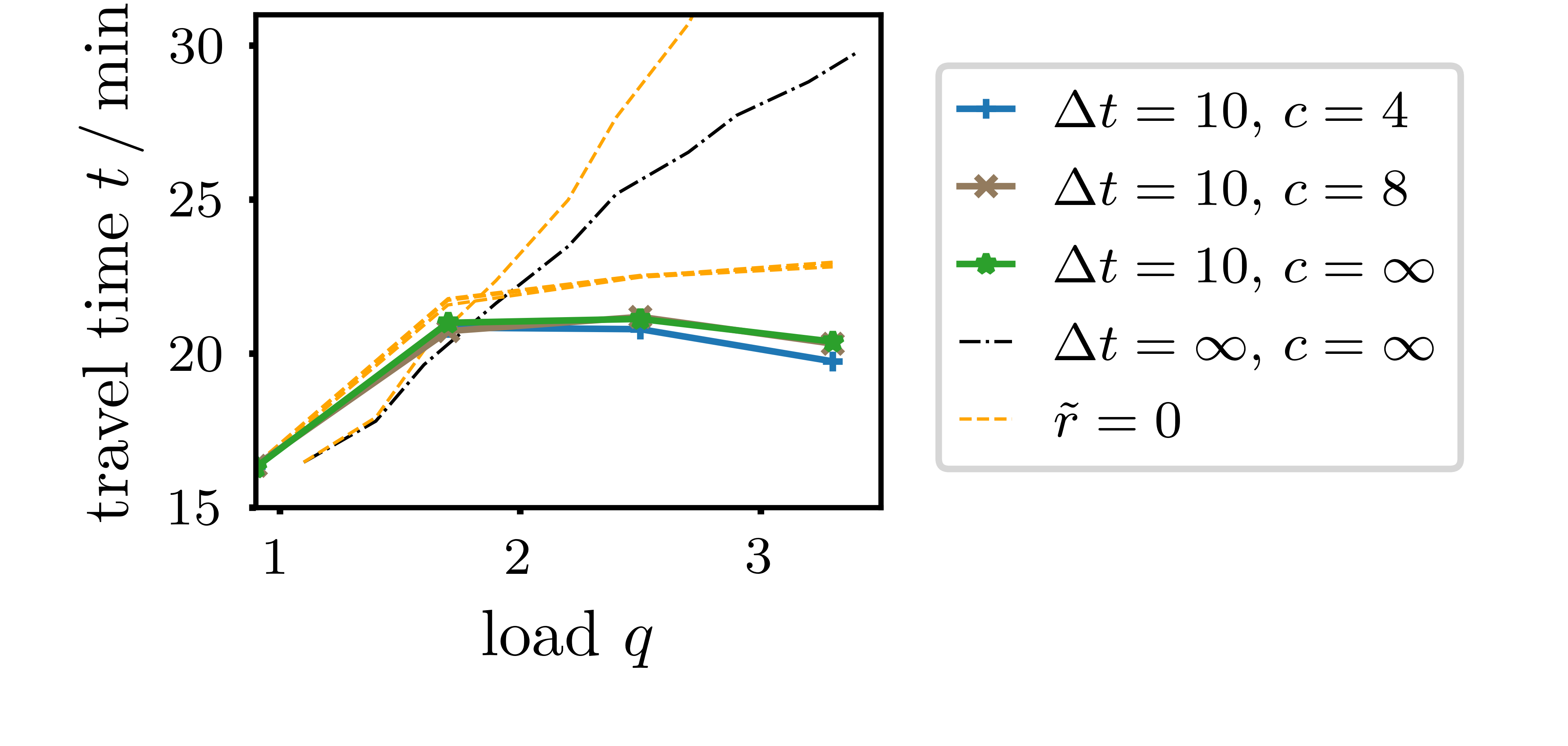}}
		\end{center}
	\end{subfigure}
\vspace{-0.5cm} 
	\caption{\small  \textbf{Best walk limit increases with load robustly for constrained algorithm.} \small 
	Stop pooling reduces the travel time $t$ with respect to to standard ride sharing (orange dashed lines, panel b).
	 The best walk limit $\rbt$ (panel a) and the magnitude of $t$ reduction (panel b) is comparable for the simple algorithm without any constraints (dashed line) and with a more complex algorithm using strong constraints like a maximal user delay $\dtm=10$ min or a bus capacity $c=4$. Generally, constraints reduce the travel time $t$ significantly compared to the simple algorithm without constraint (panel b). However, the specific strength of the constraints presented has minor influence on  $t$.}
\label{fig:dsp_manhattan_robustness_constraint}
\end{figure}
The best walk limit $\rbt$ increases with the load $q$ for more complex assignment algorithms with maximal delay $\dtm$ for the users and maximal bus capacity $c$ (see Supplementary Note 4.I). If users cannot be served within the maximal delay $\dtm$ and maximal bus occupancy $c$, the users are rejected and contribute to the travel time with their complete walk time. With the average direct travel time  $\tind=13$ min and the maximal delay $\dtm=10$ min, the maximal feasible walk time for users that does not violate the maximal delay is 23 min. The corresponding maximal feasible walk limit is thus $\rt=11.5$ min. Consequently, the best walk limit $\rbt$ does not increase linearly with the load $q$, but sublinearly as for the simple model with $\rmax=1$ (Fig.~\ref{fig:dsp_manhattan_robustness_constraint}).

\subsubsection{Stop Pooling Reduces Fluctuations of Rejection Rate in Constrained Model}
\begin{figure}[t!]   	
	\begin{subfigure}[t]{8.7cm}
		\begin{center}
			\sfl{a}{\hspace{0.1cm}
			\includegraphics{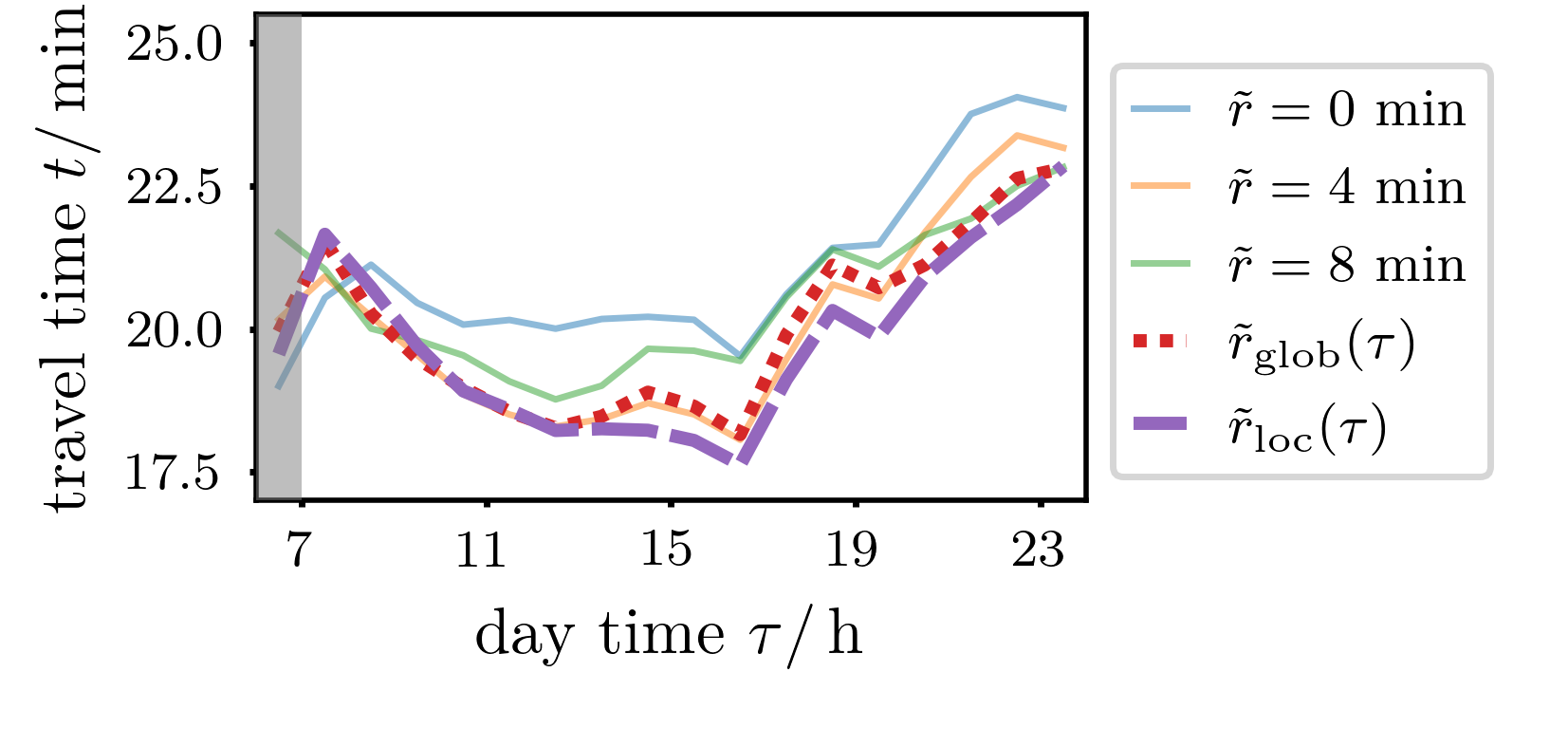}\hspace{-0.1cm}}
		\end{center}
	\end{subfigure}\hfill 	 	
	\begin{subfigure}[t]{6.5cm}
		\begin{center}
			\sfl{b}{
			\includegraphics{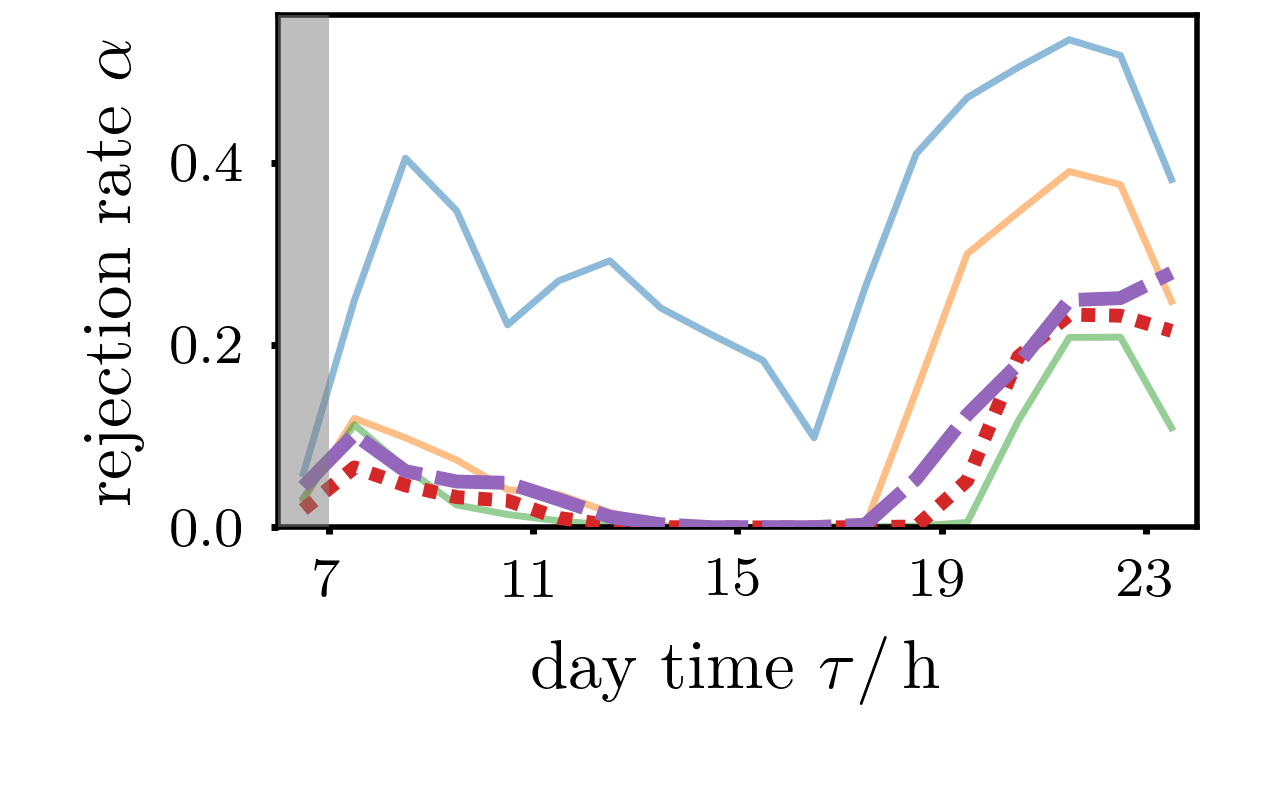}}
		\end{center}
	\end{subfigure}	
	\begin{subfigure}[t]{4.95cm}\vspace{-0.5cm}
		\begin{center}
			\sfl{c}{
			\includegraphics{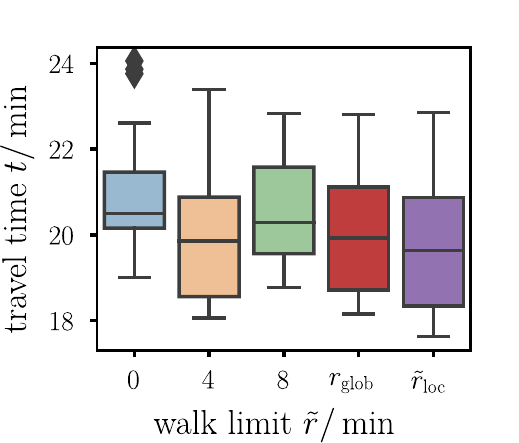}}
		\end{center}
	\end{subfigure}\hfill 	
	\begin{subfigure}[t]{4.95cm}\vspace{-0.5cm}
		\begin{center}\sfl{d}{
			\includegraphics{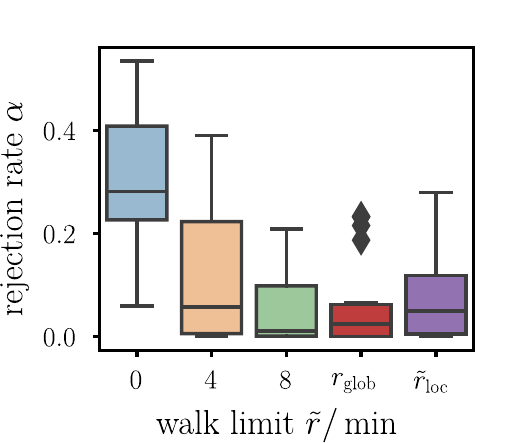}}
		\end{center}
	\end{subfigure}\hfill 	
\begin{subfigure}[t]{4.95cm}\vspace{-0.5cm}
\begin{center}\sfl{e}{
	\includegraphics{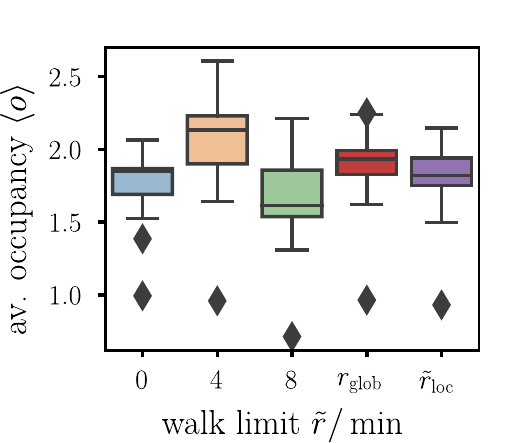}}
\end{center}
\end{subfigure}
	\caption[Adaptive stop pooling reduces rejection rate in more complex model.]{\small  \textbf{Adaptive stop pooling reduces rejection rate in more complex model.} \small  With a maximal delay of $\dtm=\SI{10}{\minute}$ the travel time $t$ is smaller than with the simple algorithm (panel a).
	Adaptive stop pooling  yields shorter travel times $t$ than fixed $\rt$ for most times $\tau$ with a global walk limit $\rglob(q)$ and even slightly better travel times $t$ with a local walk limit $\rloc(\tau)$  (panel a). At the same time, adaptive stop pooling reduces the rejection rate $\alpha$, but does not reach the minimal rejection rate $\alpha$ with $\rt=\SI{8}{\minute}$ (panel b).
	The average values of travel time have a small mean but not the smallest spread  with global walk limit $\rglob$ and local walk limit $\rloc$ (panel c). However, the spread of the rejection rate reduces (panel d) without higher average travel time $t$ (as for $\rt=\SI{8}{\minute}$.)
	The average occupancy  is only slightly higher with global walk limit $\rglob$ and local walk limit $\rloc$ than with $\rt=\SI{8}{\minute}$ (panel e).}
	\label{fig:ASP_window}
\end{figure}
A constrained algorithm changes the observables, because users are rejected if the fleet is not able to serve them within the restrictions. In this section, we provide one example setting with a strong limited delay $\dtm=\SI{10}{\minute}$. In this setting, stop pooling does not reduce fluctuations of the travel time, but of the rejection rate - a new important observable for the service quality. 

First, 
the user travel times $t$ reduce significantly with a limited delay $\dtm=\SI{10}{\minute}$ compared to the simple assignment algorithm (Fig.~\ref{fig:ASP_window}a). In return, up to half of the users is rejected (Fig.~\ref{fig:ASP_window}b). Notably, the travel time $t$ takes the walk time of rejected users into account and is still smaller than with standard ride sharing, because in the model users walk on shorter routes than buses drive (details see \cite[A.15]{Lotze.2023}). Stop pooling reduces the travel time $t$ only slightly, but strongly reduces the rejection rate $\alpha$  (Fig.~\ref{fig:ASP_window} upper panels). With stop pooling, some users with short trip length $\ell<2r$ walk their complete trip. These users are effecitvely also not served, but in difference to the rejected users, it is ensured that the completely walking users reach their destination in a feasible time by walking.
But different fixed $\rt$ yield the best travel time at different times of day. 
Adapting the walk limit  (i) globally $\rglob$ (Eq.~\eqref{eq:rglob}) and (ii) locally $\rloc$ (Eq.~\eqref{eq:rloc}) reduces the travel time $t$ to a minimum while also reducing the rejection rate. However, the procedures do not yield minimal rejection rates $\alpha$. The presented procedures minimize the travel time with the simple assignment algorithm.
To achieve minimal rejection rates one might develop a different procedure more suitable for this special purpose.
\null\newpage

\end{document}